\documentclass[a4paper,11pt]{article}

\usepackage{jheppub}
\makeatletter
\let\@fpheader      % To remove the JHEP header
\makeatother

\usepackage[utf8]{inputenc}
\usepackage[american]{babel}
\usepackage{verbatim}
\usepackage{xcolor}
\usepackage{subcaption}
\usepackage{multicol}
\usepackage{booktabs}
\usepackage{bigdelim}

\usepackage[normalem]{ulem}

\usepackage{booktabs}
\usepackage{array}
\newcolumntype{C}{>{$}c<{$}} % math-mode version of "c" column type
\newcolumntype{R}{>{$}r<{$}} % math-mode version of "r" column type
\newcolumntype{L}{>{$}l<{$}} % math-mode version of "l" column type

\usepackage{hyperref}
\hypersetup{%
    colorlinks,%
    breaklinks,%
    citecolor=blue,urlcolor=blue,linkcolor=blue,%
    hypertexnames=false,%
    pdftitle={The three-pion K-matrix at NLO in ChPT},%
    pdfauthor={J. Baeza-Ballesteros, J. Bijnens, T. Husek, F. Romero-Lopez, S.R. Sharpe, M. Sjo}%
}
\makeatletter
\renewcommand\HyPsd@CatcodeWarning[1]{}     % To remove some warnings
\makeatother

\usepackage{cleveref}
\crefname{equation}{eq.}{eqs.}
\crefname{section}{sec.}{secs.}
\newcommand{\rcite}[1]{ref.~\cite{#1}}
\newcommand{\rrcite}[1]{refs.~\cite{#1}}

\usepackage{glossaries-extra}
\setabbreviationstyle[acronym]{long-short}
\glssetcategoryattribute{acronym}{nohyperfirst}{true}

\let\Re\relax\let\Im\relax
\DeclareMathOperator{\Re}{Re}
\DeclareMathOperator{\Im}{Im}

\DeclareMathOperator*{\diag}{diag}

\usepackage{pgf,tikz}
\usetikzlibrary{%
    calc,
    decorations.pathreplacing,
    decorations.markings,
    decorations.pathmorphing,
    shapes.geometric,
    external%
}
\usepackage{pgfplots}
\pgfplotsset{compat=1.17}   % To remove some warnings
\newcommand{\arrowscale}{1.2}
\newcommand{\diagramxscale}{1.4}
\newcommand{\diagramyscale}{1}
\usepgfplotslibrary{fillbetween}
\usetikzlibrary{decorations.text}
\tikzset{
    on each segment/.style={
        decorate,
        decoration={
              show path construction,
              moveto code={},
              lineto code={
                \path [#1]
                (\tikzinputsegmentfirst) -- (\tikzinputsegmentlast);
              },
              curveto code={
                    \path [#1] (\tikzinputsegmentfirst)
                    .. controls
                    (\tikzinputsegmentsupporta) and (\tikzinputsegmentsupportb)
                    ..
                    (\tikzinputsegmentlast);
              },
              closepath code={
                    \path [#1]
                    (\tikzinputsegmentfirst) -- (\tikzinputsegmentlast);
              },
        },
    },
    mid arrow/.style={postaction={decorate,decoration={
        markings,
        mark=at position #1 with {\arrow[scale=\arrowscale]{stealth}}
      }}},
    mid arrow/.default=0.55,
    mid barrow/.style={postaction={decorate,decoration={
        markings,
        mark=at position #1 with {\arrowreversed[scale=\arrowscale]{stealth}}
      }}},
    mid barrow/.default=0.45,
    prop/.style={thick,join=round},
    dprop/.style={prop,postaction={on each segment={mid arrow=#1}}},
    bdprop/.style={prop,postaction={on each segment={mid barrow=#1}}},
    sketch onshell prop/.style={very thick},
    sketch offshell prop/.style={sketch onshell prop, densely dotted},
    sketch blob/.style={draw=black,thick, fill=#1, anchor=center},
    sketch onshell blob/.style={sketch blob=#1,
                                shape=regular polygon, regular polygon sides=4},
    sketch offshell blob/.style={sketch blob=#1,
                                shape=ellipse}
}
\newcommand{\makeexternallegcoordinates}{%
    \coordinate (k1) at (+1,+1);%
    \coordinate (k2) at (+1, 0);%
    \coordinate (k3) at (+1,-1);%
    \coordinate (p1) at (-1,+1);%
    \coordinate (p2) at (-1, 0);%
    \coordinate (p3) at (-1,-1);%
    }
\newcommand{\makeexternallegs}{%
    \makeexternallegcoordinates
    \draw (k1) node [right] {$k_1$};%
    \draw (k2) node [right] {$k_2$};%
    \draw (k3) node [right] {$k_3$};%
    \draw (p1) node [left] {$p_1$};%
    \draw (p2) node [left] {$p_2$};%
    \draw (p3) node [left] {$p_3$};%
    }\newcommand{\makeexternallegsshifted}{%
    \makeexternallegcoordinates
    \draw (k1) node [right] {$k_3$};%
    \draw (k2) node [right] {$k_1$};%
    \draw (k3) node [right] {$k_2$};%
    \draw (p1) node [left] {$p_1$};%
    \draw (p2) node [left] {$p_2$};%
    \draw (p3) node [left] {$p_3$};%
    }

\definecolor{plotI}{HTML}{0077BB}
\definecolor{plotII}{HTML}{33BBEE}
\definecolor{plotIII}{HTML}{009988}
\definecolor{plotIV}{HTML}{EE7733}
\definecolor{plotV}{HTML}{CC3311}
\definecolor{plotVI}{HTML}{DDAA33}
\definecolor{plotVII}{HTML}{AA3377}

\newcommand{\sketchoperatorscale}{1.6}
\definecolor{LOcolor_}{HTML}{6699CC}
\definecolor{NLOcolor_}{HTML}{EE99AA}
\colorlet{LOcolor}{LOcolor_!70}
\colorlet{NLOcolor}{NLOcolor_!70}

\definecolor{fitorange}{HTML}{FFA500}
\definecolor{fitblue}{HTML}{0000FF}
\definecolor{fitgray}{HTML}{808080}

\pgfmathsetlengthmacro\MajorTickLength{
      \pgfkeysvalueof{/pgfplots/major tick length} * 0.4
    }
\pgfmathsetlengthmacro\MinorTickLength{
      \pgfkeysvalueof{/pgfplots/minor tick length} * 0.3
    }

\newcommand{\addplotPartialWave}[3]{%
	\addplot[#1waveline] table[x index=1, y index=#2, col sep=tab]%
        {Data340/NumericKdf_OPEPartialWaves/#3};%
    \IfStrEq{#1}{all}{%
        \addlegendentry{Total}%
        }{%
        \pgfmathparse{int(#2-2)}%
        \addlegendentryexpanded{$\ell=\pgfmathresult$}%
        }%
    }
\newcommand{\plotPartialWaves}[2][spdf]{%
    \inelasticLine%
    \addplotPartialWave{all}{13}{#2}%
    \IfSubStr{#1}{s}{\addplotPartialWave{s}{2}{#2}}{}%
    \IfSubStr{#1}{p}{\addplotPartialWave{p}{3}{#2}}{}%
    \IfSubStr{#1}{d}{\addplotPartialWave{d}{4}{#2}}{}%
    \IfSubStr{#1}{f}{\addplotPartialWave{f}{5}{#2}}{}%
    \IfSubStr{#1}{g}{\addplotPartialWave{g}{6}{#2}}{}%
    \IfSubStr{#1}{h}{\addplotPartialWave{h}{7}{#2}}{}%
    }

\newcommand{\inelasticLine}[1][30]{%
    \addplot[black,thin,mark=none,forget plot] coordinates {(-1,0) (2,0)};%
    \addplot+[black,mark=none,dashed,forget plot] coordinates {(\inelasticthr,-#1) (\inelasticthr,+#1)};%
    }

\newcommand{\plotConvergence}[3]{
    \addplot[#1line, numline] table[x index=1, #2,  col sep=tab] {Data340/NumericKdf/#3};%
    \addplot[#1line, thrline] table[x index=1, #2,  col sep=tab] {Data340/ThresholdKdf/#3};%
}
\newcommand{\plotnonOPE}[1]{%
    \plotConvergence{nonOPE}{y index=2}{#1}%
}
\newcommand{\plotOPE}[1]{%
    \plotConvergence{OPE}{y index=4}{#1}%
}
\newcommand{\plotBH}[1]{%
    \plotConvergence{BH}{y index=3}{#1}%
}
\newcommand{\plotsOPE}[2][1]{%
    \plotConvergence{sOPE}{y expr={#1*\thisrowno{5}}}{#2}%
}
\newcommand{\plotTotal}[1]{%
    \plotConvergence{total}{y index=6}{#1}%
}

\tikzset{
    LOdash/.style={dash pattern=on 3pt off 2.5pt},
    fitdash/.style={dash pattern=on 1pt off 1.3pt},
    zeroline/.style={mark=none, black, ultra thin},   
    verticalline/.style={thin, black, dash pattern=on 3pt off 3.5pt},
    DXline/.style={ultra thick, mark=none},
    KXline/.style={DXline},
    LOline/.style={thick, mark=none},
    NLOline/.style={ultra thick, mark=none},
    errline/.style={line width=0pt, draw=none},
    errfill/.style={solid, fill, opacity=0.3},
    dash1/.style={dash pattern=on 8pt off 2pt on 2pt off 2pt},
    dash2/.style={dash pattern=on 6pt off 2pt on 2pt off 2pt on 2pt off 2pt},
    dash3/.style={dash pattern=on 5.5pt off 1.5pt on 1.5pt off 1.5pt on 1.5pt off 1.5pt on 1.5pt off 1.5pt},
    dash4/.style={dash pattern=on 5.2pt off 1.2pt on 1.2pt off 1.2pt on 1.2pt off 1.2pt on 1.2pt off 1.2pt on 1.2pt off 1.2pt},
    dash5/.style={dash pattern=on 5.2pt off 0.9pt on 0.9pt off 0.9pt on 0.9pt off 0.9pt on 0.9pt off 0.9pt on 0.9pt off 0.9pt},
    dashA/.style={dash pattern=on 4.4pt off 4.4pt},
    dashB/.style={dash pattern=on 2pt off 2pt},
    line0/.style={plotI,   solid},
    line1/.style={plotII,  dash1},
    line2/.style={plotIII, dash2},
    lineA/.style={plotIV,  dashA},
    lineB/.style={plotV,   dashB},
    line3/.style={plotVI,  dash3},
    line4/.style={plotVII, dash4},
    line5/.style={plotVII, dash5},
    D0line/.style={DXline, line0},
    D1line/.style={DXline, line1},
    D2line/.style={DXline, line2},
    DAline/.style={DXline, lineA},
    DBline/.style={DXline, lineB},
    D3line/.style={DXline, line3},
    D4line/.style={DXline, line4},
    D5line/.style={DXline, line5},
    allwaveline/.style={mark=none,line width=1.2mm, fitgray, dash pattern=on 1.2mm off 1mm},
    swaveline/.style={D0line},
    pwaveline/.style={D1line},
    dwaveline/.style={D2line},
    fwaveline/.style={D3line},
    gwaveline/.style={D4line},
    hwaveline/.style={D5line},
    numline/.style={KXline, plotI},
    thrline/.style={KXline, plotV},
	nonOPEline/.style={KXline, solid},
    BHline/.style={KXline, dash1},
	OPEline/.style={KXline, dashA},
	sOPEline/.style={KXline, dashB},
	totalline/.style={allwaveline}
}
\newcommand{\inelasticthr}{1.6666666666666666666666}
\usetikzlibrary{positioning}
\pgfkeys{/pgf/number format/1000 sep={}}
\pgfplotsset{
    general logplot/.style={
        set layers=axis on top,
        width=1.0\textwidth,
        height=0.75\textwidth,
        minor tick num=4,
        every tick/.style={
                semithick,
            },
        major tick length=\MajorTickLength,
        minor tick length=\MinorTickLength,
        every tick label/.append style={font=\footnotesize},
        axis line style={black, semithick},
        xticklabel style={inner sep=1.5pt},
        yticklabel style={inner sep=1.5pt},
            },
    general plot/.style={
        general logplot,
        y tick label style={
                /pgf/number format/.cd,
                fixed,
             fixed zerofill,
                precision=1,
                /tikz/.cd
                },
            },
    fit plot/.style={
        general plot,
        ylabel style = {rotate=-90}, ylabel shift=-1.1ex,
        legend style={font=\footnotesize}
    },
    plot MF4/.style={
        fit plot,
        width=.57\textwidth,
        height=.5\textwidth,
        xmin=-0.01, xmax=159.9, xtick distance=50, restrict x to domain=-1:180,
        legend pos=north west, legend columns=2
    },
    plot MF4 zoomed/.style={
        plot MF4, 
        scaled y ticks=false, % Remove those annoying 10^-2
        xmax=16, xtick distance=5, restrict x to domain=-1:20
    },
    plot threshold/.style={
        plot MF4, 
        wide legend,
        scaled y ticks=false, % Remove those annoying 10^-2
        xmax=2, xtick distance=0.5, restrict x to domain=-1:3,
        legend style={name=legend},
    },
    plot convergence/.style={
        plot MF4, 
        wide legend,
        scaled y ticks=false, % Remove those annoying 10^-2
        xmax=2, xtick distance=0.5, restrict x to domain=-1:3,
        legend style={name=legend},
        legend columns=1
    },
    precise y/.style={
        y tick label style={
                /pgf/number format/precision=2,
            }
    },
    cutoff plot modifications/.style={
        width=.57\textwidth,
        height=.52\textwidth,
        legend pos=north east,
         every tick label/.append style={font=\footnotesize},
        legend style={font=\footnotesize},
        ylabel style = {rotate=-90}, ylabel shift=-2ex,
        xticklabels={,$-1.0$,$-0.5$,$0.0$,$0.5$,$1.0$,$1.5$},
        xlabel={$\alpha$},
        x tick label style={
            /pgf/number format/.cd,
            fixed,
         fixed zerofill,
            precision=1,
            /tikz/.cd
            },
        major tick length=2*\MajorTickLength,
        minor tick length=2*\MinorTickLength,
    },
    no ticks/.style={
        xticklabels={},
        xlabel={}
    },
    cutoff plot/.style={
        general plot,
        cutoff plot modifications
    },
    cutoff logplot/.style={
        general logplot,
        cutoff plot modifications
    },
    legend image code/.code={
        \draw [#1] (0pt,0pt) -- (15pt,0pt);
    },
    wide legend/.style={legend image code/.code={
        \draw [#1] (0pt,0pt) -- (22pt,0pt);
    }},
    NLO legend/.style={legend image code/.code={
        \draw [#1] (0pt,0pt) -- (22pt,0pt);
        \path [#1, draw=none, errfill] (0pt,-4pt) rectangle (22pt,4pt);
    }},
    LO plus NLO legend/.style={legend image code/.code={
        \draw [#1, LOline] (0pt,5pt) -- (22pt,5pt);
        \draw [#1, NLOline] (0pt,-2pt) -- (22pt,-2pt);
        \path [#1, draw=none, errfill] (0pt,-6pt) rectangle (22pt,2pt);
    }},
    exact NLO legend/.style={legend image code/.code={
        \draw [#1] (0pt,0pt) -- (22pt,0pt);
    }},
    backbone legend/.style={legend image code/.code={
        \fill [#1, draw=none, opacity=.5] (0pt,3pt) rectangle (22pt, -3pt);
        \draw [#1] (0pt,3pt) -- (22pt,3pt);
    }}
}

\newcommand{\leftfudge}{\hspace{-.4cm}}
\newcommand{\midfudge}{\hspace{0cm}}
\newcommand{\figscale}{0.95}

\newcommand{\tikzineq}[2][]{%
	\tikz[%
		anchor=base,%
		baseline={([yshift=-1.0ex]current bounding box.center)},%
		#1]{#2}%
	}

\newcommand{\cD}{\mathcal{D}}

\newcommand{\cK}{\mathcal{K}}

\newcommand{\cM}{\mathcal{M}}
\newcommand{\cO}{\mathcal{O}}

\renewcommand{\d}{\mathrm{d}}
\newcommand{\bm}{\boldsymbol}
\newcommand{\bmh}[1]{\hat{\bm #1}}
\newcommand{\m}{\mathbf}   %  matrix

\newcommand{\inxout}{(\text{in}\leftrightarrow\text{out})}

\newcommand{\ket}[1]{|#1\rangle}
\newcommand{\pip}{\pi^+}
\newcommand{\pim}{\pi^-}
\newcommand{\pio}{\pi^0}
\newcommand{\pipm}{\pi^\pm}
\newcommand{\roo}{\rho^0}
\newcommand{\rop}{\rho^+}
\newcommand{\rom}{\rho^-}
\newcommand{\ropm}{\rho^\pm}
\newcommand{\Pio}{\Pi^0}
\newcommand{\Pip}{\Pi^+}
\newcommand{\Pim}{\Pi^-}

\newcommand{\Pipm}{\Pi^\pm}

\newcommand{\CI}{\mathcal C_{\mathrm{I}}}
\newcommand{\CS}{\mathcal C_{\mathrm{S}}}
\newcommand{\Ccom}{\mathcal C}

\newcommand{\logthree}{\log3}

\newcommand{\uu}{{(u,u)}}
\newcommand{\df}{\mathrm{df}}
\newcommand{\Mdf}{\cM_{\df,3}}
\newcommand{\Kdf}{\cK_{\df,3}}

\newcommand{\mMdf}{\m M_{\df,3}}
\newcommand{\mMdfuu}{\mMdf^\uu}
\newcommand{\mKdf}{\m K_{\df,3}}

\newcommand{\mDBH}{\m D^\BH}
\newcommand{\mDBHuu}{\m D^{\uu\BH}}

\newcommand{\on}{{\mathrm{on}}}
\newcommand{\off}{{\mathrm{off}}}

\newcommand{\Mpi}{M_\pi}

\newcommand{\Fpi}{F_\pi}
\newcommand{\Fphys}{F_{\pi,\text{phys}}}

\newcommand{\tFM}[1]{\tfrac{\Fpi^{#1}}{\Mpi^{#1}}}

\newcommand{\Kiso}{{\cK_0}}
\newcommand{\Kisoone}{{\cK_1}}
\newcommand{\Kisotwo}{{\cK_2}}
\newcommand{\KA}{{\cK_\mathrm{A}}}  
\newcommand{\KB}{{\cK_\mathrm{B}}}

\newcommand{\Diso}{{\cD_0}}
\newcommand{\Disoone}{{\cD_1}}
\newcommand{\Disotwo}{{\cD_2}}
\newcommand{\DA}{{\cD_\mathrm{A}}}
\newcommand{\DB}{{\cD_\mathrm{B}}}

\newcommand{\DeltaA}{{\Delta_\mathrm{A}}}
\newcommand{\DeltaB}{{\Delta_\mathrm{B}}}

\newcommand{\tij}[1]{\tilde t_{#1}}

\newacronym{CMF}{CMF}{center-of-momentum frame}
\newcommand{\LO}{\mathrm{LO}}
\newcommand{\NLO}{\mathrm{NLO}}
\newcommand{\OPE}{\mathrm{OPE}}
\newcommand{\sOPE}{s\text{-}\OPE}
\newcommand{\nonOPE}{{\mathrm{non}\text{-}\OPE}} % Puttng most of it in mathrm rather than text ensures is stays non-bold in headers
\newcommand{\BH}{\mathrm{BH}}

\newcommand{\mKdfI}[1]{\m{K}_{\df,3}^{[I=#1]}}
\newcommand{\mXi}[1]{\boldsymbol{\Xi}_{#1}}
\newcommand{\mDuuBH}{\m D^{\uu\mathrm{BH}}}

\newcommand{\vxi}[1][]{{\vec\xi}^{\,#1}}
\newcommand{\xip}[1][]{\xi^{\prime #1}}
\newcommand{\vxip}[1][]{\vec\xi^{\,\prime #1}}
\newcommand{\vxiS}[1][]{\vec\xi(S)^{#1}}

\newcommand{\vxiSb}[1][]{\vec\xi(\bar S)^{#1}}
\newcommand{\vxipS}[1][]{\vec\xi^{\,\prime}(S)^{#1}}

\newcommand{\vxipSb}[1][]{\vec\xi^{\,\prime}(\bar S)^{#1}}

\makeatletter
\newcommand{\presentxi}{\@ifstar{\align@presentxi}{\@presentxi}}
\newcommand{\@presentxi}[4][]{%
    \vxi[\if!#2!\else(#2)\fi #1] = \big(\ensuremath{\xi}^{\if!#2!\else(#2)\fi #1}_1,\ensuremath{\xi}^{\if!#2!\else(#2)\fi #1}_2\big)\,,%
    \qquad\text{where}%
    \quad\ensuremath{\xi}^{\if!#2!\else(#2)\fi}_1\equiv #3\,,%
    \quad\ensuremath{\xi}^{\if!#2!\else(#2)\fi}_2\equiv #4}
\newcommand{\align@presentxi}[4][]{%
    \quad\ensuremath{\xi}^{\if!#2!\else(#2)\fi}_1&\equiv #3\,,%
    &\quad\ensuremath{\xi}^{\if!#2!\else(#2)\fi}_2&\equiv #4}
\makeatother

\newcommand{\A}{\mathrm{A}}

\newcommand{\T}{\mathrm{T}}
\renewcommand{\SS}{\mathrm{SS}}
\newcommand{\SD}{\mathrm{SD}}
\newcommand{\DS}{\mathrm{DS}}
\newcommand{\DD}{\mathrm{DD}}
\newcommand{\AS}{\mathrm{AS}}

\newcommand{\DeltaAS}[1]{\Delta_\AS^{(#1)}}

\newcommand{\KX}[2][]{\cK^{#2}_{#1}}
\newcommand{\KT}[1][0]{\KX[#1]{\T}}
\newcommand{\KSS}[1][0]{\KX[#1]{\SS}}
\newcommand{\KSD}[1][0]{\KX[#1]{\SD}}

\newcommand{\KDD}[1][0]{\KX[#1]{\DD}}
\newcommand{\KSSA}{\KSS[\mathrm A]}
\newcommand{\KSSB}{\KSS[\mathrm B]}
\newcommand{\KAS}[1][0]{\KX[#1]{\AS}}

\newcommand{\DX}[2][]{\cD^{#2}_{#1}}
\newcommand{\DT}[1][0]{\DX[#1]{\T}}
\newcommand{\DSS}[1][0]{\DX[#1]{\SS}}
\newcommand{\DSD}[1][0]{\DX[#1]{\SD}}
\newcommand{\DDD}[1][0]{\DX[#1]{\DD}}
\newcommand{\DSSA}{\DSS[\mathrm A]}
\newcommand{\DSSB}{\DSS[\mathrm B]}
\newcommand{\DAS}[1][0]{\DX[#1]{\AS}}

\newcommand{\Sc}[1][]{c^\mathrm{S}_{#1}}
\newcommand{\Dc}[1][]{c^\mathrm{D}_{#1}}

\newcommand{\lr}[1]{\ell_{#1}^\mathrm{r}}
\newcommand{\kfrac}[2][1]{\tfrac{#1}{#2}\kappa}
\newcommand{\Lfrac}[2][1]{\tfrac{#1}{#2}L}
\newcommand{\lrfrac}[3][1]{\tfrac{#1}{#2}\lr{#3}}

\newcommand{\nq}[2]{n_{#1}^{(#2)}}
\newcommand{\n}[1]{n_{#1}}

\newcommand{\HSQCa}[0]{Hansen:2014eka}
\newcommand{\HSQCb}[0]{Hansen:2015zga}

\newcommand{\dwave}[0]{Blanton:2019igq}

\newcommand{\isospin}[0]{Hansen:2020zhy}

\preprint{{\small MIT-CTP/5670}}

\title{The three-pion $K$-matrix at NLO in ChPT}

\author[a]{Jorge Baeza-Ballesteros,}
\author[b]{Johan Bijnens,}
\author[c,d]{Tom\'a\v s Husek,}
\author[e]{Fernando Romero-L\'opez,}
\author[f]{Stephen R.\ Sharpe,}
\author[b,g]{and Mattias Sj\"o}

\affiliation[a]{IFIC, CSIC-Universitat de València, 46980 Paterna, Spain}
\affiliation[b]{Department of Physics, Lund University, Box 118, SE 22100 Lund, Sweden}
\affiliation[c]{Institute of Particle and Nuclear Physics, Charles University,\\
	V Hole\v sovi\v ck\'ach 2, 180 00 Prague, Czech Republic}
\affiliation[d]{School of Physics and Astronomy, University of Birmingham,\\
    Edgbaston, Birmingham, B15 2TT, UK}
\affiliation[e]{CTP, Massachusetts Institute of Technology, Cambridge, MA 02139, USA}
\affiliation[f]{Physics Department, University of Washington, Seattle, WA 98195-1560, USA}
\affiliation[g]{Aix Marseille Univ, Université de Toulon, CNRS, CPT, Marseille, France}

\emailAdd{jorge.baeza@ific.uv.es}
\emailAdd{johan.bijnens@hep.lu.se}
\emailAdd{tomas.husek@matfyz.cuni.cz}
\emailAdd{fernando@mit.edu}
\emailAdd{srsharpe@uw.edu}
\emailAdd{mattias.sjo@hep.lu.se}

\abstract{
The three-particle $K$-matrix, $\mathcal{K}_{\mathrm{df},3}$, is a scheme-dependent quantity that parametrizes short-range three-particle interactions in the relativistic-field-theory three-particle finite-volume formalism. In this work, we compute its value for systems of three pions in all isospin channels through next-to-leading order in Chiral Perturbation Theory, generalizing previous work done at maximum isospin. We obtain analytic expressions through quadratic order (or cubic order, in the case of zero isospin) in the expansion about the three-pion threshold.
}

\keywords{Chiral Lagrangian, Hadron Spectroscopy, Structure and Interactions, Lattice QCD}

\begin{document}

\maketitle
\flushbottom

\section{Introduction}

First-principles studies of three-hadron physics from Quantum Chromodynamics (QCD) are finally becoming possible after a number of theoretical, numerical, and algorithmic developments~\cite{Beane:2007qr,Detmold:2008gh,Briceno:2012rv,Polejaeva:2012ut,Hansen:2014eka,Hansen:2015zga,Briceno:2017tce,Konig:2017krd,Hammer:2017uqm,Hammer:2017kms,Mai:2017bge,Briceno:2018mlh,Briceno:2018aml,Blanton:2019igq,Pang:2019dfe,Jackura:2019bmu,Briceno:2019muc,Horz:2019rrn,Romero-Lopez:2019qrt,Hansen:2020zhy,Blanton:2020gha,Blanton:2020jnm,Pang:2020pkl,Romero-Lopez:2020rdq,Blanton:2020gmf,Muller:2020vtt,Blanton:2021mih,Muller:2021uur,Blanton:2021eyf,Jackura:2022gib,Garofalo:2022pux,Muller:2022oyw,Hansen:2021ofl,Draper:2023xvu}.
While, so far, only simple three-meson systems at maximal isospin have been studied using lattice QCD~\cite{Beane:2007es,Detmold:2011kw,Mai:2018djl,Blanton:2019vdk,Mai:2019fba,Culver:2019vvu,Fischer:2020jzp,Hansen:2020otl,NPLQCD:2020ozd,Alexandru:2020xqf,Brett:2021wyd,Blanton:2021llb,Mai:2021nul,Draper:2023boj}, it is to be expected that more complicated ones will be investigated soon.
The scattering of three generic pions constitutes a potential next milestone for lattice QCD since some relevant low-lying resonances, such as the $\omega(782)$, can be found in these processes.

The extraction of three-particle scattering amplitudes from lattice QCD utilizes the three-particle finite-volume formalism, which connects finite-volume energies obtained in lattice QCD to the three-particle scattering amplitude.
Mainly following three different approaches, formalism has been developed for a number of relevant three-hadron systems.
The approach that we will consider in this work, the so-called relativistic-field-theory (RFT) three-particle formalism~\cite{\HSQCa,\HSQCb}, has been frequently used in the literature for numerical studies~\cite{Blanton:2019vdk, Fischer:2020jzp, Hansen:2020otl, Blanton:2021llb}.
In the RFT formalism, the central object parametrizing short-range three-particle interactions is the divergence-free three-particle $K$-matrix, $\Kdf$.

The interface between lattice QCD and Chiral Perturbation Theory (ChPT) has proven to be a valuable source of insights for first-principles predictions of multi-pion quantities.
A recent example is the comparison between lattice QCD results and ChPT predictions for three pions~\cite{Baeza:2023ljl}, which has provided a useful understanding of the chiral dependence of three-pion quantities.
In particular, in \rcite{Baeza:2023ljl}, we computed the three-pion maximum-isospin $K$-matrix at next-to-leading order (NLO) in ChPT.
We showed that the previously observed tension between leading-order (LO) ChPT predictions and lattice QCD results for $\Kdf$ was significantly reduced when compared against the NLO prediction.
This improved agreement was also an important check of the RFT formalism itself.

In the present work, we generalize the NLO ChPT results of \rcite{Baeza:2023ljl} to the case of three pions in any possible isospin channel.
This result will be useful for prospective lattice QCD calculations, either by providing constraints in the near-threshold energy region of $\Kdf$ or by inspiring parametrizations of the three-particle $K$-matrix.
Note, however, that the presence of resonances when the isospin is not maximal will reduce the energy range of validity with respect to the maximal-isospin case, especially for heavier-than-physical pion masses.

The strategy followed in this work is similar to that used for the computation at maximal isospin~\cite{Baeza:2023ljl}.
We make use of the six-pion amplitude computed in \rrcite{Bijnens:2021hpq,Bijnens:2022zsq} at NLO in ChPT.
We relate this amplitude to the $K$-matrix of the three-pion generalization of the RFT formalism, derived in \rcite{Hansen:2020zhy}.
Several complications due to the presence of nonidentical pions are present in this calculation.
These include additional structures in the threshold expansion of $\Kdf$, the presence of odd partial waves in certain channels, a more complicated symmetrization procedure needed to account for all diagrammatic contributions, and the presence of an $s$-channel diagram of the form $3\pi \to \pi \to 3\pi$, which contributes to the isospin-1 three-pion $K$-matrix.

With these results in hand, several important issues can be addressed.
The first is the convergence of the chiral expansion, which we can address by comparing the sizes of LO and NLO terms.
The second is how quickly the threshold expansion converges to the true answer for the various contributions to the three-particle $K$-matrix.
And the third is the sensitivity of the results to the form of the cutoff function intrinsic to the formalism.
In general terms, we find qualitatively similar results to those we obtained for maximal isospin~\cite{Baeza:2023ljl}, but with some exceptions, to be discussed below.

This paper is organized as follows.
In \cref{sec:background}, we briefly summarize the theoretical background and proceed to describe the various isospin channels and the form of the threshold-expanded $K$-matrix.
In \cref{sec:calculation}, we describe the calculation, first at LO and then at NLO.
Lastly, we present and analyze the results in \cref{sec:results}, and close up with some conclusions in \cref{sec:conclusions}.
This paper contains three appendices detailing the bull's head subtraction and the resulting cutoff dependence of the $K$-matrix (\cref{app:bullhead}), deriving the number of terms at each order in the threshold expansions by group-theoretical means (\cref{app:group}), and detailing the kinematic configurations used to evaluate numerically $\mKdf$ (\cref{app:families}).

A preliminary version of this work appears in Mattias Sjö's doctoral thesis \cite{Sjo:2023phd}.

\section{Theoretical background}\label{sec:background}

\subsection{The three-particle $K$-matrix from ChPT}
% Wherein we refer to sec. 2 of previous paper and restate the most important points

\begin{figure}[thp]
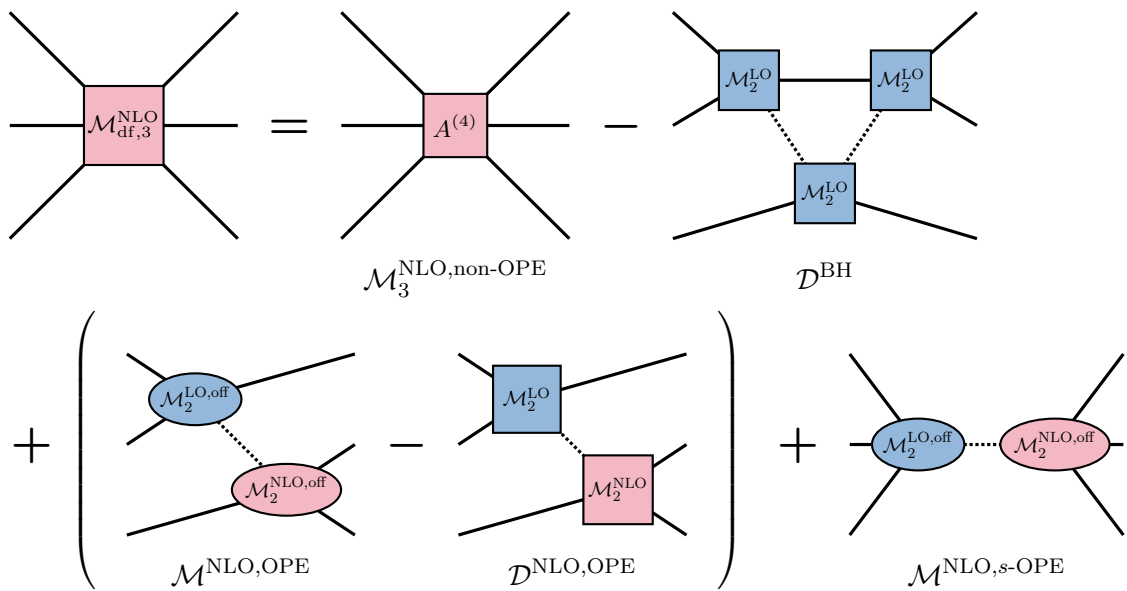

    \begin{multline*}
            \tikzineq[xscale=1.5,yscale=1.5]{
                \makeexternallegcoordinates
                \coordinate (v) at (0,0);
                \draw[sketch onshell prop] (k1) -- (v);
                \draw[sketch onshell prop] (k2) -- (v);
                \draw[sketch onshell prop] (k3) -- (v);
                \draw[sketch onshell prop] (v) -- (p1);
                \draw[sketch onshell prop] (v) -- (p2);
                \draw[sketch onshell prop] (v) -- (p3);
                \node[sketch onshell blob=NLOcolor, inner sep=-3pt]
                    (M) at (0,0) {\footnotesize $\Mdf^\NLO$};
            }
        \quad\scalebox{\sketchoperatorscale}{$\bm=$}\quad
        \underset{\rule{0pt}{1.5em}\displaystyle\cM_3^{\NLO,\nonOPE}}{%
            \tikzineq[xscale=1.5,yscale=1.5]{
                \makeexternallegcoordinates
                \coordinate (v) at (0,0);
                \draw[sketch onshell prop] (k1) -- (v);
                \draw[sketch onshell prop] (k2) -- (v);
                \draw[sketch onshell prop] (k3) -- (v);
                \draw[sketch onshell prop] (v) -- (p1);
                \draw[sketch onshell prop] (v) -- (p2);
                \draw[sketch onshell prop] (v) -- (p3);
                \node[sketch onshell blob=NLOcolor, inner sep=0pt]
                    (M) at (0,0) {\footnotesize $A^{(4)}$};
            }
        }
        \quad\scalebox{\sketchoperatorscale}{$\bm-$}\quad
        \underset{\rule{0pt}{1.5em}\displaystyle\cD^\BH}{%
            \tikzineq[xscale=2,yscale=1.5]{
                \makeexternallegcoordinates
                \coordinate (v1) at (+.5,+.4);
                \coordinate (v2) at (-.5,+.4);
                \coordinate (v3) at (  0,-.6);
                \draw[sketch onshell prop] (k1) -- (v1);
                \draw[sketch onshell prop] (k2) -- (v1);
                \draw[sketch onshell prop] (k3) -- (v3);
                \draw[sketch onshell prop] (v2) -- (p1);
                \draw[sketch onshell prop] (v2) -- (p2);
                \draw[sketch onshell prop] (v3) -- (p3);
                \draw[sketch onshell prop] (v1) -- (v2);
                \draw[sketch offshell prop] (v1) -- (v3);
                \draw[sketch offshell prop] (v2) -- (v3);
                \foreach \i in {1,2,3}
                    \node[sketch onshell blob=LOcolor, inner sep=-1pt]
                        (M\i) at (v\i) {\scalebox{.7}{$\cM_2^\LO$}};
            }
        }
        \\
        \quad\scalebox{\sketchoperatorscale}{$\bm+$}\;\,
        \left(\quad
        \underset{\rule{0pt}{1.5em}\displaystyle\cM^{\NLO,\OPE}}{%
            \tikzineq[xscale=1.5,yscale=1.2]{
                \makeexternallegcoordinates
                \coordinate (v1) at (+.4,-.5);
                \coordinate (v2) at (-.4,+.5);
                \draw[sketch onshell prop] (k1) -- (v2);
                \draw[sketch onshell prop] (k2) -- (v1) -- (k3);
                \draw[sketch onshell prop] (p3) -- (v1);
                \draw[sketch onshell prop] (p1) -- (v2) -- (p2);
                \draw[sketch offshell prop] (v1) -- (v2);
                \node[sketch offshell blob=NLOcolor, inner ysep=2pt, inner xsep=-1pt]
                    (NLO) at (v1) {\scalebox{.7}{$\cM_2^{\NLO,\off}$}};
                \node[sketch offshell blob=LOcolor, inner ysep=2pt, inner xsep=-1pt]
                    (LO) at (v2) {\scalebox{.7}{$\cM_2^{\LO,\off}$}};
            }
        }
        \quad\scalebox{\sketchoperatorscale}{$\bm-$}\quad
        \underset{\rule{0pt}{1.5em}\displaystyle\cD^{\NLO,\OPE}}{%
            \tikzineq[xscale=1.5,yscale=1.2]{
                \makeexternallegcoordinates
                \coordinate (v1) at (+.4,-.5);
                \coordinate (v2) at (-.4,+.5);
                \draw[sketch onshell prop] (k1) -- (v2);
                \draw[sketch onshell prop] (k2) -- (v1) -- (k3);
                \draw[sketch onshell prop] (p3) -- (v1);
                \draw[sketch onshell prop] (p1) -- (v2) -- (p2);
                \draw[sketch offshell prop] (v1) -- (v2);
                \node[sketch onshell blob=LOcolor, inner sep=0pt]
                    (LO) at (v2) {\scalebox{.7}{$\cM_2^{\LO}$}};
                \node[sketch onshell blob=NLOcolor, inner sep=-2pt]
                    (NLO) at (v1) {\scalebox{.7}{$\cM_2^{\NLO}$}};
            }
        }
        \quad\right)
        \quad\scalebox{\sketchoperatorscale}{$\bm+$}\quad
        \underset{\rule{0pt}{1.5em}\displaystyle\cM^{\NLO,\sOPE}}{%
            \tikzineq[xscale=1.8,yscale=1.2]{
                \makeexternallegcoordinates
                \coordinate (v1) at (+.5,0);
                \coordinate (v2) at (-.5,0);
                \draw[sketch onshell prop] (k1) -- (v1);
                \draw[sketch onshell prop] (k2) -- (v1) -- (k3);
                \draw[sketch onshell prop] (p3) -- (v2);
                \draw[sketch onshell prop] (p1) -- (v2) -- (p2);
                \draw[sketch offshell prop] (v1) -- (v2);
                \node[sketch offshell blob=NLOcolor, inner ysep=2pt, inner xsep=-1pt]
                    (NLO) at (v1) {\scalebox{.7}{$\cM_2^{\NLO,\off}$}};
                \node[sketch offshell blob=LOcolor, inner ysep=2pt, inner xsep=-1pt]
                    (LO) at (v2) {\scalebox{.7}{$\cM_2^{\LO,\off}$}};
            }
        }
    \end{multline*}
    \caption{
        Sketch of \cref{eq:mastereq} (cf.\ fig.~2 in \rcite{Baeza:2023ljl}).
        Solid lines represent on-shell pions, while dotted lines are off-shell propagators.
        The incoming particles are on the right, outgoing on the left.
        Square boxes indicate fully on-shell amplitudes, while oval boxes have one leg off shell (factors of $G^\infty$ ensure only on-shell amplitudes are needed  in $\cD$).
        Finally, blue and pink fillings indicate, respectively, LO and NLO quantities.
        For the $\OPE$ and $\sOPE$ contributions, there are additional diagrams, not shown, in which the $\LO$ and $\NLO$ amplitudes are exchanged.
        We leave implicit that only the real parts of all quantities are to be taken.}
    \label{fig:sketch}
\end{figure}

In order to compute $\Kdf$ at NLO, we use the same master equation as in \rcite{Baeza:2023ljl}, \mbox{$\mKdf = \Re \mMdf$}, with the main novelty being that quantities are matrices in flavor space, as described in the next section.
We denote such matrix quantities using boldface throughout the paper, even if the matrices are one-dimensional, which is the case for $I=0,3$.
We also stress that, following the usual index convention in the RFT formalism, rows (columns) label final (initial) states.
In the case of generic three-pion isospin, the calculation can be split into several parts:
\begin{equation}
     \Re \mMdf =  \Re \mMdf^{\nonOPE} - \Re {\m D}^{\rm BH} + \left( \Re \mMdf^{\rm OPE} - \Re \m D^{\NLO,\OPE} \right) +  \Re \mMdf^{\sOPE}.
     \label{eq:mastereq}
\end{equation}
This decomposition, schematically shown in \cref{fig:sketch}, is similar to the one in \rcite{Baeza:2023ljl}, except that now we explicitly account for the `$s$-channel one-particle exchange' ($\sOPE$) contribution, which is only present for $I=1$ since the entire 3-particle isospin is transferred to a single pion.
The other contributions are, respectively, the non-OPE part, the real part of which does not require subtraction, the bull's head subtraction, and the OPE part with its subtraction, which are labeled accordingly.

\subsection{States and channels}\label{sec:states}
% Wherein we describe the isospin channels and their transformation properties

Single pion states are typically described in the charge basis, as they are readily combined into multi-pion states of definite isospin using Clebsch--Gordan coefficients.
However, one usually needs to relate single-pion states to those in the flavor basis (in which the matrix of pseudo-Nambu--Goldstone boson states is $\ket{\phi} = \sum_i \sigma^i\ket{i}$, with $\sigma^i$ the Pauli matrices) in order to determine scattering amplitudes from effective models.
The two bases are related through
\begin{equation}
    \ket{\pipm} = \mp\,\frac{\ket{1} \pm i\ket{2}}{\sqrt{2}}\,,\qquad \ket{\pio} = \ket{3}\,,
\end{equation}
where we use the Condon--Shortley sign convention.

To study three-pion states, we will follow the approach presented in \rcite{Hansen:2020zhy} and use states with zero electric charge since these occur for all three-particle isospins.
Assuming isospin is an exact symmetry, the same results would be obtained from states of different charge within the same isospin multiplet.%
\footnote{
    We have used higher-charge states for some cross-checks, including, of course, the maximum-isospin $\ket{\pip\pip\pip}$ state studied in \rcite{Baeza:2023ljl}.}
In the \emph{charge basis}, we order the seven zero-charge states as
\begin{equation}
    \ket{\pi\pi\pi}_{\mathrm C} = 
    \begin{pmatrix}
        \ket{\pim\pio\pip}  \\
        \ket{\pio\pim\pip}  \\
        \ket{\pim\pip\pio}  \\
        \ket{\pio\pio\pio}  \\
        \ket{\pip\pim\pio}  \\
        \ket{\pio\pip\pim}  \\
        \ket{\pip\pio\pim}  \\
    \end{pmatrix}.
    \label{eq:chargebasis}
\end{equation}
The three particles have different momenta, respectively $k_1$, $k_2$, and $k_3$ in the initial state and $p_1$, $p_2$, and $p_3$ in the final state.
All quantities appearing in this derivation, unless otherwise stated, are
$7\times7$ matrices in the space of three-pion states.

In many parts of the calculation, it is more useful to rotate to a basis of states with definite three-particle isospin.
This rotation is not unique;
the choice made for the most part in \rcite{Hansen:2020zhy} is to let the first two particles form states of definite two-particle isospin, $I_{\pi\pi}$, which we label $\ket{\sigma}$ ($I_{\pi\pi}=0$, the channel where the $\sigma$ resonance is present), $\ket{\rho}$ $(I_{\pi\pi}=1$, where the $\rho$ resonance is present), and $\ket{\Pi}$ ($I_{\pi\pi}=2$, with no resonances).
In this \emph{isospin basis},
\begin{equation}
    \ket{\pi\pi\pi}_{\mathrm I} = 
    \left(
    \begin{array}{l}
        \ket{\Pi\pi}_3   \vphantom{\Big(}\\
        \ket{\Pi\pi}_2   \vphantom{\Big(}\\
        \ket{\rho\pi}_2  \vphantom{\Big(}\\
        \ket{\Pi\pi}_1   \vphantom{\Big(}\\
        \ket{\rho\pi}_1  \vphantom{\Big(}\\
        \ket{\sigma\pi}_1\vphantom{\Big(}\\
        \ket{\rho\pi}_0\vphantom{\Big(}\\
    \end{array}
    \right)
    = 
    \left(
    \begin{array}{l}
        \tfrac1{\sqrt5}   \Big(\ket{\Pip\pim} + \sqrt3\ket{\Pio\pio} + \ket{\Pim\pip}\Big)          \\
        \tfrac1{\sqrt2}   \Big(\ket{\Pip\pim} - \ket{\Pim\pip}\Big)                                 \\
        \tfrac1{\sqrt6}   \Big(\ket{\rop\pim} + 2\ket{\roo\pio} + \ket{\rom\pip}\Big)               \\
        \tfrac1{\sqrt{10}}\Big(\sqrt3\ket{\Pip\pim} - 2\ket{\Pio\pio} + \sqrt3\ket{\Pim\pip}\Big)   \\
        \tfrac1{\sqrt2}   \Big(\ket{\rop\pim} - \ket{\rom\pip}\Big)                                 \\
        \vphantom{\Big(}       \ket{\sigma\pio}                                                     \\
        \tfrac1{\sqrt3}   \Big(\ket{\rop\pim} - \ket{\roo\pio} + \ket{\rom\pip}\Big)                \\
    \end{array}
    \right),
    \label{eq:isobasis}
\end{equation}
where the subscripts indicate three-particle isospin, and the specific two-pion states are
\begin{equation}
    \begin{gathered}
        \ket{\sigma}    =  \frac{\ket{\pip\pim} + \ket{\pim\pip} - \ket{\pio\pio}}{\sqrt3}\,,     \qquad
        \ket{\Pio}      =  \frac{\ket{\pip\pim} + \ket{\pim\pip} + 2\ket{\pio\pio}}{\sqrt6}\,,    \\
        \ket{\roo}      =  \frac{\ket{\pip\pim} - \ket{\pim\pip}}{\sqrt2}\,,                      \qquad
        \ket{\Pipm}     =  \frac{\ket{\pipm\pio} + \ket{\pio\pipm}}{\sqrt2}\,,                    \qquad
        \ket{\ropm}     =  \pm\frac{\ket{\pipm\pio} - \ket{\pio\pipm}}{\sqrt2}\,.
    \end{gathered}
    \label{eq:twopi}
\end{equation}
$\mKdf$ and all other relevant quantities block-diagonalize in this basis, as is described in detail in \rcite{Hansen:2020zhy}.

Yet another basis, which relates more directly to the threshold expansion, aligns the states with irreps of the $S_3$ group describing permutations of the three particles, still with definite isospin.
States are denoted $\ket{\chi_s}$ for the trivial (symmetric) irrep, $\ket{\chi_a}$ for the alternating irrep, and $\ket{\chi_1},\ket{\chi_2}$ for the two-dimensional standard irrep; the details of the irreps are given in appendix~C of \rcite{Hansen:2020zhy}.
Of these irreps, $I=3$ is in the trivial, $I=0$ in the alternating, $I=2$ in the standard, and $I=1$ in a direct sum of the trivial and the standard.
Thus, only for $I=1$ does the isospin basis differ from this \emph{symmetric basis}, where%
\footnote{
    This summarizes eqs.~(C.11) to~(C.19) of \rcite{Hansen:2020zhy}, while \cref{eq:isobasis} corresponds to eqs.~(C.1) to~(C.7).}
\begin{equation}
    \ket{\pi\pi\pi}_{\mathrm S} = 
    \left(
    \begin{array}{l}
        \ket{\chi_s}_3  \vphantom{\tfrac{\sqrt5}3}\\
        \ket{\chi_1}_2  \vphantom{\tfrac{\sqrt5}3}\\
        \ket{\chi_2}_2  \vphantom{\tfrac{\sqrt5}3}\\
        \ket{\chi_s}_1  \vphantom{\tfrac{\sqrt5}3}\\
        \ket{\chi_1}_1  \vphantom{\tfrac{\sqrt5}3}\\
        \ket{\chi_2}_1  \vphantom{\tfrac{\sqrt5}3}\\
        \ket{\chi_a}_0  \vphantom{\tfrac{\sqrt5}3}\\
    \end{array}
    \right)
    =
    \left(
    \begin{array}{l}
        \ket{\Pi\pi}_3  \vphantom{\tfrac{\sqrt5}3}\\
        \ket{\Pi\pi}_2  \vphantom{\tfrac{\sqrt5}3}\\
        \ket{\rho\pi}_2 \vphantom{\tfrac{\sqrt5}3}\\
        \hphantom{+}\tfrac23\ket{\Pi\pi}_1 + \tfrac{\sqrt5}3\ket{\sigma\pi}_1   \\
        -\tfrac{\sqrt5}3\ket{\Pi\pi}_1 + \tfrac23\ket{\sigma\pi}_1  \\
        \ket{\rho\pi}_1 \vphantom{\tfrac{\sqrt5}3}\\
        \ket{\rho\pi}_0  \vphantom{\tfrac{\sqrt5}3}\\
    \end{array}
    \right).
    \label{eq:symbasis}
\end{equation}
The rotation matrices needed to transform from the charge to the isospin and symmetric bases are denoted by 
$\CI$ and $\CS$, respectively (the former stated in eq.~(2.60) of \rcite{Hansen:2020zhy}, with eq.~(2.59) of that work explaining the precise action of the rotation matrices), and are given by
\begin{equation}
    \CI =
        \Ccom
        \begin{pmatrix}
            1   &   1   &   1   &   2   &   1   &   1   &   1   \\
            -1  &   -1  &   0   &   0   &   0   &   1   &   1   \\
            -1  &   1   &   -2  &   0   &   2   &   -1  &   1   \\
            3   &   3   &   -2  &   -4  &   -2  &   3   &   3   \\
            \sqrt3   &   -\sqrt3  &   0   &   0   &   0   &   -\sqrt3  &   \sqrt3   \\
            0   &   0   &   2   &   -2  &   2   &   0   &   0   \\
            -1  &   1   &   1   &   0   &   -1  &   -1  &   1   \\
        \end{pmatrix},\quad
    \CS =
        \Ccom
        \begin{pmatrix}
            1   &   1   &   1   &   2   &   1   &   1   &   1   \\
            -1  &   -1  &   0   &   0   &   0   &   1   &   1   \\
            -1  &   1   &   -2  &   0   &   2   &   -1  &   1   \\
            2   &   2   &   2   &   -6  &   2   &   2   &   2   \\ % singlet
            -1  &   -1  &   2   &   0   &   2   &   -1  &   -1  \\ % doublet 1
            \sqrt3   &   -\sqrt3  &   0   &   0   &   0   &   -\sqrt3  &   \sqrt3   \\ % doublet 2
            -1  &   1   &   1   &   0   &   -1  &   -1  &   1   \\            
        \end{pmatrix},
\end{equation}
with $\Ccom = \diag \Bigl(
            \tfrac1{\sqrt{10}}  ,
            \tfrac12            ,
            \tfrac1{\sqrt{12}}  ,
            \tfrac1{\sqrt{60}}  ,
            \tfrac1{\sqrt{12}}  ,
            \tfrac1{\sqrt{12}}  ,
            \tfrac1{\sqrt6}     
        \Bigr)$
pulling out common coefficients.

\subsection{The threshold expansion}
\label{sec:threxp}
% Wherein we state the form of the threshold expansion, à la 2003.10974 + quadratic_isovector + Steve's new operator

Here, we write down the parametrization of the threshold expansion for each isospin channel in terms of different kinematic operators that have the correct transformation properties under the action of $S_3$.
We thus work in the symmetric basis.
In addition to the initial and final momenta, $\{k_i\}$ and $\{p_i\}$, respectively, and the total momentum $P=p_1+p_2+p_3=k_1+k_2+k_3$, the fundamental building blocks of this parametrization are, following \rcite{Blanton:2019igq},
\begin{equation}
    \Delta \equiv \frac{P^2 - 9\Mpi^2}{9\Mpi^2}\,,\qquad
    \Delta_i\equiv \frac{(P-k_i)^2 - 4\Mpi^2}{9\Mpi^2}\,,\qquad
    \tij{ij} \equiv \frac{(p_i-k_j)^2}{9\Mpi^2}\,,
    \label{eq:threxp-def}
\end{equation}
plus $\Delta'_i$, which is the analogue of $\Delta_i$ obtained by substituting $k_i\to p_i$.
(Throughout the following, a prime always refers to this substitution.)
All of these are considered to be $\cO(\Delta)$ in the expansion.
They are related through
\begin{equation}
    \Delta = -\tfrac12\sum_{i,j}\tij{ij}\,,\qquad
    \Delta_j = \Delta + \sum_i\tij{ij}\,,\qquad
    \Delta'_i = \Delta + \sum_j\tij{ij}\,,
    \label{eq:threxp-rel}
\end{equation}
where all sums, both here and in the remainder of this section, run from $1$ to $3$.

The threshold expansions were derived in \rrcite{Blanton:2019igq,Hansen:2020zhy}, working up to quadratic order for $I=2,3$, linear order for $I=1$, and cubic order for $I=0$.
We have extended the expansion for $I=1$ to quadratic order.
In addition, we have checked the enumeration of operators using a group-theoretic method described in \cref{app:group}, finding one additional operator at cubic order for $I=0$.
As in \rcite{Baeza:2023ljl}, we somewhat simplify the notation for the $\cK$ coefficients.
Furthermore, we depart from \rcite{Hansen:2020zhy} in defining all building-block operators ($\Delta$, $\vec\xi$, etc.) to be dimensionless.

\subsubsection{\texorpdfstring{$I_{\pi\pi\pi}=3$}{I = 3}}

Here, the flavor space is one-dimensional.
Through quadratic order in $\Delta$, we have the five terms computed in \rcite{Baeza:2023ljl},
\begin{equation}
    \Mpi^2\mKdfI{3} = \Kiso + \Kisoone\Delta + \Kisotwo\Delta^2 + \KA\DeltaA + \KB\DeltaB\ + \cO(\Delta^3)\,,
    \label{eq:threxp-I3}
\end{equation}
where
\begin{equation}
    \DeltaA \equiv \sum_{i}\big(\Delta_i^2 + \Delta_i^{\prime 2}\big) - \Delta^2\,,\qquad
    \DeltaB \equiv \sum_{i,j}\tij{ij}^{\,2} - \Delta^2\,.
\end{equation}

\subsubsection{\texorpdfstring{$I_{\pi\pi\pi}=2$}{I = 2}}

This channel involves a two-dimensional flavor space, so all operators need to be doublets that transform under the standard representation of $S_3$.
Following the basis choice of \rcite{Hansen:2020zhy}, the initial-state doublet at linear order in momenta is
\begin{equation}
    \presentxi[\mu]{}{\frac{2k_3-k_1-k_2}{\sqrt6 \Mpi}}{\frac{k_2-k_1}{\sqrt2 \Mpi}}\,.
    \label{eq:xi}
\end{equation}
Still following \rcite{Hansen:2020zhy}, quadratic order introduces three Lorentz-tensor doublets, of which only the following two are relevant here,
\begin{equation}
        \vxiS[\mu\nu] \equiv \frac{\vxi[\mu]P^\nu + \vxi[\nu]P^\mu}{\Mpi}\,,\qquad
        \vxiSb[\mu\nu] \equiv \Big(\xi_2^\mu\xi_2^\nu - \xi_1^\mu\xi_1^\nu,\; \xi_1^\mu\xi_2^\nu + \xi_2^\mu\xi_1^\nu \Big)\,,
    \label{eq:xiS}
\end{equation}
and one Lorentz-scalar doublet,
\begin{equation}
    \presentxi{2}{\frac{2\Delta_3-\Delta_1-\Delta_2}{\sqrt6}}{\frac{\Delta_2-\Delta_1}{\sqrt2}}\,,
    \label{eq:xi2}
\end{equation}
which has the property $\vxi[(2)] = -\frac2{9\Mpi}\vxi[\mu]P_\mu$.
The $\xi$'s and their primed counterparts form four independent tensors in isospin space (one at linear order and three at quadratic);
to simplify the notation, we label these using
\begin{equation}
    \begin{alignedat}{2}
        \mXi{1} &\equiv \vxip[\mu]\otimes\vxi_\mu\,,  \qquad&
        \mXi{2} &\equiv \vxip[(2)]\otimes\vxi[(2)]\,, \\
        \mXi{3} &\equiv \tfrac{1}{\sqrt6}\big[\vxipSb[\mu\nu]\otimes\vxiS_{\mu\nu} + \vxipS[\mu\nu]\otimes\vxiSb_{\mu\nu}\big]\,,   \qquad&
        \mXi{4} &\equiv \vxipSb[\mu\nu]\otimes\vxiSb_{\mu\nu}\,,
    \end{alignedat}
\end{equation}
where $\otimes$ indicates a tensor product like
\begin{equation}
    \vxip[\mu]\otimes\vxi_\mu =
    \begin{pmatrix}
        \xip_1\cdot\xi_1\quad &   \xip_1\cdot\xi_2 \\
        \xip_2\cdot\xi_1\quad &   \xip_2\cdot\xi_2 \\
    \end{pmatrix}
    = \mXi{1}\,,
\end{equation}
and we have pulled out a factor of $\sqrt{6}$ in the definition of $\mXi{3}$ since this would otherwise appear in all our results.
This allows the threshold expansion to be written as
\begin{equation}
    \Mpi^2 \mKdfI{2} = \Big(\KT + \KT[1]\Delta\Big)\:\mXi{1} + \sum_{n=2,3,4} \KT[n]\: \mXi{n} + \cO(\Delta^3)\,,
    \label{eq:threxp-I2}
\end{equation}
where the `$\T$' superscript stands for `isotensor'.

\subsubsection{\texorpdfstring{$I_{\pi\pi\pi}=1$}{I = 1}}

Here, the flavor space is three-dimensional.
Following \rcite{Hansen:2020zhy}, we decompose the states into a singlet and a doublet, transforming under the trivial and standard representations of $S_3$, respectively, and put the singlet as the first component.
Thus, in block form, we have
\begin{equation}
    \mKdfI{1} = 
    \begin{pmatrix}
        \mKdfI{1,\SS}    &   \mKdfI{1,\SD}    \\
        \mKdfI{1,\DS}    &   \mKdfI{1,\DD}    \\
    \end{pmatrix}.
    \label{eq:threxp-I1}
\end{equation}
The singlet-singlet ($\SS$) sector is similar to the $I=3$ case, \cref{eq:threxp-I3},
\begin{equation}
    \Mpi^2\mKdfI{1,\SS} = \KSS + \KSS[1]\Delta + \KSS[2]\Delta^2 + \KSSA\DeltaA + \KSSB\DeltaB\ + \cO(\Delta^3)\,,
    \label{eq:threxp-I1SS}
\end{equation}
whereas the doublet-doublet ($\DD$) sector is similar to the $I=2$ case, \cref{eq:threxp-I2},
\begin{equation}
    \Mpi^2\mKdfI{1,\DD} = \Big(\KDD + \KDD[1]\Delta\Big)\:\mXi{1} + \sum_{n=2,3,4} \KDD[n]\: \mXi{n} + \cO(\Delta^3)\,.
    \label{eq:threxp-I1DD}
\end{equation}
At $\cO(\Delta)$, the sole operator that fits the singlet-doublet-mixing ($\SD$) sector is $\vxi[(2)]$, defined in \cref{eq:xi2}.
At $\cO(\Delta^2)$, new operators are needed that are not included in \rcite{Hansen:2020zhy}.
They are constructed by taking the following building blocks, which are singlets under permutations of the final-state momenta,
\begin{equation}
    \Delta\Delta_i\,,\qquad 
    \Delta_i\Delta_j\,,\qquad 
    \sum_i\tij{ij}\tij{ik}\,,
\end{equation}
and forming doublets under permutations of the initial-state momenta.
From $\Delta\Delta_i$, we simply obtain $\Delta\vxi[(2)]$,
while $\Delta_i\Delta_j$ and $\sum_i \tij{ij}\tij{ik}$ yield the following operators:%
\footnote{
    It is also possible to form the analog of $\vxi[(4,3)]$ using $\Delta_i\Delta_j$, but that is equal to $\vxi[(2)]\Delta - \vxi[(4,2)]$ and is therefore redundant.}
\begin{multline}
    \vxi[(4,n)] = \big(\xi^{(4,n)}_1,\;\xi^{(4,n)}_2\big)\,,\qquad\text{where}\quad n=2,3,4\quad\text{and}\\[1em]
    \left\{
    \begin{alignedat}{2}
        % \presentxi*{4,1}%
        %     {\frac{(\Delta_1+\Delta_2)\Delta_3 - 2\Delta_1\Delta_2}{\sqrt6}}%
        %     {\frac{(\Delta_2-\Delta_1)\Delta_3}{\sqrt2}}\,,\\
        \presentxi*{4,2}%
            {\frac{2\Delta_3^2 - \Delta_1^2 - \Delta_2^2}{\sqrt6}}%
            {\frac{\Delta_2^2-\Delta_1^2}{\sqrt2}}\,,\\
        \presentxi*{4,3}%
            {\sum_i\frac{(\tij{i1}+\tij{i2})\tij{i3} - 2\tij{i1}\tij{i2}}{\sqrt6}}%
            {\sum_i\frac{(\tij{i2}-\tij{i1})\tij{i3}}{\sqrt2}}\,,\\
        \presentxi*{4,4}%
            {\sum_i\frac{2\tij{i3}^{\,2} - \tij{i1}^{\,2} - \tij{i2}^{\,2}}{\sqrt6}}%
            {\sum_i \frac{\tij{i2}^{\,2}-\tij{i1}^{\,2}}{\sqrt2}}\,.
    \end{alignedat}
    \right.
    \label{eq:xi4}
\end{multline}
Thus,
\begin{equation}
    \frac{\Mpi^2}{\sqrt{30}}\mKdfI{1,\SD} = \Big(\KSD + \KSD[1]\Delta\Big)\vxi[(2)] + \sum_{n=2,3,4}\KSD[n]\vxi[(4,n)] + \cO(\Delta^3)\,.
    \label{eq:threxp-I1SD}
\end{equation}
$\mKdfI{1,\DS}$ is obtained from this by exchanging $p_i\leftrightarrow k_i$ and taking the transpose.
We have pulled out a factor of $\sqrt{30}$ in the definition of $\mKdfI{1,\SD}$ since this would otherwise appear in all our results.
The counting of operators---in particular, the appearance of four of them at $\cO(\Delta^2)$---is confirmed by the group-theoretic analysis in \cref{app:group}.

\subsubsection{\texorpdfstring{$I_{\pi\pi\pi}=0$}{I = 0}}

Here, flavor space is one-dimensional.
All operators must be totally antisymmetric under permutations of the momenta, which puts the leading order at $\cO(\Delta^2)$ and makes $\cO(\Delta^3)$ contributions simple enough to include, unlike in the other channels (see \cref{app:group}).
The threshold expansion is
\begin{equation}
    \Mpi^2\mKdfI{0} = \Big(\KAS + \KAS[1]\Delta\Big)\DeltaAS{2} + \KAS[3]\DeltaAS{3}  + \KAS[4]\DeltaAS{4} + \cO(\Delta^4)\,,
    \label{eq:threxp-I0}
\end{equation}
where `$\AS$' stands for `antisymmetric' and the operators are
\begin{equation}
    \begin{gathered}
        \DeltaAS{2} \equiv \sum_{\substack{i,j,k\\m,n,r}} \epsilon_{ijk}\epsilon_{mnr} \tij{im}\tij{jn}\,,\\
        \DeltaAS{3} \equiv \sum_{\substack{i,j,k\\m,n,r}} \epsilon_{ijk}\epsilon_{mnr} \tij{im}\tij{jn}\tij{kr}\,,\qquad
        \DeltaAS{4} \equiv \sum_{\substack{i,j,k\\m,n,r}} \epsilon_{ijk}\epsilon_{mnr} \tij{im}\tij{jn}\big(\tij{im} + \tij{jn}\big)\,,
    \end{gathered}
\end{equation}
of which $\DeltaAS{4}$ was missed in the analysis of \rcite{Hansen:2020zhy}.

\section{Calculation of $\Kdf$}\label{sec:calculation}
% Which is essentially an updated version of sec. 2.5 in previous paper

Here, we describe the calculation of $\Kdf$.
It largely follows the same lines as that performed in \rcite{Baeza:2023ljl}, and we refer the reader there for most of the procedural details.

\subsection{Leading-order calculation}

We start with the calculation at LO.
The bull's head subtraction in \cref{eq:mastereq} is absent at LO, so we thus split the calculation into three parts:
the OPE contribution, the $s$-channel OPE contribution, and the non-OPE part.

\subsubsection{OPE contribution}\label{sec:LO-OPE}

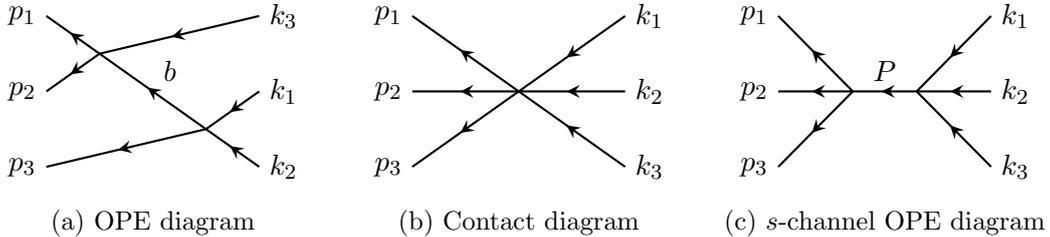
\begin{figure}[thp]
    \centering
    \begin{subfigure}{0.31\textwidth}
        \centering
        \begin{tikzpicture}[xscale=\diagramxscale,yscale=\diagramyscale]
            \makeexternallegsshifted
            \coordinate (v1) at (+.5,-.5);
            \coordinate (v2) at (-.5,+.5);
            \draw[dprop] (k1) -- (v2) -- (p1);
            \draw[dprop] (k2) -- (v1) -- (v2) node [midway, above right] {$b$} -- (p2);
            \draw[dprop] (k3) -- (v1) -- (p3);
        \end{tikzpicture}
        \caption{OPE diagram}
        \label{fig:FeynmanLO:OPE}
    \end{subfigure}
    \begin{subfigure}{0.31\textwidth}
        \centering
        \begin{tikzpicture}[xscale=\diagramxscale,yscale=\diagramyscale]
            \makeexternallegs
            \coordinate (v) at (0,0);
            \draw[dprop] (k1) -- (v) -- (p1);
            \draw[dprop] (k2) -- (v) -- (p2);
            \draw[dprop] (k3) -- (v) -- (p3);
        \end{tikzpicture}
        \caption{Contact diagram}
        \label{fig:FeynmanLO:contact}
    \end{subfigure}
    \begin{subfigure}{0.31\textwidth}
        \centering
        \begin{tikzpicture}[xscale=\diagramxscale,yscale=\diagramyscale]
            \makeexternallegs
            \coordinate (v1) at (+.3,0);
            \coordinate (v2) at (-.3,0);
            \draw[dprop] (k1) -- (v1);
            \draw[dprop] (k3) -- (v1);
            \draw[dprop] (k2) -- (v1) -- (v2) node[midway,above] {$P$} -- (p2);
            \draw[dprop] (v2) -- (p1);
            \draw[dprop] (v2) -- (p3);
        \end{tikzpicture}
        \caption{$s$-channel OPE diagram}
        \label{fig:FeynmanLO:sOPE}
    \end{subfigure}
    \caption{
        Feynman diagrams contributing to $\cM_3$ at LO.
        For diagram (a), there are an additional eight diagrams corresponding to the symmetrization of initial and final momenta.
        Diagram (c) only contributes at $I=1$.}
    \label{fig:FeynmanLO}
\end{figure}

At LO, the OPE contribution comes from the symmetrization of \cref{fig:FeynmanLO:OPE} and from the corresponding subtraction term.
Thus, we require the four-particle amplitude with a single leg potentially off shell, corresponding to the intermediate propagator.
In terms of flavor indices, the four-pion amplitude for $\phi^a(k_1)\phi^b(k_2)\to\phi^c(p_3)\phi^d(b)$ decomposes as
\begin{equation}
    \cM_2(s,t,u) = \delta^{ab}\delta^{cd}A(s,t,u) + \delta^{ac}\delta^{bd}A(t,u,s) + \delta^{ad}\delta^{bc}A(u,s,t)\,,
    \label{eq:M2}
\end{equation}
where
\begin{equation}
    s \equiv (k_1+k_2)^2\,,\qquad t \equiv (k_1-p_3)^2\,,\qquad u \equiv (k_2-p_3)^2
    \label{eq:mandelstam}
\end{equation}
are the usual Mandelstam variables, and the function $A(s,t,u)$ is symmetric in its last two arguments.
The leg with momentum $b$ may be off shell, in which case we use the off-shell convention of \rcite{Bijnens:2021hpq} to express $A$;
the explicit form is given in eqs.~(17), (18), and (23) of that work.
Due to its symmetries, we can abbreviate $A(s) \equiv A(s,t,u)$, and similarly for $A(t)$ and $A(u)$.

The OPE contribution to the unsymmetrized and divergence-free amplitude, $\mMdf^\uu$, is then
\begin{equation}
    \mMdf^{\uu,\OPE} = -\m M_{2,\off} \frac{\m T_G}{\bar b^2+i\epsilon}\m M_{2,\off}
    + \sum_{\ell' \ell} \m M^{\ell'}_{2,\on} \m T_G G^\infty_{\ell' \ell} \m M^{\ell}_{2,\on}\,,
    \label{eq:OPE-LO}
\end{equation}
where $\bar b^2\equiv b^2 - \Mpi^2$, and $\m M^{\ell}_{2,\on}$ is the partial-wave-projected on-shell scattering amplitude.%
\footnote{
    We refer to pages~5--6 of \rcite{Baeza:2023ljl} for a concise definition of $G^\infty_{\ell'\ell}$ and other standard RFT notation.}
Note that in this equation, the momentum dependence of the four-pion amplitudes is left implicit: the amplitudes to the left%
\footnote{
    Strictly speaking, they should appear transposed in \cref{eq:OPE-LO}, but we have chosen our bases such that it is symmetric.}
depend on the outgoing $\{p_i\}$ momenta and those to the right on the incoming momenta $\{k_i\}$.
We also emphasize that in $\m M_{2,\off}$, the momentum of the exchanged particle $b$ is in general off shell, while in $\m M_{2,\on}$, everything is kept on shell.

We now explain the bold-face quantities appearing in \cref{eq:OPE-LO}.
First, $\m M_2$ contains the two-to-two scattering amplitudes that contribute to allowed transitions between three-pion states, considering the third particle as the spectator.
In the charge basis, \cref{eq:chargebasis}, it has a block-diagonal form,
\begin{comment}
Letting $\cM_2$ act on the first two particles in a three-pion state with the third particle as spectator, considering which particle exchanges are valid reveals a block-diagonal form in the charge basis, \cref{eq:chargebasis}:
\end{comment}
\begin{gather}
    \label{eq:M2-block}
    \m M_2 = 
        \begin{pmatrix}
            \m M_2^+   &   &   \\
            &   \m M_2^0   &   \\
            &   &   \m M_2^-   \\
        \end{pmatrix},\qquad\text{with}\\
    \m M_2^+=\m M_2^- = 
        \begin{pmatrix}
            A(t)    &   A(u)    \\
            A(u)    &   A(t)    \\
        \end{pmatrix},\quad
    \m M_2^0 =
        \begin{pmatrix}
            A(s) + A(t) &   -A(s)   &   A(s) + A(u) \\
            -A(s)   &   A(s) + A(t) + A(u)  &    -A(s)  \\
            A(s) + A(u) &   -A(s)   &   A(s) + A(t) \\
        \end{pmatrix}.
    \notag
\end{gather}
Second, the matrix $\m T_G$ indicates valid exchanges between states; in the charge basis, it is
\begin{equation}
    \m T_G = 
        \begin{pmatrix}
            \square      &\square      &\square      &\square      &\square      &\square      &\blacksquare \\
            \square      &\square      &\square      &\square      &\blacksquare &\square      &\square      \\
            \square      &\square      &\square      &\square      &\square      &\blacksquare &\square      \\
            \square      &\square      &\square      &\blacksquare &\square      &\square      &\square      \\
            \square      &\blacksquare &\square      &\square      &\square      &\square      &\square      \\
            \square      &\square      &\blacksquare &\square      &\square      &\square      &\square      \\
            \blacksquare &\square      &\square      &\square      &\square      &\square      &\square      \\
        \end{pmatrix},\qquad \square=0\,,\quad\blacksquare=1\,,
    \label{eq:Tgdef}
\end{equation}
using squares rather than numbers for legibility.

Expanding $A$ to leading order in ChPT, we find
\begin{equation}
    \m M_2^\LO = \m k_0 + \m k_1\bar s + \m k_2 (t+u) + \m k_3 (t-u)\, ,
    \label{eq:M2-LO}
\end{equation}
where $\bar s\equiv s-4\Mpi^2$, and the coefficients $\m k_i$ are straightforward to compute in the charge basis for a given amplitude $A$.

In the subtraction, we need to separate the on-shell four-pion amplitude into partial waves.
At LO, only $s$ and $p$ waves appear.
The separation can thus be performed simply by dividing the amplitude into symmetric and  antisymmetric parts,
\begin{equation}
    \begin{aligned}
        \m M^s_2(s,t,u) &= \tfrac12 \big[ \m M_2(s,t,u) + \m M_2(s,u,t) \big] = \m k_0 + \m k_2\bar s + \m k_2 (t+u)\,,\\
        \m M^p_2(s,t,u) &= \tfrac12 \big[ \m M_2(s,t,u) - \m M_2(s,u,t) \big] = \m k_3 (t-u)\,.\\
    \end{aligned}
    \label{eq:s-and-p}
\end{equation}
At both LO and NLO in ChPT, 
the $p$-wave amplitude is proportional to $t-u$, which can be expanded using the addition theorem for spherical harmonics:
\begin{equation}
    t-u = 4\bm p^*_k\cdot\bm a^*_k 
        = 4 p^*_k q^*_{2k} \bigg[
            \frac{4\pi}{3}\sum_m Y^*_{1m}(\bmh a^*_k) Y_{1m}(\bmh p^*_k)
        \bigg]\,;
        \label{eq:p-wave}
\end{equation}
see section~2 of \rcite{Baeza:2023ljl} for the definitions of the kinematic quantities used here and below.
In our off-shell prescription, this implies that---as is the case for $d$-waves \cite{Baeza:2023ljl}---the difference between the on- and off-shell $p$-wave amplitudes is entirely given by barrier factors:
\begin{equation}
    \m M^p_{2,\off}(\{k_i\}) = \m M^p_{2,\on}(\{k_i\}) \biggl(\frac{p^*_k}{q^*_{2k}}\biggr)\,,\qquad
    \m M^p_{2,\off}(\{p_i\}) = \m M^p_{2,\on}(\{p_i\})\biggl(\frac{k^*_p}{q^*_{2p}}\biggr)\,.
    \label{eq:on-off}
\end{equation}

In order to compute the subtracted result, we separate the $s$-wave part into on-shell and  off-shell parts, the latter being proportional to $\bar b^2$, as can be seen by applying the off-shell relation $t+u=4M_\pi^2-s+\bar b^2$.
Taking this into account, we write
\begin{equation}
    \m M_{2,\off} = \m M^s_{2,\on} + \m M^p_{2,\off} + \bar b^2\:\delta\m M^s_2\,.
\end{equation}
The unsymmetrized divergence-free OPE amplitude is then
\begin{equation}
    \mMdf^{\uu,\OPE,\LO} =  - \bar b^2\delta\m M^s_2 \frac{1}{\bar b^2}\m T_G   \m M_{2, \off}  - \m M_{2, \off}  \frac{1}{\bar b^2}\m T_G  \,\bar b^2\delta\m M^s_2  + \bar b^2\delta\m M^s_2 \frac{1}{\bar b^2}\m T_G  \,\bar b ^2\delta\m M^s_2\,.
\end{equation}
Although we do not show this intermediate result, it can be computed easily.

To obtain the complete LO OPE contribution, the symmetrization procedure must be performed.
In the case of general three-pion isospin, it is slightly more complicated than in \rcite{Baeza:2023ljl}, since it must be done for momentum and flavor simultaneously,
as discussed in \rcite{Hansen:2020zhy}.
This is achieved by using
\begin{equation}\label{eq:symmetrize}
    \mMdf = \sum_{m=0}^2\sum_{n=0}^2 \big(\m R^m\big)^{\!\T} \mMdfuu\big(R^m\{p_i\},R^n\{k_i\}\big) \m R^n\,,
\end{equation}
where $R\{p_1,p_2,p_3\} = \{p_2,p_3,p_1\}$ is a cyclic permutation (due to the symmetry of the interacting pair, only the cyclic subgroup of $S_3$ needs to be considered) and $\m R$ is the representation of that permutation on the space of three-pion states.
Its form in the charge basis is~\cite{Hansen:2020zhy}
\begin{equation}
    \m R = 
    \begin{pmatrix}
        \square      &\square      &\square      &\square      &\blacksquare &\square      &\square      \\
        \square      &\square      &\square      &\square      &\square      &\square      &\blacksquare \\
        \square      &\blacksquare &\square      &\square      &\square      &\square      &\square      \\
        \square      &\square      &\square      &\blacksquare &\square      &\square      &\square      \\
        \square      &\square      &\square      &\square      &\square      &\blacksquare &\square      \\
        \blacksquare &\square      &\square      &\square      &\square      &\square      &\square      \\
        \square      &\square      &\blacksquare &\square      &\square      &\square      &\square      \\
    \end{pmatrix}
    \,,\qquad \square=0\,,\quad\blacksquare=1\,.
\end{equation}
In the symmetric basis, it instead takes the block-diagonal form
\begin{equation}
    \m R = \diag\bigl(1,\m R_2,1,\m R_2,1\bigr)\,,\qquad
    \m R_2 = \frac12
    \begin{pmatrix}
        -1      &   -\sqrt3 \\
        +\sqrt3 &   -1      \\
    \end{pmatrix},
\end{equation}
corresponding to the distribution of one- and two-dimensional irreps.

After symmetrization and conversion of the kinematic variables to $\tij{ij}$, we can identify the terms in the threshold expansion.
At LO, this can be done by inspection since there is only one term per order in the threshold expansion in each isospin sector; at higher orders, it requires solving systems of equations.
The LO results are listed in \cref{tab:LO-results}.
Note that most of the contributions are purely $s$-wave; all pure $p$-wave contributions cancel, and only $\KT$ and $\KDD$ get contributions from mixed $s$- and $p$-wave diagrams (amounting to $9$ out of the total $21/2$ in both cases).

We have checked these results (and also those for the $s$-channel OPE and some of the non-OPE contributions, both at LO and NLO) using an alternative method in which symmetrization is implemented in the charge basis by simply including all possible exchanged pions, allowing all pions in the states to be the spectator.
One then rotates to the symmetric basis.

\subsubsection[$s$-channel OPE contributions]{$\bm s$-channel OPE contributions}\label{sec:LO-sOPE}

The $s$-channel OPE diagram, \cref{fig:FeynmanLO:sOPE}, needs no subtraction since the exchanged momentum $P=k_1+k_2+k_3$ is off shell in the kinematic range of interest.
This contribution appears only in the $I=1$ channel.
For zero-charge states, the exchanged particle must be a $\pi^0$.
Thus, the $s$-channel OPE amplitude can be factorized as
\begin{equation}
    \mMdf^{\sOPE,\LO} = -\bm v_\LO\big(\{p_i\}\big)\frac{1}{P^2 - \Mpi^2}\bm v_\LO^\dag\big(\{k_i\}\big)\,,
    \label{eq:sOPE-LO}
\end{equation}
where $\bm v$ is a column vector of the $\pi\pi\pi\to\pio$ amplitudes from all seven states in the charge basis, with the exchanged $\pio$ off shell.
It can be computed from the amplitude introduced in \cref{eq:M2}, which gives
\begin{equation}
    \begin{aligned}
        \Fpi^2\cM^\LO_2[\pio(k_1)\pio(k_2)\pio(k_3)\to\pio(P)] &= s_{12} + s_{23} + s_{13} - 3\Mpi^2\,,\\
        \Fpi^2\cM^\LO_2[\pip(k_1)\pio(k_2)\pim(k_3)\to\pio(P)] &= \Mpi^2 - s_{13}\,,       
    \end{aligned}
\end{equation}
where $s_{ij}\equiv (k_i+k_j)^2$.
After taking the relevant permutations, rotating to the symmetric basis, and converting to threshold expansion parameters, we get in the $I=1$ sector of the symmetric basis
\begin{equation}
    \bm v_\LO\big(\{k_i\}\big) = \frac{\Mpi^2}{\Fpi^2}
    \begin{pmatrix}
        -3\sqrt{15}\,(1+ \Delta)  \\
        81\,\xi^{(2)}_1/\sqrt2    \\
        81\,\xi^{(2)}_2/\sqrt2    \\
    \end{pmatrix},
\end{equation}
where $\vxi[(2)]$ is given in \cref{eq:xi2}.
We also need the threshold expansion of the single-particle propagator:
\begin{equation}
    \frac{1}{P^2 - \Mpi^2} = \frac{1}{ \Mpi^2 (8  - 9 \Delta)}  = \frac{1}{8\Mpi^2}\left[1 - \frac{9}{8}\Delta + \frac{81}{64}\Delta^2 + \cO(\Delta^3)\right].
\end{equation}
Note that this expansion formally sets the radius of convergence of the threshold expansion at $|\Delta| \leq 8/9$.
This effect will be numerically explored below.

Using these expressions, we can directly identify the coefficients in the threshold expansion.
The results up to quadratic order are listed in \cref{tab:LO-results}.
Note that some terms will appear at higher orders in the threshold expansion, but we will not consider them;
we only check their effect numerically in \cref{sec:validity}.

\subsubsection{Non-OPE contributions}\label{sec:LO-nonOPE}

At leading order, the non-OPE part constitutes the remainder of the six-pion amplitude once the OPE and $s$-OPE parts have been singled out and subtracted; see also \rrcite{Bijnens:2021hpq,Baeza:2023ljl} for details.
It thus includes the contribution from the contact diagram in \cref{fig:FeynmanLO:contact} and has a simple form given in \rcite{Bijnens:2021hpq}.
In the charge basis, the amplitude matrix follows by crossing from
\begin{equation}
    \begin{aligned}
        \cM_3[\pio\pio\pio\to\pio\pio\pio] &= 27\Mpi^2\,,\\
        \cM_3[\pio\pio\pio\to\pip\pio\pim] &= 5\Mpi^2 - 3s'_{13} - t_{12}-t_{22}-t_{32}\,,\\
        \cM_3[\pip\pio\pim\to\pip\pio\pim] &= -6\Mpi^2 + s_{13}+s'_{13} + t_{11}+2t_{22}+t_{33}\,,
    \end{aligned}
\end{equation}
where the momenta are $k_1,k_2,k_3\to p_1,p_2,p_3$ (in that order), and
\begin{equation}
    s_{ij} = (k_i+k_j)^2\,,\qquad
    s'_{ij} = (p_i+p_j)^2\,,\qquad
    t_{ij} = (k_i-p_j)^2\,.
\end{equation}
The result is not divergent, and no subtraction is needed.
Rotating to the symmetric basis and identifying the coefficients of the threshold expansion yields the results collected in \cref{tab:LO-results}.%
\footnote{
    Note that there was a typo in eq.~(4.18) of \rcite{Baeza:2023ljl}; the correct contribution to $\Kisoone$ is proportional to $-36$ (rather than the $-26$ quoted in \rcite{Baeza:2023ljl}).
    This typo did not propagate into final results.}

\begin{table}[htp]
    \centering
    {\renewcommand{\arraystretch}{1.2}
    \begin{tabular}{cR@{$\tFM4$}L|R|R|R|R}
        \toprule
        &       \multicolumn{2}{c|}{}
            &   \multicolumn{1}{c|}{Total}
            &   \multicolumn{1}{c|}{OPE}
            &   \multicolumn{1}{c|}{$s$-channel OPE}
            &   \multicolumn{1}{c}{non-OPE}
        \\
        \midrule
        \multirow{2}{*}{\rotatebox{90}{$I{=}3$}}
        &&   \Kiso       &   18                  &   36              &   0                   &   -18 \\
        &&   \Kisoone    &   27                  &   63              &   0                   &   -36 \\
        \midrule
        \multirow{1}{*}{\rotatebox{90}{\footnotesize$I{=}2$}}
        &&   \KT         &   \tfrac{9}{2}        &   \tfrac{21}{2}   &   0                   &   -6  \\
        \midrule
        \multirow{7}{*}{\rotatebox{90}{$I=1$}}
        &&   \KSS        &   -\tfrac{111}{8}     &   -54             &   -\tfrac{135}{8}     &   57  \\
        &&   \KSS[1]     &   -\tfrac{1137}{64}   &   -27             &   -\tfrac{945}{64}    &   24  \\
        &&   \KSS[2]     &   -\tfrac{135}{512}   &   0               &   -\tfrac{135}{512}   &   0   \\
        \cmidrule{3-7}
        &&   \KSD        &   -\tfrac{3}{8}       &   -9              &   -\tfrac{27}{8}      &   12  \\
        &&   \KSD[1]     &   \tfrac{27}{64}      &   0               &   \tfrac{27}{64}      &   0   \\
        \cmidrule{3-7}
        &&   \KDD        &   \tfrac{1}{2}        &   \tfrac{21}{2}   &   0                   &   -10 \\
        &&   \KDD[2]     &   -\tfrac{81}{4}       &   0               &   -\tfrac{81}{4}       &   0   \\
        \midrule
        \multirow{1}{*}{\rotatebox{90}{\footnotesize$I{=}0$}}
        & \multicolumn{2}{c|}{} & \multicolumn{4}{c}{(there are no $I=0$ contributions at this order)}\\
        \bottomrule
    \end{tabular}}
    \caption{
        LO contributions to $\mKdf$ and the contributions from different parts.
        There is no bull's head subtraction or cutoff dependence at this order.
        Note that `$s$-channel OPE' is not a part of `OPE' but a separate contribution.}
    \label{tab:LO-results}
\end{table}

\subsection{Next-to-leading-order calculation}

Unlike at LO, the NLO amplitude depends on the low-energy constants (LECs) of ChPT.
The four LECs that are relevant to our calculations are denoted $\lr{i}$, $i=1,2,3,4$, with the `r' indicating that they are renormalized, as described in more detail in \rcite{Baeza:2023ljl}.
The renormalization scale $\mu$ appears through the quantity $L\equiv \kappa\log(\Mpi^2/\mu^2)$, with $\kappa\equiv 1/(16\pi^2)$.
As in the LO calculation,  as a cross-check the results in this section have also been obtained using an alternative method, keeping the flavors throughout and projecting onto the symmetric basis at the end.

\subsubsection{OPE contributions}\label{sec:NLO-OPE}

These contributions are calculated in a similar way to that described in \cref{sec:LO-OPE}, except that one of the $\pi\pi$ scattering amplitudes is promoted to NLO.
This amplitude can be expressed in a similar fashion to \cref{eq:M2-LO},
\begin{equation}
    \begin{split}
        \m M^\NLO_2 
            &= \m a_1 +  \m b_1 \bar s + \m  b_2 (t+u) +  \m b_3 (t-u) \\
            & +  \m c_1 \bar s^2 +  \m c_2 \bar s (t+u)   +  \m c_3 \bar s (t-u)  +  \m c_4  (t+u)^2 +  \m c_5 (t+u)(t-u)  +  \m c_6 tu \\
            &+ \m d_1 \bar s^3  + \m d_2 \bar s^2 (t+u)+ \m d_3 \bar s^2 (t-u) + \m d_4 \bar s (t+u)^2 + \m d_5 \bar s (t+u)(t-u) \\&+ \m d_6 \bar s  tu + \m d_7 (t+u)^3 + \m d_8 (t+u)^2(t-u) + \m d_9 (t+u)tu + \m d_{10} (t-u)tu\\
            &+\cO(\bar s^4,t^4,u^4)\,.
    \end{split}
\end{equation}
All the coefficients $\m a_i, \m b_i, \m c_i, \m d_i $  can be computed analytically from \rcite{Bijnens:2011fm}.
The central equation for this computation is \cref{eq:OPE-LO}, inserting the incoming amplitude at LO and the outgoing amplitude at NLO, and vice versa: 
\begin{equation}
    \mMdf^{\uu,\OPE,\NLO} = -\m M^\NLO_{2,\off} \frac{\m T_G}{\bar b^2+i\epsilon}\m M^\LO_{2,\off}
    + \sum_{\ell' \ell} \m M^{\NLO,\ell'}_{2,\on} \m T_G G^\infty_{\ell' \ell} \m M^{\LO,\ell}_{2,\on}\, + \ (\LO \leftrightarrow \NLO).
    \label{eq:OPE-NLO}
\end{equation}
For simplicity, we will subdivide the calculation into multiple parts based on their contributions to different partial waves, including up to $\ell = 3$.
As for other quantities, primed Mandelstam variables refer to the final state.

Terms with $\m a_1$, $\m b_1$, and $\m c_1$ are completely on-shell and purely $s$-wave.
The only contribution that survives after subtraction comes from the $(t+u)$ part of the LO amplitude; specifically,
\begin{equation}
    \mMdfuu \supset - \big(\m a_1 + \m b_1 \bar s' + \m c_1 \bar s'^2 \big) \,\m G \,\m k_2 \bar b^2  + \inxout\,,
\end{equation}
where, for brevity, $\m G=\frac{1}{\bar b^2-i\epsilon}{\m T_G}$.

Terms with $\m b_2$, $\m c_2$, and $\m d_2$ contain both on- and off-shell parts and are purely $s$-wave.
All terms in the LO amplitude contribute, but the cubic term $\m d_2$ only survives in combination with $\m k_0$.
Specifically,
\begin{equation}
    \begin{split}
        \mMdfuu \supset 
            &- \big(\m b_2 + \m c_2 \bar s'  \big)  \bar b^2 \,\m G \, \big( \m k_0 +  \m k_1 \bar s +  \m k_3  (t-u) \big)\\
            &- \big(\m b_2 + \m c_2 \bar s'\big) \,\m G \, \m k_2 \bar b^2 \big(\bar b^2 - \bar s' - \bar s\big) 
            - \m d_2  \bar s'^2 \bar b^2  \,\m G \, \m k_0 \\
            &+ \inxout\,.
    \end{split}
\end{equation}

Terms with $\m b_3$, $\m c_3$, and $\m d_3$ are purely $p$-wave, and no off-shellness remains after accounting for barrier factors.
Terms with $\m d_3$ do not contribute at quadratic order, and those with $\m b_3$ and $\m c_3$ contribute only in combination with $\m k_2$ leading to contributions with $s$--$p$ wave mixing; specifically,
\begin{equation}
    \begin{split}
        \mMdfuu \supset 
            &- \big(\m b_3 + \m c_3 \bar s'  \big)  (t'-u') \,\m G \,\m  k_2 \bar b^2
            + \inxout\,.
    \end{split}
\end{equation}

The term with $\m c_4$ contains both on- and off-shell parts, and is purely $s$-wave, with all terms in the LO amplitude contributing.
Terms with $\m d_4$ and $\m d_7$ are purely $s$-wave and have off-shell parts that contribute only in combination with $\m k_0$.
Specifically,
\begin{equation}
    \begin{split}
        \mMdfuu \supset 
            &- \m c_4   \bar b^2 ( \bar b^2 -2 \bar s') \,\m G \, \big( \m k_0 +  \m k_1 \bar s +  \m k_3  (t-u) \big)\\
            & - \m c_4 \,\m G \, \m k_2 \bar b^2 \big( (\bar b^2)^2 - \bar b^2 \bar s +2 \bar s' \bar s - 2 \bar s' \bar b^2 + \bar s'^2\big) \\
            &- \m d_4 \bar s'  \bar b^2 ( \bar b^2 -2 \bar s')  \,\m G \, \m k_0  - \m d_7 \bar b^2 \big( (\bar b^2)^2 -3 \bar b^2 \bar s' +3 \bar s'^2 \big)  \,\m G \, \m k_0 
            + \inxout\,.
    \end{split}
\end{equation}

Terms with $\m c_5$, $\m d_5$, and $\m d_8$ are purely $p$-wave and they contain off-shell parts even after accounting for barrier factors.
They contribute as
\begin{equation}
    \begin{split}
    \mMdfuu \supset 
        &- \m c_5  \bar b^2 (t'-u') \,\m G \, \big( \m k_0 +  \m k_1 \bar s +  \m k_3  (t-u) \big)
        - \m c_5 (t'-u') \bar b^2 \,\m G \, \m k_2 ( \bar b^2 - \bar s' -  \bar s)\\
        &- \big(\m d_5 \bar s' + \m d_8 (\bar b^2 - 2\bar s') \big) \bar b^2 ( t'-u')  \,\m G \, \m k_0
        + \inxout\,.
    \end{split}
\end{equation}

Terms with $\m c_6$, $\m d_6$, and $\m d_9$ contain $s$- and $d$-waves and contribute
\begin{equation}
    \begin{split}
    \mMdfuu \supset 
        &- \m c_6  \bar b^2 \left( \tfrac{1}{4}\bar b^2 - \tfrac{1}{3}\bar s' -\tfrac{1}{48} \bar s' \bar b^2   \right) \,\m G \, \big( \m k_0 +  \m k_1 \bar s +  \m k_3  (t-u) \big) \\
        &- \m c_6 [t'u']_s \,\m G \, \m k_2(\bar b^2 - \bar s) 
        +  \m c_6 [t'u']_s^\on \,\m G \, \m k_2(- \bar s)  \\
        &- \m c_6 [t'u']_d \,\m G \, \m k_2(\bar b^2)   
        - \m d_9 [t'u']_d \,(\bar b^2)\,\m G \, \m k_0    
        - \m d_6 \bar s' \bar b^2 \left( \tfrac{1}{4}\bar b^2 - \tfrac{1}{3}\bar s'  \right) \,\m G \,  \m k_0 \\
        &- \m d_9 \bar b^2 \left( \tfrac{1}{4}(\bar b^2)^2 - \tfrac{7}{12}\bar s' \bar b^2 +\tfrac{1}{2} \bar s'^2  \right) \,\m G \,  \m k_0 
        + \inxout\,,
    \end{split}
\end{equation}
where  $tu = [tu]_s + [tu]_d$, 
\begin{equation}
    [tu]_s = \frac{1}{4}(\bar s - \bar b^2)^2 - \frac{4}{3} q_{2,p}^{*2} k_p^{*2}\,,
    \quad [tu]_d = q_{2,p}^{*2} k_p^{*2} \frac{8\pi}{15} \sum_m Y_{2m}^*( \bmh a^*_p)  Y_{2m}( \bmh k^*_p)\,,
\end{equation}
and $[tu]_s^\on =\frac{1}{4}(\bar s)^2 - \frac{4}{3} q_{2,p}^{*4}$.

The term with $\m d_{10}$ is cubic, so it only contributes in combination with $\m k_0$, and since $\m k_0$ terms are on shell, only the off-shellness of the $\m d_{10}$ term survives after subtraction.
It contains both $p$- and $f$-waves ($\ell=3$), which requires the decomposition
\begin{equation}
    (\bm a^*_p \cdot \bm k_p^* )^3 = q_{2,p}^{*3} k^{*3} \left[ 
        \frac{3}{5}  \frac{4\pi}{3} Y^*_{1m}( \bmh a^*_p) Y_{1m}( \bmh k^*_p) 
        + \frac{2}{5} \frac{4\pi}{7} Y^*_{3m}( \bmh a^*_p) Y_{3m}( \bmh k^*_p)  
        \right],
    \label{eq:f-wave}
\end{equation}
which is the $f$-wave counterpart of \cref{eq:p-wave}.
Guided by this, we split the coefficient as $(t-u)ut = [(t-u)ut]_p + [(t-u)ut]_f$, of which the latter cancels exactly in the subtraction.
The remaining $p$-wave part is
\begin{equation}
    [ (t-u)ut ]_p =  \frac{1}{4}\bigg( (\bar s - \bar b^2)^2 - \frac{48}{5} q_{2,p}^{*2} k^{*2} \bigg)  q_{2,p}^{*} k^{*}  \frac{16\pi}{3} Y^*_{1m}( \bmh a^*_p) Y_{1m}( \bmh k^*_p)\,.
\end{equation}
Analogously to \cref{eq:on-off}, the on-shell version is obtained by setting $\bar b=0$ and multiplying by a factor of $k^*_p/q^*_{2p}$, which the $G$ in the subtraction changes to $(k^*)^2$.
Thus, the subtracted result is
\begin{equation}
    \begin{split}
        \mMdfuu \supset & \tfrac{1}{4} \m d_{10} \left( \bar b^2-  \tfrac{4}{5} \bar s' \right) \bar b^2 (t-u) \,\m G \,  \m k_0 +
        \inxout\,.
    \end{split}
\end{equation}

Lastly, the cubic-order terms for $I=0$ require only the $p$-wave part of the LO amplitude, namely $\m k_3(t-u)$,
and we likewise require only the part of the NLO amplitude that is proportional to $(t-u)$, namely
\begin{equation}
    \m M_2^\NLO \supset   \m d_3 \bar s^2 (t-u) + \m d_5 \bar s (t+u)(t-u)   + \m d_8 (t+u)^2(t-u) + \m d_{10} (t-u)ut\,,
\end{equation}
of which $\m d_3$ vanishes after subtraction, and the coefficient $\m d_5$ is found to be zero at NLO.
The contributions from $\m d_5$ and $\m d_8$ are simple to evaluate,
\begin{equation}
    \mMdfuu \supset 
        - \big[(\m d_5 \bar s' + \m d_8 (\bar b^2 - 2\bar s') \big] \bar b^2 ( t'-u')  \,\m G \, \m k_3 (t-u)
        + \inxout\,,
\end{equation}
while for $\m d_{10}$ we obtain, after similar manipulations to those above,
\begin{equation}
    \mMdfuu \supset 
        \tfrac{1}{4} \m d_{10} \big( \bar b^2-  \tfrac{4}{5} \bar s' \big) \bar b^2 (t'-u') \,\m G \,  \m k_3 (t-u) 
        + \inxout\,.
\end{equation}

After computing the full $\mMdfuu$, we symmetrize it using \cref{eq:symmetrize} and rotate to the symmetric basis, where the different coefficients are identified.
The complete NLO OPE contributions, including those of cubic order for $I=0$, are listed in \cref{tab:OPE}.

\begin{table}[th!]
    \centering
    {\renewcommand{\arraystretch}{1.28}
    \begin{tabular}    {cR@{$\tFM6$}L|R@{\:}C@{\:}R@{\:}C@{\:}R@{\:}C@{\:}R@{\:}C@{\:}R@{\:}C@{\:}R}
        \toprule
        \multirow{5}{*}{\rotatebox{90}{$I=3$}}
        &&\Kiso             &   25\kappa                &+& 78L                 &-& 576\lr{1}           &-&  432\lr{2}              &-& 72\lr{3}    &+&  144\lr{4}  \\
        &&\Kisoone          &   \kfrac[6831]{20}        &+& 372L                &-& 1332\lr{1}          &-&  1206\lr{2}             & &             &+&  252\lr{4}  \\
        &&\Kisotwo          &   \kfrac[230\,481]{280}     &+& 576L                &-& 1080\lr{1}          &-& 1188\lr{2}              & &             & &             \\
        &&\KA               &   -\kfrac[53\,199]{560}     &+& 45L                 &+& 189\lr{1}           &-& \lrfrac[459]{2}{2}      & &             & &             \\
        &&\KB               &   \kfrac[54\,171]{140}      &+& 216L                &-& 648\lr{1}           &-& 324\lr{2}               & &             & &             \\
        \midrule
        \multirow{5}{*}{\rotatebox{90}{$I=2$}}
        &&\KT               &   \kfrac[207]{40}         &-& 2L                  &-& 210\lr{1}           &-& 15\lr{2}                & &             &+& 42\lr{4}    \\
        &&\KT[1]            &   \kfrac[351\,251]{3360}     &+& \Lfrac[125]{2}      &-& \lrfrac[483]{2}{1}           &-& \lrfrac[267]{4}{2}     & &             & &             \\
        &&\KT[2]            &   -\kfrac[47\,109]{160} &-& \Lfrac[387]{2}     &+& \lrfrac[837]{2}{1}  &+& \lrfrac[1485]{4}{2}    & &             & &             \\
        &&\KT[3]            &   \kfrac[138\,043]{20\,160}   &+& \Lfrac[27]{4}     &-& \lrfrac[45]{4}{1}  &-& \lrfrac[117]{8}{2}    & &             & &             \\
        &&\KT[4]            &   \kfrac[2693]{630}  &+& \Lfrac[11]{3}     &-& 17\lr{1}   &-& \lrfrac[5]{2}{2}    & &             & &             \\
        \midrule
        \multirow{15}{*}{\rotatebox{90}{$I=1$\hspace{.6cm}~}}
        &&\KSS              &   -\kfrac[1475]{6}        &+& 303L                &-& 96\lr{1}            &-& 312\lr{2}               &-& 132\lr{3}   &-& 216\lr{4}   \\
        &&\KSS[1]           &   -\kfrac[12\,773]{40}       &+& 362L                &-& 522\lr{1}           &-& 501\lr{2}               & &             &-& 108\lr{4}   \\
        &&\KSS[2]           &   \kfrac[304\,767]{560}      &+& 516L                &-& 1170\lr{1}          &-& 963\lr{2}               & &             & &             \\
        &&\KSSA             &   \kfrac[489\,117]{1120}     &+& \Lfrac[1365]{4}                &-& \lrfrac[1917]{2}{1}          &-& \lrfrac[1089]{2}{2}               & &             & &             \\
        &&\KSSB             &   \kfrac[95\,097]{280}      &+& 351L                &-& 648\lr{1}           &-& 729\lr{2}              & &             & &             \\
        \cmidrule{3-14}
        &&\KSD              &   \kfrac[154]{5}          &+& 59L                 &-& 72\lr{1}            &-& 33\lr{2}                & &             &-& 36\lr{4}    \\
        &&\KSD[1]           &   -\kfrac[53\,775]{896}     &-& \Lfrac[99]{4}       &-& \lrfrac[171]{8}{1}  &+& \lrfrac[1359]{16}{2}    & &             & &             \\
        &&\KSD[2]           &   \kfrac[24\,123]{224}      &+& \Lfrac[237]{4}      &-& \lrfrac[837]{4}{1}  &-& \lrfrac[585]{8}{2}      & &             & &             \\
        &&\KSD[3]           &   \kfrac[3729]{320}       &+& \Lfrac[75]{4}       &-& \lrfrac[351]{4}{1}  &-& \lrfrac[99]{8}{2}       & &             & &             \\
        &&\KSD[4]           &   -\kfrac[61\,143]{1120}    &-& \Lfrac[3]{2}        &-& \lrfrac[351]{4}{1}  &+& \lrfrac[387]{8}{2}      & &             & &             \\
        \cmidrule{3-14}
        &&\KDD              &   -\kfrac[857]{120}       &-& 18 L                &-& 126\lr{1}           &-& 9\lr{2}                 & &             &+& 42\lr{4}    \\
        &&\KDD[1]           &   \kfrac[926\,543]{10\,080}      &+& \Lfrac[305]{6}    &-& \lrfrac[309]{2}{1}  &-& \lrfrac[301]{4}{2}     & &             & &             \\
        &&\KDD[2]           &   -\kfrac[134\,797]{1120}  &-& \Lfrac[93]{2}    &-& \lrfrac[405]{2}{1}&+& \lrfrac[963]{4}{2}   & &             & &             \\
        &&\KDD[3]           &   \kfrac[398\,287]{60\,480}   &+& \Lfrac[149]{36}     &+& \lrfrac[13]{4}{1} &-& \lrfrac[337]{24}{2}    & &             & &             \\
        &&\KDD[4]           &   \kfrac[37\,577]{7560} &+& \Lfrac[13]{3}    &-& 17\lr{1} &-& \lrfrac[9]{2}{2}   & &             & &             \\
        \midrule
        \multirow{4}{*}{\rotatebox{90}{$I=0$}}
        &&\KAS              &   \kfrac[693]{20}         &+& 54L        &-& 324 \lr{1}            & &                         & &             & &             \\
        &&\KAS[1]           &   0                       & &                     & &                     & &                         & &             & &             \\
        &&\KAS[3]         &   -\kfrac[1215]{32}          & &                     & &                     & &                         & &             & &             \\
        &&\KAS[4]         &   -\kfrac[26\,487]{1120}      & &                     & &                     & &                         & &             & &             \\
        \bottomrule
    \end{tabular}}
    \caption{
        All NLO OPE contributions (excluding the $s$-channel OPE) up to quadratic order in the threshold expansion.
        The cubic-order contributions to $I=0$ are also included.}
    \label{tab:OPE}
\end{table}

\clearpage

\subsubsection[$s$-channel OPE contributions]{$\bm s$-channel OPE contributions}\label{sec:NLO-sOPE}

This calculation is done as described in \cref{sec:LO-sOPE}, but instead of \cref{eq:sOPE-LO} we use
\begin{equation}
    \mMdf^{\sOPE,\NLO} 
        = -\bm v_\NLO^\dag\big(\{p_i\}\big)\frac{1}{P^2 - \Mpi^2}\bm v_\LO\big(\{k_i\}\big) + (\LO \leftrightarrow \NLO)\,,
    \label{eq:sOPE-NLO}
\end{equation}
where, in the $I=1$ sector of the symmetric basis,
\begin{equation}
    \bm v_\NLO\big(\{k_i\}\big) = \frac{\Mpi^2}{\Fpi^2}
    \begin{pmatrix}
        \Sc[0] + \Sc[1]\Delta + \Sc[2]\Delta^2 + \Sc[\mathrm A]\Delta_\A  \\
        \Dc[1]\xi^{(2)}_1 + \Dc[21]\Delta\xi^{(2)}_1 + \Dc[22]\xi^{(4,2)}_1    \\
        \Dc[1]\xi^{(2)}_2 + \Dc[21]\Delta\xi^{(2)}_2 + \Dc[22]\xi^{(4,2)}_2    \\
    \end{pmatrix},
\end{equation}
and
\begin{equation}
    \begin{split}
        \tfrac{1}{\sqrt{15}}\Sc[0]  &= -\tfrac{1}{2} (19\kappa-19 L+ 16\ell_1+16\ell_2+4\ell_3+12\ell_4)\,,\\
        \tfrac{1}{\sqrt{15}}\Sc[1]  &= -\tfrac{1}{12} (181 \kappa - 168 L+288\ell_1+144\ell_2+72\ell_4)\,, \\
        \tfrac{1}{\sqrt{15}}\Sc[2]  &= \tfrac{201}{16} \kappa+9 L-27 \ell_1-\tfrac{27}{2}\ell_2\,,\\
        \tfrac{1}{\sqrt{15}}\Sc[\mathrm A]  &= \tfrac{405}{16}\kappa+\tfrac{27}{2} L-27 \ell_1-27 \ell_2\,, \\
        \tfrac{1}{\sqrt{2}}\Dc[1]   &= \tfrac{125}{4}\kappa+42 L-72 \ell_1-36 \ell_2-18 \ell_4\,,\\
        \tfrac{1}{\sqrt{2}}\Dc[21]  &= -\tfrac{9}{2} \kappa+27  L-81 \ell_2\,,\\
        \tfrac{1}{\sqrt{2}}\Dc[22]  &= \frac{513}{8}\kappa-162 \ell_1+81 \ell_2 \,.
    \end{split}
\end{equation}
Combining these results and expanding the propagator, the contributions up to quadratic order in the threshold expansion have been computed.
They are summarized in \cref{tab:s-OPE}.

\begin{table}[tph]
    \centering
    {\renewcommand{\arraystretch}{1.28}
    \begin{tabular}{cR@{$\tFM6$}L|R@{\:}C@{\:}R@{\:}C@{\:}R@{\:}C@{\:}R@{\:}C@{\:}R@{\:}C@{\:}R}
        \toprule
        \multirow{15}{*}{\rotatebox{90}{$I=1$}\hspace{.6cm}}
        &&\KSS              &   -\kfrac[855]{8}        &+& \Lfrac[855]{8}            &-& 90\lr{1}            &-& 90\lr{2}               &-&  \lrfrac[45]{2}{3}  &-& \lrfrac[135]{2}{4}    \\
        &&\KSS[1]           &   -\kfrac[10005]{64}       &+& \Lfrac[9225]{64}            &-&  \lrfrac[1035]{4}{1}          &-&  \lrfrac[495]{4}{2}            &+&   \lrfrac[45]{16}{3}           &-& \lrfrac[945]{16}{4}    \\
        &&\KSS[2]           &   \kfrac[75525]{512}      &+& \Lfrac[49455]{512}             &-&\lrfrac[9045]{32}{1}           &-& \lrfrac[4725]{32}{2}                     &-&     \lrfrac[405]{128}{3}     &-&      \lrfrac[135]{128}{4}            \\
        &&\KSSA          &   \kfrac[18225]{64}     &+& \Lfrac[1215]{8}      &-& \lrfrac[1215]{4}{1}            &-& \lrfrac[1215]{4}{2}               & &             & &             \\
        &&\KSSB          &   0  & &    & & & &   & &             & &         \\
        \cmidrule{3-14}
        &&\KSD              &   \kfrac[33]{32}          &+& \Lfrac[423]{16}                &-& 36\lr{1}            &-& \lrfrac[45]{2}{2}                &-&   \lrfrac[9]{4}{3}           &-& \lrfrac[27]{2}{4}     \\
        &&\KSD[1]           &   -\kfrac[2073]{256}     &+& \Lfrac[1521]{128}       &-& \lrfrac[27]{2}{1}  &-& \lrfrac[513]{16}{2}    &+& \lrfrac[81]{32}{3}   &+&    \lrfrac[27]{16}{4}    \\
        &&\KSD[2]           &   \kfrac[1539]{64}      & &   &-& \lrfrac[243]{4}{1}  &+&  \lrfrac[243]{8}{2}     & &             & &             \\
        &&\KSD[3]           &   0  & &    & & & &   & &             & &              \\
        &&\KSD[4]          &   0  & &    & & & &   & &             & &                    \\
        \cmidrule{3-14}
        &&\KDD             &   0  & &    & & & &   & &             & &          \\
        &&\KDD[1]           &   0  & &    & & & &   & &             & &                  \\
        &&\KDD[2]           &   \kfrac[1125]{8}  &+& 189L   &-& 324\lr{1}&-& 162\lr{2}   & &             &-& 81\lr{4}            \\
        &&\KDD[3]           &  0  & &    & & & &   & &             & &          \\
        &&\KDD[4]           &  0  & &    & & & &   & &             & &             \\
        \bottomrule
    \end{tabular}}
    \caption{
        NLO $\sOPE$ contributions up to quadratic order in the threshold expansion for the $I=1$ channel.
        These contributions are not present in the other channels.}
    \label{tab:s-OPE}
\end{table}

\subsubsection{Non-OPE contributions}\label{sec:NLO-non-OPE}

At NLO, the non-OPE contribution encompasses the large number of diagrams not covered by the OPE or $s$-channel OPE parts, including the ``bull's head'' triangle diagrams shown in \cref{fig:BHNLO}.
These contributions are all regular in the real part, so they can be added directly to $\mMdfuu$ without any additional treatment regarding the subtraction.
The highly nontrivial threshold expansion of the loop integral functions is described in section~4.2 of \rcite{Baeza:2023ljl};
it is done in the same way here, although a larger number of cases must be considered.
The complete contributions are summarized in \cref{tab:non-OPE}.
We have also checked the results numerically.
For this purpose, the kinematic configurations listed in \cref{app:families} can be used.

\begin{table}[tph]
    \centering
    \begin{tabular}{cR@{$\tFM6$}L|R@{\:}C@{\:}R@{\:}C@{\:}R@{\:}C@{\:}R@{\:}C@{\:}R@{\:}C@{\:}R}
        \toprule
        \multirow{5}{*}{\rotatebox{90}{$I=3$}}
        &&\Kiso            &   14\kappa                 &+& 33L             &+& 288\lr1             &&                      &+& 36\lr3      &-& 72\lr4      \\
        &&\Kisoone         &   -\kfrac[35]{2}           &+& 12L             &+& 720\lr1             &+& 36\lr2              &&              &-& 144\lr4     \\
        &&\Kisotwo         &   -\kfrac[9747]{50}        &-& 216L            &+& 648\lr1             &+& 324\lr2             &&              &&              \\
        &&\KA              &   \kfrac[576]{5}           &-& 54L             &-& 162\lr1             &+& 243\lr2             &&              &&              \\
        &&\KB              &   -\kfrac[13\,797]{50}     &-& 162L            &+& 486\lr1             &+& 243\lr2             &&              &&              \\
        \midrule
        \multirow{5}{*}{\rotatebox{90}{$I=2$}}
        &&\KT              &    \kfrac[85]{12}          &+& 2L              &+& 120\lr1             &+& 6\lr2               &&              &-& 24\lr4      \\
        &&\KT[1]           &    -\kfrac[988]{25}        &-& 36L             &+& 144\lr1             &+& 36\lr2              &&              &&              \\
        &&\KT[2]           &    \kfrac[2052]{25}        &+& 108L            &&                      &-& 324\lr2             &&              &&              \\
        &&\KT[3]           &    \kfrac[501]{50}         &&                  &&                      &&                      &&              &&              \\
        &&\KT[4]           &    \kfrac[451]{150}        &-& 2L              &+& 12\lr1              &&                      &&              &&              \\
        \midrule
        \multirow{15}{*}{\rotatebox{90}{$I=1$\hspace{2cm}}}
        &&\KSS             &    \kfrac[1522]{3}         &-& \Lfrac[1129]{2} &+& {528}\lr1             &+& 840\lr2             &+& 126\lr3     &+& 228\lr4     \\
        &&\KSS[1]          &    545\kappa               &-& 888L            &+& 1440\lr1            &+& 1656\lr2            &&              &+& 96\lr4      \\
        &&\KSS[2]          &    -\kfrac[30\,441]{25}    &-& 846L            &+& 1728\lr1            &+& 1674\lr2            &&              &&              \\
        &&\KSSA            &    -\kfrac[22\,461]{20}    &-& 459L            &+& 1188\lr1            &+& 783\lr2             &&              &&              \\
        &&\KSSB            &    -\kfrac[63\,039]{100}   &-& 387L            &+& 756\lr1             &+& 783\lr2             &&              &&              \\
        \cmidrule{3-14}
        &&\KSD             &    -\kfrac[23]{2}          &-& 84L             &+& 144\lr1              &+& 36\lr2              &&              &+& 48\lr4      \\
        &&\KSD[1]          &    \kfrac[597]{10}          &&                  &+& 108\lr1             &-& 54\lr2              &&              &&              \\
        &&\KSD[2]          &    -\kfrac[1041]{10}         &-& 54L             &+& 270\lr1             &+& 27\lr2              &&              &&              \\
        &&\KSD[3]          &    \kfrac[231]{20}           &&                  &+& 108\lr1             &-& 54\lr2              &&              &&              \\
        &&\KSD[4]          &    -\kfrac[4179]{100}       &-& 18L             &+& 162\lr1             &-& 27\lr2              &&              &&              \\
        \cmidrule{3-14}
        &&\KDD             &    -\kfrac[239]{4}         &+& 46L             &+& 72\lr1              &-& 54\lr2              &&              &-& 40\lr4      \\
        &&\KDD[1]          &    -\kfrac[21\,158]{225}   &-& \Lfrac[20]{3}   &+& 80\lr1              &-& 20\lr2              &&              &&              \\
        &&\KDD[2]          &    -\kfrac[7607]{25}       &-& 204L            &+& 1152\lr1            &+& 36\lr2              &&              &&              \\
        &&\KDD[3]          &    -\kfrac[7897]{1350}     &+& \Lfrac[64]{9}   &-& \lrfrac[64]{3}{1}   &-& \lrfrac[32]{3}{2}   &&              &&              \\
        &&\KDD[4]          &    -\kfrac[6409]{1350}     &-& \Lfrac[14]{9}   &+& \lrfrac[44]{3}{1}   &-& \lrfrac[8]{3}{2}    &&              &&              \\
        \midrule
        \multirow{4}{*}{\rotatebox{90}{$I=0$}}
        &&\KAS             &    \kfrac[1017]{5}         &-& 54L             &+& 162\lr1             &+& 81\lr2              &&              &&              \\
        &&\KAS[1]          &    -\kfrac[972]{5}         &&                  &&                      &&                      &&              &&              \\
        &&\KAS[3]          &    -\kfrac[14\,499]{70}    &&                  &&                      &&                      &&              &&              \\
        &&\KAS[4]          &    \kfrac[88\,371]{2240}   &&                  &&                      &&                      &&              &&              \\
        \bottomrule
    \end{tabular}
    \caption{
        All NLO non-OPE contributions up to quadratic order in the threshold expansion, including cubic order for $I=0$.
        Note that the non-OPE contributions do not include the bull's head subtraction.}
    \label{tab:non-OPE}
\end{table}

\subsubsection{Bull's head subtraction}\label{sec:NLO-BH}

\begin{figure}[tph]
    \centering
    \begin{subfigure}{0.48\textwidth}
    \centering
        \begin{tikzpicture}[xscale=\diagramxscale,yscale=\diagramyscale]
            \makeexternallegs
            \coordinate (v1) at (+.5,+.5);
            \coordinate (v2) at (-.5,+.5);
            \coordinate (v3) at (  0,-.7);
            \draw[dprop] (k1) -- (v1) -- (v2) node[midway, above] {\footnotesize$r$} -- (p1);
            \draw[dprop] (k2) -- (v1) -- (v3) -- (v2) -- (p2);
            \draw[dprop] (k3) -- (v3) -- (p3);
        \end{tikzpicture}
        \caption{The ``bull's head'' diagram.}
        \label{fig:BHNLO:regular}
    \end{subfigure}
    \begin{subfigure}{0.48\textwidth}
    \centering
        \begin{tikzpicture}[xscale=\diagramxscale,yscale=\diagramyscale]
            \makeexternallegs
            \coordinate (v1) at (+.5,+.5);
            \coordinate (v2) at (-.5,+.5);
            \coordinate (v3) at (  0,-.7);
            \draw[dprop=.8] (v1) .. controls ($ (k1)!.5!(v1) $) .. (p1);
            \draw[line width=3pt, white] (k1) .. controls ($ (p1)!.5!(v2) $) .. (v2);
            \draw[dprop=.2] (k1) .. controls ($ (p1)!.5!(v2) $) .. (v2);
            \draw[dprop] (v2) -- (v3) -- (v1);
            \draw[dprop] (k2) -- (v1) -- (v2) node[midway, above] {\footnotesize$r$}-- (p2)  ;
            \draw[dprop] (k3) -- (v3) -- (p3);
        \end{tikzpicture}
        \caption{The ``crossed bull's head'' diagram.}
        \label{fig:BHNLO:crossed}
    \end{subfigure}
    \caption{
        Two configurations of the triangle-loop diagrams.
        There are a total of 15 diagrams with the triangle topology, of which 9 correspond to the configuration (a) [so their sum corresponds to the symmetrization of~(a)]  and 6 to the configuration~(b).
        Neither diagram is singular in the real part, and only (a) is singular in the imaginary part, which cancels against $\Im\mDBH$.}
   \label{fig:BHNLO}
\end{figure}
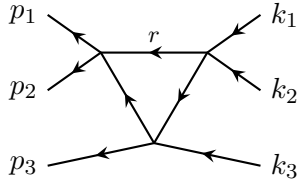
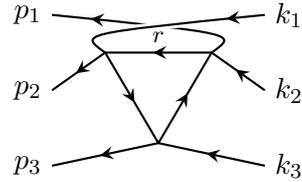

\noindent 
The bull's head subtraction term, shown schematically in \cref{fig:sketch} and corresponding to the topology in \cref{fig:BHNLO:regular}, is given by
\begin{equation}
    \label{eq:DBHuu}
    \mDBHuu(\bm p_3,\bm k_3) = \int_r \m M^\LO_{2,\on}(\bm p_3) \m T_G G^\infty(\bm p_3,\bm r) \m M^\LO_{2,\on}(\bm r) \m T_G G^\infty(\bm r,\bm k_3) \m M^\LO_{2,\on}(\bm k_3)\,,
\end{equation}
where $\int_r\equiv \int\d^3r/[2\omega_r(2\pi)^3]$ is the Lorentz-invariant integral over the on-shell loop momentum $\bm r$, with $\omega_r=\sqrt{\bm r^2 + \Mpi^2}$, and partial-wave indices are implicitly summed over.
We recall that $\m T_G$ is defined in \cref{eq:Tgdef}, while $G^\infty$ is defined in eq.~(2.12) of \rcite{Baeza:2023ljl}.
Unlike the OPE contributions, the bull's head part lacks singularities in the real part, so there is no need to cancel $\mDBH$ against a matching off-shell expression.

\Cref{eq:DBHuu} can be rewritten as
\begin{align}\label{eq:mDuuBH}
    \mDuuBH(\bm p_3, \bm k_3) &= \frac1{\Fpi^6}\int_r H^2(x_r)
    D(\bm p_3,\bm k_3,\bm r) \m N(\bm p_3,\bm k_3) \,,\\
    D(\bm p_3,\bm k_3,\bm r) &\equiv  \frac1{\big[(P-p_3-r)^2-\Mpi^2+i\epsilon\big]\big[(P-k_3-r)^2-\Mpi^2+i\epsilon\big]}\,,
\end{align}
where $x_r \equiv (P-r)^2/(4\Mpi^2)$ and $H(x)$ is a smooth cutoff function that is $0$ when $x\leq 0$ and $1$ when $x \geq 1$; it is discussed further in \cref{app:cutoff}.%
\footnote{
    We stress that at this order, $\mDBH$ is the only part of $\mKdf$ that depends on this cutoff.}
The numerator matrix $\m N(\bm p_3,\bm k_3)$ captures the isospin dependence.
Its elements can be expressed in terms of the quantities
\begin{equation}\label{eq:nijk}
    \n{ijk}\equiv \nq{i}{p_3}\nq{j}{r}\nq{k}{k_3}\,,\qquad i,j,k\in\{0,1,2\}\,,
\end{equation}
where each $\nq{i}{q}$ incorporates one component of $\m M^\LO_{2,\on}(\bm q)$ and the associated barrier factors coming from $G^\infty$.
Following \cref{eq:M2-block,eq:M2-LO}, $\nq{0}{q}$ comes from $A(s_q)$, $\nq{1}{q}$ from \mbox{$A(t_q)-A(u_q)$}, and $\nq{2}{q}$ from \mbox{$A(t_q)+A(u_q)$};%
\footnote{
    Note that the subscript on $\nq{i}{q}$ denotes neither partial-wave index (since $\nq{2}{q}$ is $s$-wave) nor two-particle isospin (since $\nq{0}{q}$ is not purely $I_{\pi\pi}=0$), but is simply constructed for convenience.}
see below for their explicit forms.
In the symmetric basis, the components of $\m N$ in each sector are [cf.\ \cref{eq:threxp-I3,eq:threxp-I2,eq:threxp-I1,eq:threxp-I0}] 
{\allowdisplaybreaks
\begin{subequations}
\begin{align}
    \m N^{[I=3]} &= \n{222}\,,\\
    \m N^{[I=2]} &= \frac{1}{4}
        \begin{pmatrix}
            \n{222} + 3\n{212}      &   \sqrt3(\n{221}-\n{211}) \\
            \sqrt3(\n{122}-\n{112}) &   \n{111} + 3\n{121}      \\
        \end{pmatrix},\\
    \m N^{[I=1,\SS]} &= \n{222} + \tfrac53\big(n_{000}+\n{010}+\n{200}+\n{002}+2\n{020}+\n{220}+\n{022}+\n{202}\big)\,,\\
    \m N^{[I=1,\SD]} &=
        \begin{pmatrix}
             \tfrac{\sqrt{5}}{3} N_1\quad&
            -\tfrac{1}{2}\sqrt{\tfrac{5}{3}} N_2\\
        \end{pmatrix},\qquad
    \m N^{[I=1,\DS]} =
        \begin{pmatrix}
             \tfrac{\sqrt{5}}{3} N'_1 \\
            -\tfrac{1}{2}\sqrt{\tfrac{5}{3}} N'_2\\
        \end{pmatrix},\\
    &N_1\equiv    2\n{000} - \n{002} + 2\n{010} + \tfrac32 \n{012} + 4\n{020} + \tfrac12 \n{022} + 2\n{200} - \n{202} + 2\n{220}\,,\notag\\
    &N_2\equiv   2\n{001}+\n{011}-\n{021}+2\n{201}\,,\notag\\
    \m N^{[I=1,\DD]} &= 
        \begin{pmatrix}
            N_{11}+N'_{11} \quad&
            -\tfrac{1}{\sqrt3} N_{12} \\
            -\tfrac{1}{\sqrt3} N'_{12} \quad&
            N_{22}
        \end{pmatrix},\\
    &N_{11}\equiv 2\n{000} - 2\n{002} + 2\n{010} + 3\n{012} + 4\n{020} + \n{022} + \tfrac12\n{202} + \tfrac38 \big(3\n{212} + \n{222}\big)\,,\notag\\
    &N_{12}\equiv 2\n{001}+\n{011}-\n{021}-\n{201}+\tfrac34\big(\n{211}-\n{221}\big)\,,\notag\\
    &N_{22}\equiv \n{101} + \tfrac14\big(\n{111} + 3\n{121}\big)\,,\\
    \m N^{[I=0]} &= \n{111}\,,
\end{align}
\label{eq:Nijk}%
\end{subequations}}
where, as in \cref{sec:threxp}, a prime like that on $N'_1$ indicates swapping $p$ and $k$, i.e., $\n{ijk}\to\n{kji}$.

Being $s$-wave, $\smash{\nq{0,2}{q}}$ lack barrier factors, and are therefore simply $\nq{0}{q} = s_q-\Mpi^2$ and $\smash{\nq{2}{q}} = t_q+u_q-2\Mpi^2$; that is,
\begin{alignat}{3}
    \nq{0}{p_3}&=(p_1+p_2)^2-\Mpi^2\,,
    &\quad
    \nq{0}{r}&=(P-r)^2-\Mpi^2\,,
    &\quad
    \nq{0}{k_3}&=(k_1+k_2)^2-\Mpi^2\,,
    \\
    \nq{2}{p_3}&=-2p_1\cdot p_2\,,
    &\quad
    \nq{2}{r}&=2\Mpi^2-(P-r)^2\,,
    &\quad
    \nq{2}{k_3}&=-2k_1\cdot k_2\,.
\end{alignat}
The $p$-wave $\nq{1}{q}$ is proportional to $t_q-u_q$, which can be re-expressed using the on-shell version of  \cref{eq:p-wave}:
\begin{equation}
    (t-u)_q
        = 4\bm a^*_q\cdot\bm a^{\prime*}_q
        = 4\big(q^*_{2q}\big)^2\bigg[
            \frac{4\pi}{3}\sum_m Y^*_{1m}(\bmh a^*_q) Y_{1m}(\bmh a^{\prime*}_q)
        \bigg]\,.
\end{equation}
After including the relevant parts of $G^\infty$, $(q^*_{2q})^2$ cancels the denominators of the barrier factors, and the sum rule for spherical harmonics leaves expressions of the kind $\bm p^*_q\cdot\bm k^*_q$ for some $\bm p,\bm k$.
In $\nq{1}{p_3}$ and $\nq{1}{k_3}$, where one of $\bm p$ and $\bm k$ remains a pair momentum, this can be evaluated using 
\begin{equation}
    {\bm a}_{k_3}^*\cdot{\bm r}_{k_3}^*
    =\tfrac12(\bm k_1^*-\bm k_2^*)\cdot{\bm r}_{k_3}^*
    =\tfrac12 (k_{10}^*-k_{20}^*){r}_{k_30}^*-\tfrac12(k_1-k_2)\cdot r
    =-\tfrac12(k_1-k_2)\cdot r\,,
\end{equation}
and similarly for $p_3$.
However, this does not work for $\nq{1}{r}$, where we get $\bm p=\bm p_3$ and $\bm k=\bm k_3$, neither of which is a pair momentum with $\bm r$ as spectator.
There, we must instead use the general formula for the product of 3-momenta in a given rest frame, leaving%
\footnote{
    Note that  $\n{ij1}$  and $\n{1jk}$ are antisymmetric in the initial and final pair momenta, respectively, while the other $\n{ijk}$ are symmetric.
    This makes the symmetry properties of $\m N$ relatively manifest.}
\begin{equation}
    \begin{gathered}
        \nq{1}{p_3}=-2(p_1-p_2)\cdot r\,,
        \qquad
        \nq{1}{k_3}=-2(k_1-k_2)\cdot r\,,
        \\
        \nq{1}{r}= 4\bigg[\frac{p_{3}\cdot(P-r)\;k_{3}\cdot(P-r)}{(P-r)^2} - p_3\cdot k_3\bigg]\,.
    \end{gathered}
\end{equation}

Following \cref{eq:Nijk}, we can write $\mDBHuu$ [and $\mDBH$, fully symmetrized using \cref{eq:symmetrize}] in terms of the 27 integrals
\begin{equation}
    I_{ijk}(\bm p,\bm k) \equiv \frac{1}{\Fpi^6}\int_r H^2(x_r) D(\bm p,\bm k,\bm r) n_{ijk}(\bm p,\bm k) \,,\qquad i,j,k\in\{0,1,2\}\,,
\end{equation}
which can be evaluated using the methods of \rcite{Baeza:2023ljl}, where only the single integral that is needed at maximum isospin (essentially, $I_{222}$) was treated.
The only substantial difference here is the inclusion of $1/(P-r)^2$ factors due to $\nq{1}{r}$.
After threshold expansion and the change of integration variable to $z$ such that $\omega_r=\Mpi(1+2z^2)$, this leaves a power series in $1/x$ where $x\equiv1-3z^2$.
Thus, the class of integrals needed here is extended to
\begin{equation}
    H_{m,n,p} \equiv \frac{1}{\pi^2}\int_0^{1/\sqrt3} \d z\,\frac{\sqrt{1+z^2}}{z^m x^p} \frac{\d^n}{\d x^n}\big[H^2(x)\big]\,,
\end{equation}
where $H_{m,n,0}$ correspond to the $H_{m,n}$ of \rcite{Baeza:2023ljl}.
The Hadamard finite-part prescription can still be used to regularize the integrals, leaving
\begin{equation}\label{eq:Hmnp}
    H_{m,n,p} \equiv
    \begin{cases}
        \displaystyle\frac{1}{\pi^2}\int_0^{1/\sqrt3}\d z\,\frac{\sqrt{1+z^2}}{z^m x^p}\frac{\d^n}{\d x^n}\big[H^2(x)\big]\,, & n>0\,,\\
        \displaystyle\rule{0pt}{2em}\int_0^{1/\sqrt3}\d z\; 6zf_{m,p}(z) \frac{\d}{\d x}\big[H^2(x)\big]\,, & n=0\,,
    \end{cases}
\end{equation}
where now
\begin{equation}
    \frac{\d}{\d z}f_{m,p}(x) = \frac{1}{\pi^2}\frac{\sqrt{1 + z^2}}{z^m x^p}\,.
\end{equation}
The resulting expressions can be simplified by using the algebraic relations
\begin{equation}\label{eq:alg-rel}
    H_{m,n,p} = 3H_{m-2,n,p} + H_{m,n,p-1} = \tfrac13\big(H_{m+2,n,p}-H_{m+2,n,p-1}\big)\,,
\end{equation}
and also the integration-by-parts relations derived in \rcite{Baeza:2023ljl} when $p=0$.

As a last step in \rcite{Baeza:2023ljl}, the $H_{m,n,p}$ were approximated by setting $H(x)=1$, reducing all $H_{m,n,0}$ to $f_{m,0}(1/\sqrt3)\delta_{n,0}$ and allowing analytic evaluation of the results.
This analytic approximation becomes more complicated here, since $H_{m,n,p}$ with $H(x)=1$ does not converge if $p>0$ due to the pole at $x=0$ in the upper integration limit.%
\footnote{
    Physically, the $x=0$ pole corresponds to $P-r$ becoming a lightlike momentum, rendering quantities like $\bm p^*_r$, which appears in $\nq{1}{r}$, ill-defined.}
Regularizing this divergence with the Hadamard finite-part prescription is equivalent to simply dropping terms containing $p>0$.%
\footnote{
    It is also conceivable to approximate $H(x)=x^p$ to remove these divergences, where $p$ can either be set to different values as needed in each term or fixed throughout the entire expression.
    However, all variants of this approach we have tried result in a worse approximation, in the sense that $|\cD_X|$ as defined in \cref{eq:DX} become larger in all but a few cases.}
Thus, we effectively let
\begin{equation}
    H_{m,n,p} = \widetilde H_{m,n,p} + f_{m,0}(1/\sqrt3)\delta_{n,0}\delta_{p,0}\,,
\end{equation}
where only $\widetilde H_{m,n,p}$ is cutoff-dependent; setting $H(x)=1$ corresponds to $\widetilde H_{m,n,p}=0$.
$f_m(1/\sqrt3)$ can be expressed entirely in terms of rational numbers and $\log3$.
Thus, each coefficient in the threshold expansion can be separated like
\begin{equation}\label{eq:DX}
    \cK_X = \cK_X^{[\tilde H_{m,n,p}=0]} - \cD_X\,,
\end{equation}
where the cutoff-independent $\cK_X^{[\tilde H_{m,n,p}=0]}$ can be expressed similarly to the OPE and other contributions.
The cutoff-dependent term $\cD_X$ must be evaluated numerically, but is well-behaved and typically small, even in most of the cases where $p>0$ is present.
The contributions from the bull's head subtraction are listed in \cref{tab:BH} decomposed as in \cref{eq:DX}, and in \cref{tab:BH-exact} in terms of $H_{m,n,p}$.
The cutoff dependence of $\cD_X$ is studied in \cref{app:cutoff}.

As in \rcite{Baeza:2023ljl}, we have verified these results using direct numerical integration of \cref{eq:mDuuBH}, which also allows $\mDuuBH$ to be studied further away from threshold, as is used in \cref{sec:validity}.
Unlike in \rcite{Baeza:2023ljl}, it is no longer well-defined to set $H(x)=1$ globally, so the analytic approximation cannot be studied, at least using the prescription described above.

\begin{table}[tp]
    \centering
    \begin{tabular}{cR@{$\tFM6$}L|R@{\:$($}R@{\:}C@{\:}R@{\:$)$\:\:}C@{\:\:}R@{$.$}L}
        \toprule
        \multirow{5}{*}{\rotatebox{90}{$I=3$}}
        &&\Kiso            &   - \kappa &   144                      &+&  36\logthree                   &+&     0&056\,347\,6589        \\
        &&\Kisoone         &   - \kappa &   424                      &+&  96\logthree                   &-&     0&129\,589\,681         \\
        &&\Kisotwo         &   - \kappa &   \tfrac{9801}{50}         &+&  \tfrac{621}{10}\logthree     &-&     0&432\,202\,370         \\
        &&\KA              &     \kappa &   \tfrac{3303}{10}         &-&  \tfrac{135}{8}\logthree      &-&     0&000\,907\,273\,890    \\
        &&\KB              &     \kappa &   \tfrac{909}{50}          &-&  \tfrac{189}{40}\logthree     &-&     0&000\,162\,394\,747    \\
        \midrule
        \multirow{5}{*}{\rotatebox{90}{$I=2$}}
        &&\KT              &   - \kappa &   \tfrac{9883}{324}        &+&  \tfrac{1009}{144}\logthree   &+&     0&007\,042\,111\,64     \\
        &&\KT[1]           &   - \kappa &   \tfrac{714497}{16200}    &+&  \tfrac{989}{480}\logthree    &+&     0&095\,869\,747\,4      \\
        &&\KT[2]           &     \kappa &   \tfrac{140449}{3600}     &-&  \tfrac{5641}{320}\logthree   &+&     0&264\,963\,303         \\
        &&\KT[3]           &   - \kappa &   \tfrac{58169}{3600}      &+&  \tfrac{317}{960}\logthree    &-&     0&021\,650\,723\,1      \\
        &&\KT[4]           &   - \kappa &   \tfrac{31069}{10800}     &+&  \tfrac{59}{960}\logthree     &+&     0&001\,531\,207\,94     \\
        \midrule
        \multirow{15}{*}{\rotatebox{90}{$I=1$\hspace{2cm}}}
        &&\KSS             &   - \kappa &   399                      &+&  \tfrac{369}{4}\logthree      &-&     1&213\,748\,64          \\
        &&\KSS[1]          &   - \kappa &   \tfrac{1333}{2}          &+&  \tfrac{993}{8}\logthree      &-&     4&737\,727\,30          \\
        &&\KSS[2]          &     \kappa &   \tfrac{7169}{50}         &-&  \tfrac{33957}{320}\logthree  &-&     2&098\,947\,60          \\
        &&\KSSA            &     \kappa &   \tfrac{142983}{160}      &-&  \tfrac{19575}{128}\logthree  &+&     2&393\,448\,70          \\
        &&\KSSB            &     \kappa &   \tfrac{158139}{800}      &-&  \tfrac{13419}{640}\logthree  &+&     1&089\,982\,49          \\
        \cmidrule{3-10}
        &&\KSD             &     \kappa &   \tfrac{95}{2}            &-&  \tfrac{255}{8}\logthree      &-&     0&060\,539\,453\,1      \\
        &&\KSD[1]          &     \kappa &   \tfrac{1283}{16}         &-&  \tfrac{3543}{64}\logthree    &-&     0&558\,130\,406         \\
        &&\KSD[2]          &     \kappa &   \tfrac{7503}{160}        &+&  \tfrac{513}{128}\logthree    &+&     0&105\,910\,881         \\
        &&\KSD[3]          &     \kappa &   \tfrac{10179}{160}       &-&  \tfrac{3699}{128}\logthree   &+&     0&135\,426\,533         \\
        &&\KSD[4]          &     \kappa &   \tfrac{46377}{400}       &-&  \tfrac{11097}{320}\logthree  &+&     0&349\,051\,891         \\
        \cmidrule{3-10}
        &&\KDD             &     \kappa &   \tfrac{26585}{324}       &+&  \tfrac{1259}{144}\logthree   &-&     0&048\,482\,775\,8      \\
        &&\KDD[1]          &     \kappa &   \tfrac{433507}{16200}    &+&  \tfrac{4279}{480}\logthree   &+&     0&316\,388\,524         \\
        &&\KDD[2]          &     \kappa &   \tfrac{2975701}{3600}    &-&  \tfrac{11869}{320}\logthree  &+&     1&906\,375\,12          \\
        &&\KDD[3]          &     \kappa &   \tfrac{5071}{400}        &+&  \tfrac{449}{320}\logthree    &-&     0&034\,410\,564\,7      \\
        &&\KDD[4]          &     \kappa &   \tfrac{27919}{10800}     &+&  \tfrac{83}{320}\logthree     &+&     0&017\,668\,886\,1      \\
        \midrule
        \multirow{4}{*}{\rotatebox{90}{$I=0$}}
        &&\KAS             &   - \kappa &   102                      &+&  \tfrac{81}{2}\logthree       &+&     0&301\,063\,917         \\
        &&\KAS[1]          &     \kappa &   \tfrac{1632}{5}          & &                                &-&     0&881\,880\,013         \\
        &&\KAS[3]          &     \kappa &   \tfrac{27459}{280}       &+&  \tfrac{2187}{32}\logthree    &-&     0&607\,228\,425         \\
        &&\KAS[4]          &     \kappa &   \tfrac{54459}{1120}      &-&  \tfrac{3645}{128}\logthree   &+&     0&227\,122\,084         \\
        \bottomrule
    \end{tabular}
    \caption{
        Contributions from the bull's head subtraction up to quadratic order in the threshold expansion, including cubic order for $I=0$.
        The contributions are separated into a cutoff-independent analytic part (containing $\kappa$ and $\log3$) and a cutoff-dependent numerical part according to \cref{eq:DX}.
        The latter is computed using the standard cutoff choice, shown in \cref{eq:H-standard}.}
    \label{tab:BH}
\end{table}

\section{Results}\label{sec:results}

\begin{figure}
    % \begin{comment} % Uncomment for fast version with pre-rendered plots
    \phantom{\footnotesize$0$}%
    \leftfudge
    \includegraphics[scale=\figscale]{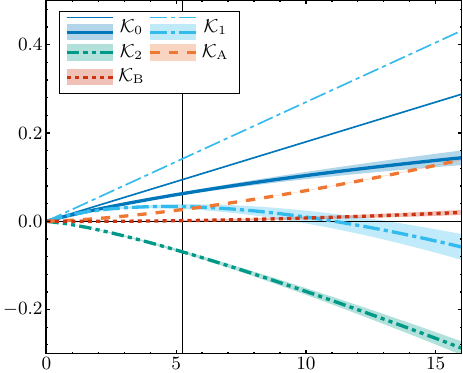}
    \includegraphics[scale=\figscale]{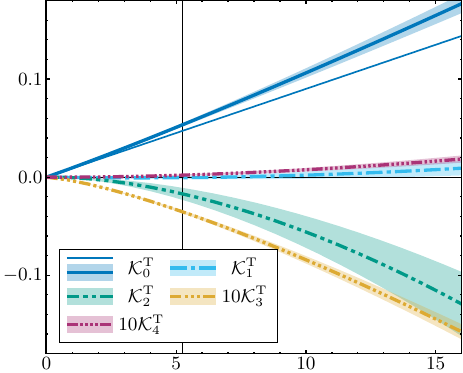}
    
    \phantom{\footnotesize$0$}%
    \leftfudge
    \includegraphics[scale=\figscale]{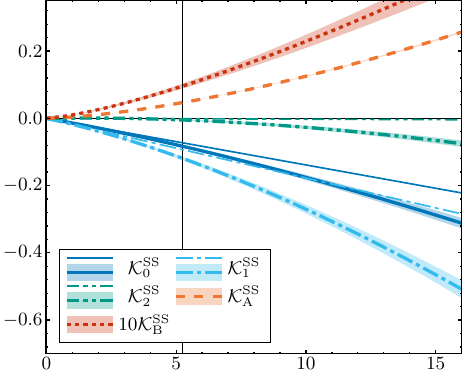}
    \includegraphics[scale=\figscale]{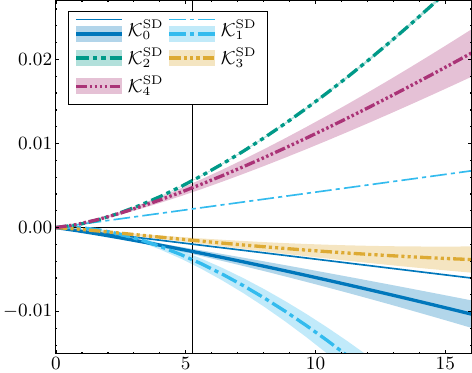}

    \leftfudge
    \includegraphics[scale=\figscale]{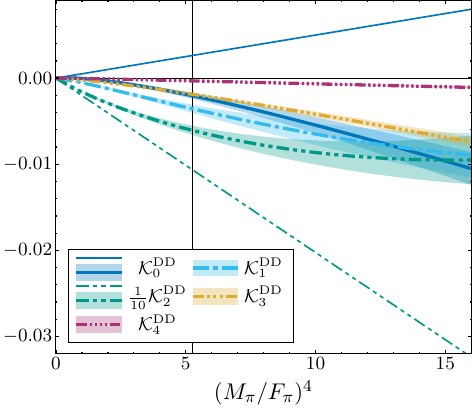}
    \includegraphics[scale=\figscale]{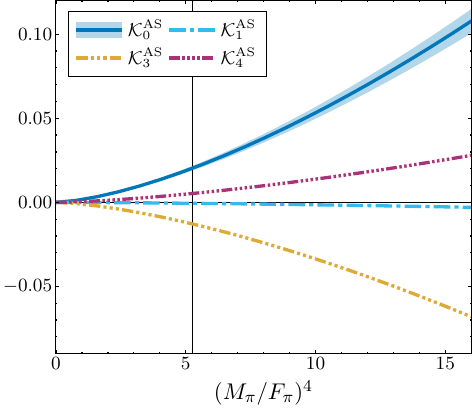}
    \caption{
        $\LO{+}\NLO$ ChPT predictions for $\Kdf$ as functions of $(\Mpi/\Fpi)^4$, with the physical point $(\Mpi/\Fpi)^4\approx 5.25$ shown as a vertical line.
        Colored bands represent uncertainties inherited from the LECs (see the main text), and thin lines represent $\LO$-only contributions when present.
        The legends unambiguously indicate whether uncertainty bands and $\LO$ contributions are absent or just too small to see.
        The coefficients are grouped by isospin, and the lines are drawn so that the number of dots reflects the numeric index on $\cK_X$ when applicable.
        Some coefficients have been rescaled for legibility.}
    \label{fig:results-zoomed}
\end{figure}

\begin{figure}
    % \begin{comment} % Uncomment for fast version with pre-rendered plots
    \phantom{\footnotesize$0$}%
    \leftfudge
    \includegraphics[scale=\figscale]{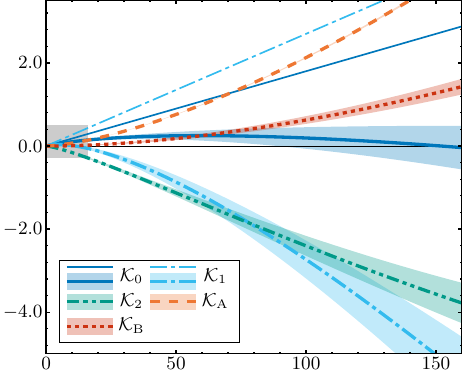}
    \includegraphics[scale=\figscale]{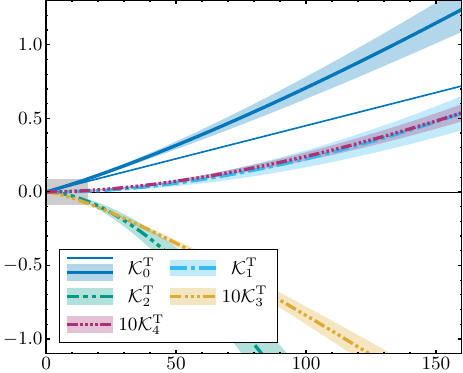}

    \leftfudge
    \includegraphics[scale=\figscale]{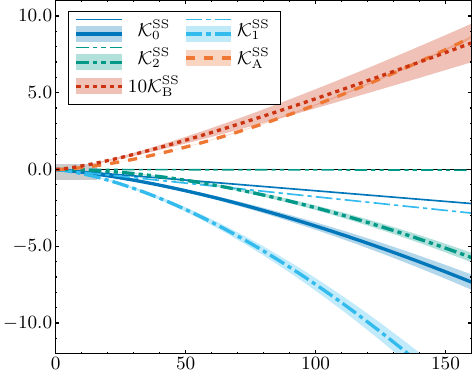}
    \includegraphics[scale=\figscale]{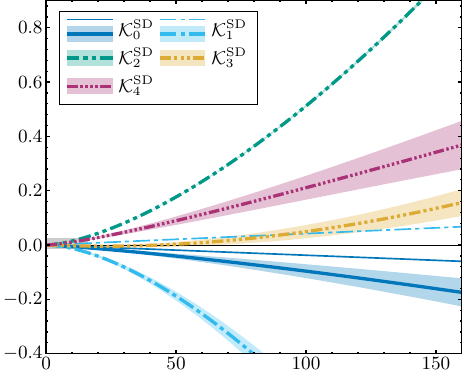}

    \phantom{\footnotesize$0$}%
    \leftfudge
    \includegraphics[scale=\figscale]{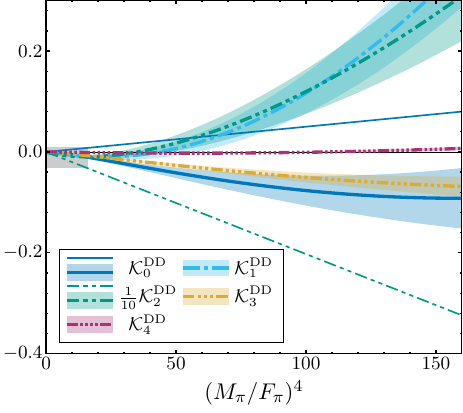}
    \includegraphics[scale=\figscale]{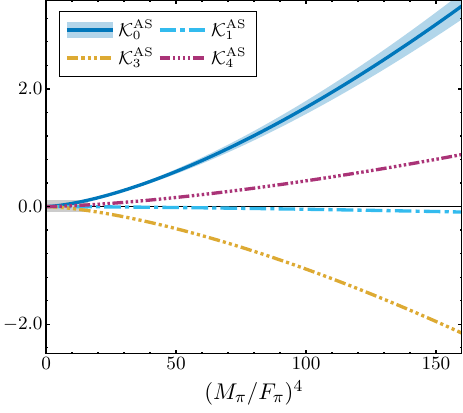}
    \caption{
        The same quantities as in \cref{fig:results-zoomed} but shown over the much wider range of $\Mpi/\Fpi$ used in \rcite{Baeza:2023ljl};
        the top left panel corresponds to figs.~3 and~4 in that work, although no fits to lattice data are shown.
        The regions covered in \cref{fig:results-zoomed} are shown as gray boxes.}
    \label{fig:results}
\end{figure}

Our full results are stated in \cref{tab:NLO-results}, supplemented by \cref{tab:LO-results,tab:BH} for the LO contributions and cutoff-dependent remainders, respectively.
A numerical comparison of the different contributions to $\mKdf$ is given in \cref{tab:numeric-results}.
The results are plotted as functions of $\Mpi/\Fpi$ at a scale $\mu\approx4\pi\Fphys$, as was done in \rcite{Baeza:2023ljl}, in \cref{fig:results,fig:results-zoomed}.
The former shows the region close to the physical pion mass, while the latter shows an extended region that includes $\Mpi/\Fpi$ values used in recent three-pion lattice calculations.

At fixed $\Mpi/\Fpi$, there are two sources of uncertainty in the results: the LECs of ChPT and higher-order corrections.
We make no attempt to estimate the latter, so the errors shown in \cref{tab:numeric-results,fig:results-zoomed,fig:results} are entirely due to the LECs; see \rcite{Baeza:2023ljl} for details and the specific values that we use.
Note in \cref{fig:results,fig:results-zoomed} that some lines lack error bands since they do not depend on the LECs (see \cref{tab:NLO-results}).
Other error bands are too narrow to make out.

In \rcite{Baeza:2023ljl}, we found, for maximal isospin, poor convergence of the chiral expansion for pion masses near the upper end of the displayed range ($M_\pi \approx 350$\;MeV).
This can be seen from the results for $\cK_0$ and $\cK_1$ in the top left panels of the figures.
We find here that this result is generic, with large corrections seen for most of the coefficients in the threshold expansion.
Convergence at the physical pion mass is, however, reasonable in all cases, with the exception of $\KDD[0]$.

\begin{table}[tp]
    \centering
    \hspace{-.3cm}
    \begin{tabular}{cR@{$\tFM6$}L|R@{\:$\big($}R@{\:}C@{\:}R@{\:$\big)$\:\:}C@{\:}R@{\:}C@{\:}R@{\:}C@{\:}R@{\:}C@{\:}R@{\:}C@{\:}R@{\:}}
        \toprule
        \multirow{5}{*}{\rotatebox{90}{$I=3$}}
        &&\Kiso            &    - \kappa & 105                       &+& 36 \logthree                    &+& 111L                    &-& 288\lr1                 &-& 432\lr2                 &-& 36\lr3              &+& 72\lr4              \\
        &&\Kisoone         &    - \kappa & \tfrac{1999}{20}          &+& 96 \logthree                    &+& 384L                    &-& 612\lr1                 &-& 1170\lr2                &&                      &+& 108\lr4             \\
        &&\Kisotwo         &      \kappa & \tfrac{605061}{1400}      &-& \tfrac{621}{10}\logthree       &+& 360L                    &-& 432\lr1                 &-& 864\lr2                 &&                      &&                      \\
        &&\KA              &      \kappa & \tfrac{196281}{560}       &-& \tfrac{135}{8}\logthree        &-& 9L                      &+& 27\lr1                  &+& \lrfrac[27]{2}2         &&                      &&                      \\
        &&\KB              &      \kappa & \tfrac{90423}{700}        &-& \tfrac{189}{40}\logthree       &+& 54L                     &-& 162\lr1                 &-& 81\lr2                  &&                      &&                      \\
        \midrule
        \multirow{5}{*}{\rotatebox{90}{$I=2$}}
        &&\KT              &    - \kappa & \tfrac{59113}{3240}       &+& \tfrac{1009}{144}\logthree     &&                          &-& 90\lr1                  &-& 9\lr2                   &&                      &+& 18\lr4              \\
        &&\KT[1]           &      \kappa & \tfrac{9486697}{453600}   &-& \tfrac{989}{480}\logthree      &+& \Lfrac[53]{2}           &-& \lrfrac[195]{2}1        &-& \lrfrac[123]{4}2        &&                      &&                      \\
        &&\KT[2]           &    - \kappa & \tfrac{1248031}{7200}     &+& \tfrac{5641}{320}\logthree     &-& \Lfrac[171]{2}          &+& \lrfrac[837]{2}1        &+& \lrfrac[189]{4}2        &&                      &&                      \\
        &&\KT[3]           &      \kappa & \tfrac{23833}{33600}      &-& \tfrac{317}{960}\logthree      &+& \Lfrac[27]{4}           &-& \lrfrac[45]{4}1         &-& \lrfrac[117]{8}2        &&                      &&                      \\
        &&\KT[4]           &      \kappa & \tfrac{332981}{75600}     &-& \tfrac{59}{960}\logthree       &+& \Lfrac[5]{3}            &-& 5\lr1                   &-& \lrfrac[5]{2}2          &&                      &&                      \\
        \midrule
        \multirow{15}{*}{\rotatebox{90}{$I=1$\hspace{2cm}}}
        &&\KSS             &    - \kappa & \tfrac{1955}{8}           &+& \tfrac{369}{4}\logthree        &-& \Lfrac[1237]{8}         &+& 342\lr1                 &+& 438\lr2                 &-& \lrfrac[57]{2}3     &-& \lrfrac[111]{2}4    \\
        &&\KSS[1]          &    - \kappa & \tfrac{191089}{320}       &+& \tfrac{993}{8}\logthree        &-& \Lfrac[24\,439]{64}     &+& \lrfrac[2637]{4}1       &+& \lrfrac[4125]{4}2       &+& \lrfrac[45]{16}3    &-& \lrfrac[1137]{16}4  \\
        &&\KSS[2]          &    - \kappa & \tfrac{34274101}{89600}   &+& \tfrac{33957}{320}\logthree    &-& \Lfrac[119\,505]{512}   &+& \lrfrac[8811]{32}1      &+& \lrfrac[18\,027]{32}2   &-& \lrfrac[405]{128}3  &-& \lrfrac[135]{128}4  \\
        &&\KSSA            &      \kappa & \tfrac{1102239}{2240}     &-& \tfrac{19575}{128}\logthree    &+& \Lfrac[273]{8}          &-& \lrfrac[297]{4}1        &-& \lrfrac[261]{4}2        &&                      &&                      \\
        &&\KSSB            &    - \kappa & \tfrac{521271}{5600}      &+& \tfrac{13419}{640}\logthree    &-& 36L                     &+& 108\lr1                 &+& 54\lr2                  &&                      &&                      \\
        \cmidrule{3-17}
        &&\KSD             &      \kappa & \tfrac{10853}{160}        &-& \tfrac{255}{8}\logthree        &+& \Lfrac[23]{16}          &+& 36\lr1                  &-& \lrfrac[39]{2}2         &-& \lrfrac[9]{4}3      &-& \lrfrac[3]{2}4      \\
        &&\KSD[1]          &      \kappa & \tfrac{643087}{8960}      &-& \tfrac{3543}{64}\logthree      &-& \Lfrac[1647]{128}       &+& \lrfrac[585]{8}1        &-& \lrfrac[9]{8}2          &+& \lrfrac[81]{32}3    &+& \lrfrac[27]{16}4    \\
        &&\KSD[2]          &      \kappa & \tfrac{166953}{2240}      &+& \tfrac{513}{128}\logthree      &+& \Lfrac[21]{4}           &&                          &-& \lrfrac[63]{4}2         &&                      &&                      \\
        &&\KSD[3]          &      \kappa & \tfrac{27783}{320}        &-& \tfrac{3699}{128}\logthree     &+& \Lfrac[75]{4}           &+& \lrfrac[81]{4}1         &-& \lrfrac[531]{8}2        &&                      &&                      \\
        &&\KSD[4]          &      \kappa & \tfrac{109539}{5600}      &-& \tfrac{11097}{320}\logthree    &-& \Lfrac[39]{2}           &+& \lrfrac[297]{4}1        &+& \lrfrac[171]{8}2        &&                      &&                      \\
        \cmidrule{3-17}
        &&\KDD             &      \kappa & \tfrac{49121}{3240}       &+& \tfrac{1259}{144}\logthree     &+& 28L                     &-& 54\lr1                  &-& 63\lr2                  &&                      &+& 2\lr4               \\
        &&\KDD[1]          &      \kappa & \tfrac{11178103}{453600}  &+& \tfrac{4279}{480}\logthree     &+& \Lfrac[265]{6}          &-& \lrfrac[149]{2}1        &-& \lrfrac[381]{4}2        &&                      &&                      \\
        &&\KDD[2]          &      \kappa & \tfrac{27345737}{50400}   &-& \tfrac{11869}{320}\logthree    &-& \Lfrac[123]{2}          &+& \lrfrac[1251]{2}1       &+& \lrfrac[459]{4}2        &&                      &-& 81\lr4              \\
        &&\KDD[3]          &      \kappa & \tfrac{150229}{11200}     &+& \tfrac{449}{320}\logthree      &+& \Lfrac[45]{4}           &-& \lrfrac[217]{12}1       &-& \lrfrac[593]{24}2       &&                      &&                      \\
        &&\KDD[4]          &      \kappa & \tfrac{212299}{75600}     &+& \tfrac{83}{320}\logthree       &+& \Lfrac[25]{9}           &-& \lrfrac[7]{3}1          &-& \lrfrac[43]{6}2         &&                      &&                      \\
        \midrule
        \multirow{4}{*}{\rotatebox{90}{$I=0$}}
        &&\KAS             &      \kappa & \tfrac{2721}{20}          &-& \tfrac{81}{2}\logthree         &&                          &-& 162\lr1                 &+& 81\lr2                  &&                      &&                      \\
        &&\KAS[1]          &      \kappa & 132                       & &                                 &&                          &&                          &&                          &&                      &&                      \\
        &&\KAS[3]          &    - \kappa & \tfrac{164673}{1120}      &-& \tfrac{2187}{32}\logthree      &&                          &&                          &&                          &&                      &&                      \\
        &&\KAS[4]          &      \kappa & \tfrac{28863}{448}        &-& \tfrac{3645}{128}\logthree     &&                          &&                          &&                          &&                      &&                      \\
        \bottomrule
    \end{tabular}
    \caption{
        The full $\NLO$ results for $\Kdf$ up to quadratic order in the threshold expansion (cubic for $I=0$), combining the $\OPE$, $\sOPE$, $\nonOPE$, and bull's head contributions, i.e., \cref{tab:OPE,tab:s-OPE,tab:non-OPE,tab:BH}.
        The corresponding $\LO$ results are given in \cref{tab:LO-results}.
        For compactness, we omit the cutoff-dependent remainders $\cD_X$; they are given in \cref{tab:BH} and further studied in \cref{app:cutoff}.}
    \label{tab:NLO-results}
\end{table}

\begin{table}[tp]
    \centering
    \newcommand{\nodot}[1]{\multicolumn{1}{R@{\,}}{#1}&}
    \begin{tabular}{cL|R@{.}L|R@{.}L@{ }R@{.}L|R@{.}L@{ }R@{.}L@{ }R@{.}L@{ }R@{.}L@{ }R@{.}L}
        \toprule
            &
            &   \multicolumn{2}{c|}{\multirow{2}{*}{Total}}
            &   \multicolumn{2}{c@{}}{\multirow{2}{*}{$\LO{\times}\tFM4$}}
            &   \multicolumn{2}{c|}{\multirow{2}{*}{$\NLO{\times}\tFM6$}}
            &   \multicolumn{1}{c@{}}{} &
                \multicolumn{7}{l}{\hspace{-2.5mm}\raisebox{-1ex}{$\overbrace{\rule{6.0cm}{0pt}}^{\NLO\times\Fpi^6/\Mpi^6}$}}
        \\
            &
            &   \multicolumn{2}{c|}{}
            &   \multicolumn{2}{C@{}}{}
            &   \multicolumn{2}{C|}{}
            &   \multicolumn{2}{C@{}}{\OPE}
            &   \multicolumn{2}{C@{}}{\sOPE}
            &   \multicolumn{2}{C@{}}{\nonOPE}
            &   \multicolumn{2}{C@{}}{\BH}
            \\
        \midrule
        \multirow{5}{*}{\rotatebox{90}{$I=3$}}
        &\Kiso      &   61&6(3.0)   &   \nodot{18}  &   -2&65(26)   &   0&50(53)        &   \nodot{}    &   -2&04(28)   &   -1&11  \\
        &\Kisoone   &   33&4(5.3)   &   \nodot{27}  &   -9&04(46)   &   -1&8(1.0)       &   \nodot{}    &   -3&75(61)   &   -3&48  \\
        &\Kisotwo   &   -67&4(2.8)  &   \nodot{}    &   -5&79(24)   &   -5&11(58)       &   \nodot{}    &   1&43(37)    &   -2&11  \\
        &\KA        &   25&77(18)   &   \nodot{}    &   2&21(2)     &   -2&76(15)       &   \nodot{}    &   3&00(14)    &   1&97   \\
        &\KB        &   1&4(1.1)    &   \nodot{}    &   0&12(9)     &   -0&22(37)       &   \nodot{}    &   0&25(28)    &   0&08   \\
        \midrule
        \multirow{5}{*}{\rotatebox{90}{$I=2$}}
        &\KT        &   26&12(88)   &   4&5         &   0&26(8)     &   1&05(18)        &   \nodot{}    &   -0&56(10)   &   -0&23  \\
        &\KT[1]     &   -0&28(67)   &   \nodot{}    &   -0&02(6)    &   0&16(14)        &   \nodot{}    &   0&02(9)     &   -0&20  \\
        &\KT[2]     &   -8&2(3.0)   &   \nodot{}    &   -0&70(26)   &   1&33(23)        &   \nodot{}    &   -2&42(7)    &   0&39   \\
        &\KT[3]     &   -1&72(7)    &   \nodot{}    &   -0&147(6)   &   -0&085(6)       &   \nodot{}    &   -0&064      &   -0&13  \\
        &\KT[4]     &   -0&10(3)    &   \nodot{}    &   -0&008(3)   &   -0&01(1)        &   \nodot{}    &   0&014(7)    &   -0&02  \\
        \midrule
        \multirow{15}{*}{\rotatebox{90}{$I=1$}}
        &\KSS       &   -81&2(2.8)  &   -13&88      &   -0&85(24)   &   -9&21(63)       &   -3&06(20)   &   15&81(72)   &   -4&38  \\
        &\KSS[1]    &   -116&0(4.8) &   -17&77      &   -2&12(41)   &   -9&10(42)       &   -3&52(22)   &   20&32(82)   &   -9&82  \\
        &\KSS[2]    &   -4&5(1.8)   &   -0&26       &   -0&27(15)   &   -4&80(65)       &   -0&29(16)   &   6&75(94)    &   -1&93  \\
        &\KSSA      &   45&25(48)   &   \nodot{}    &   3&88(4)     &   -1&77(55)       &   -0&83(17)   &   -0&51(67)   &   6&99   \\
        &\KSSB      &   9&31(72)    &   \nodot{}    &   0&80(6)     &   -4&19(35)       &   \nodot{}    &   2&80(41)    &   2&20   \\
        \cmidrule{2-16}
        &\KSD       &   -2&77(30)   &   -0&38       &   -0&07(3)    &   -0&99(11)       &   -0&52(4)    &   1&42(16)    &   0&02   \\
        &\KSD[1]    &   -3&65(54)   &   0&42        &   -0&50(5)    &   0&40(3)         &   -0&31(1)    &   -0&16(7)    &   -0&44  \\
        &\KSD[2]    &   5&39(4)     &   \nodot{}    &   0&463(4)    &   0&11(12)        &   0&45(4)     &   -0&53(17)   &   0&43   \\
        &\KSD[3]    &   -1&48(28)   &   \nodot{}    &   -0&13(2)    &   -0&00(5)        &   \nodot{}    &   -0&46(7)    &   0&34   \\
        &\KSD[4]    &   4&59(51)    &   \nodot{}    &   0&39(4)     &   0&13(6)         &   \nodot{}    &   -0&58(10)   &   0&84   \\
        \cmidrule{2-16}
        &\KDD       &   -1&97(35)   &   0&5         &   -0&39(3)    &   1&00(14)        &   \nodot{}    &   -1&92(13)   &   0&53   \\
        &\KDD[1]    &   -3&48(47)   &   \nodot{}    &   -0&30(4)    &   -0&04(9)        &   \nodot{}    &   -0&81(5)    &   0&55   \\
        &\KDD[2]    &   -59&8(5.2)  &   -20&25      &   3&80(44)    &   1&51(16)        &   -2&51(30)   &   -2&09(72)   &   6&88   \\
        &\KDD[3]    &   -1&82(11)   &   \nodot{}    &   -0&16(1)    &   -0&087(5)       &   \nodot{}    &   -0&12(1)    &   0&06   \\
        &\KDD[4]    &   -0&31(2)    &   \nodot{}    &   -0&027(2)   &   -0&002(10)      &   \nodot{}    &   -0&06(1)    &   0&04   \\
        \midrule
        \multirow{4}{*}{\rotatebox{90}{$I=0$}}
        &\KAS       &   19&6(1.3)   &   \nodot{}    &   1&68(11)    &   0&36(20)        &   \nodot{}    &    1&95(9)    &   -0&63  \\
        &\KAS[1]    &   -0&54       &   \nodot{}    &   -0&046      &   \nodot{}        &   \nodot{}    &   -1&231      &   1&19   \\
        &\KAS[3]    &   -12&38      &   \nodot{}    &   -1&063      &   -0&240          &   \nodot{}    &   -1&312      &   0&49   \\
        &\KAS[4]    &   5&090       &   \nodot{}    &   0&437       &   -0&150          &   \nodot{}    &   0&250       &   0&34   \\
        \bottomrule
    \end{tabular}
    \caption{
        Numerical values of the expressions given in \cref{tab:LO-results,tab:OPE,tab:s-OPE,tab:non-OPE,tab:BH,tab:NLO-results}.
        Note that only the ``Total'' column depends on the ratio $\Mpi/\Fpi$ (here evaluated at the physical point, $\Mpi/\Fpi\approx1.50$), and only the ``BH'' (bull's head) column depends on the cutoff [here using the standard choice, \cref{eq:H-standard}].
        Note also that the ``BH'' column shows $\mKdf^\BH=-\mDBH$.
        Numbers in parentheses indicate errors inherited from the LECs (see the main text); entries without errors are exact up to rounding.
        The small uncertainty in $\Fpi/\Mpi$ is not taken into account.
        Identically zero entries are left blank.}
    \label{tab:numeric-results}
\end{table}

\newcommand{\addplotswithoutLO}[3][1]{%
    \addplot[NLOline, line#3, NLO legend]                 table[x=MF4, y expr={#1*\thisrow{NLO}},  col sep=tab] {PlotData/K#2#3.dat};%
    \addplot[errline, line#3, name path global=upperK#2#3] table[x=MF4, y expr={#1*\thisrow{errU}}, col sep=tab] {PlotData/K#2#3.dat};%
    \addplot[errline, line#3, name path global=lowerK#2#3] table[x=MF4, y expr={#1*\thisrow{errL}}, col sep=tab] {PlotData/K#2#3.dat};%
    \addplot[line#3, errfill] fill between[of=upperK#2#3 and lowerK#2#3];}
\newcommand{\addplotswithLO}[3][1]{%
    \addplot[LOline, line#3]                              table[x=MF4, y expr={#1*\thisrow{LO}},  col sep=tab] {PlotData/K#2#3.dat};%
    \addplot[NLOline, line#3, LO plus NLO legend]         table[x=MF4, y expr={#1*\thisrow{NLO}},  col sep=tab] {PlotData/K#2#3.dat};%
    \addplot[errline, line#3, name path global=upperK#2#3] table[x=MF4, y expr={#1*\thisrow{errU}}, col sep=tab] {PlotData/K#2#3.dat};%
    \addplot[errline, line#3, name path global=lowerK#2#3] table[x=MF4, y expr={#1*\thisrow{errL}}, col sep=tab] {PlotData/K#2#3.dat};%
    \addplot[line#3, errfill] fill between[of=upperK#2#3 and lowerK#2#3];}
\newcommand{\addexactplot}[3][1]{%
    \addplot[NLOline, line#3, exact NLO legend]           table[x=MF4, y expr={#1*\thisrow{NLO}},  col sep=tab] {PlotData/K#2#3.dat};}

\newcommand{\physpoint}{5.25}
\newcommand{\plotKmin}{-.3}
\newcommand{\plotKmax}{+.5}
\newcommand{\plotKTmin}{-.09}
\newcommand{\plotKTmax}{+.09}
\newcommand{\plotKSSmin}{-.5}
\newcommand{\plotKSSmax}{+.42}
\newcommand{\plotKSDmin}{-.025}
\newcommand{\plotKSDmax}{+.027}
\newcommand{\plotKDDmin}{-.032}
\newcommand{\plotKDDmax}{+.009}
\newcommand{\plotKASmin}{-.07}
\newcommand{\plotKASmax}{+.12}

% This is [BLAH]
\subsection{Range of validity of the threshold expansion}\label{sec:validity}

In this section, we compare the threshold-expanded results with the full $\mKdf$, the latter numerically evaluated using the single-parameter kinematic configurations described in \cref{app:families}.
This allows us to verify the validity of the threshold expansion.
We use $\Mpi=340$\;MeV throughout, which is the heaviest of the masses used in \rcite{Blanton:2021llb}; for brevity, we omit physical-mass results, which are qualitatively similar and typically converge slightly better.

\renewcommand{\plotConvergence}[3]{
    \addplot+[#1line, color={#1col}, name path=#1num, backbone legend] table[x index=1, #2, col sep=tab] {Data340/NumericKdf/#3};%
    \addplot+[draw=none, mark=none, name path=#1thr, forget plot] table[x index=1, #2, col sep=tab] {Data340/ThresholdKdf/#3};%
    \addplot[color={#1col}, opacity=.5, forget plot] fill between [of=#1num and #1thr] {};%
}

\newcommand{\plotIthreemin}{-20}
\newcommand{\plotIthreemax}{+20}

\newcommand{\plotItwooneonemin}{-5}
\newcommand{\plotItwooneonemax}{+4}

\newcommand{\plotIoneSSmin}{-45}
\newcommand{\plotIoneSSmax}{+35}

\newcommand{\plotIoneSDonemin}{-1.2}
\newcommand{\plotIoneSDonemax}{+0.8}

\newcommand{\plotIoneDDoneonemin}{-4}
\newcommand{\plotIoneDDoneonemax}{+5}

\newcommand{\plotIzeromin}{-0.8}
\newcommand{\plotIzeromax}{+5}

\begin{figure}
    % \begin{comment} % Uncomment for fast version with pre-rendered plots
    \leftfudge
    \includegraphics[scale=\figscale]{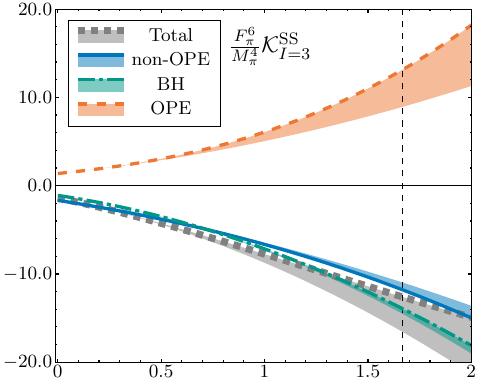}
    \includegraphics[scale=\figscale]{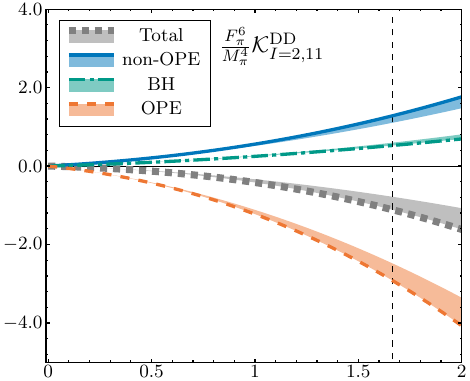}

    \leftfudge
    \includegraphics[scale=\figscale]{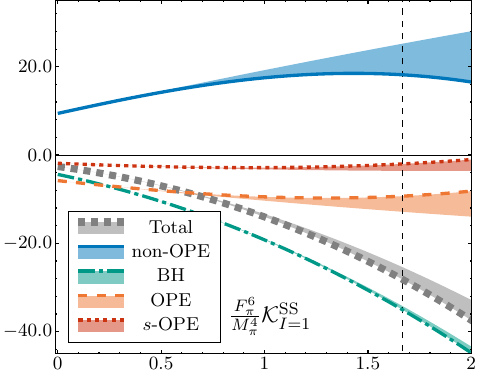}
    \includegraphics[scale=\figscale]{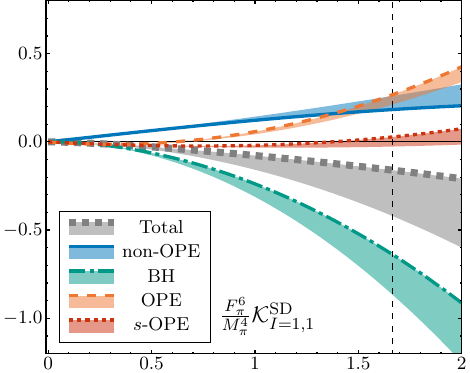}

    \leftfudge
    \phantom{\footnotesize0}%
    \includegraphics[scale=\figscale]{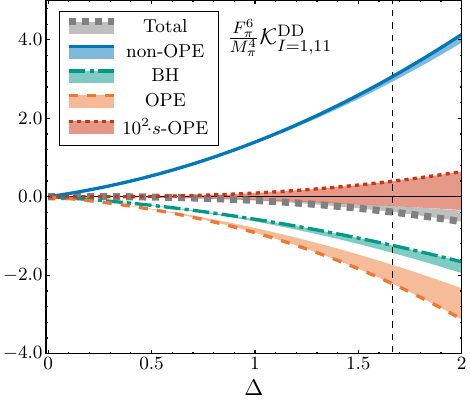}
    \phantom{\footnotesize$-$}%
    \includegraphics[scale=\figscale]{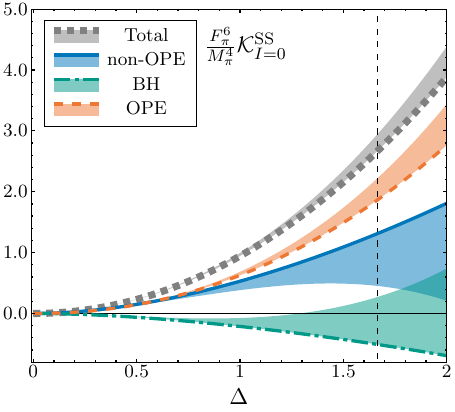}
    \caption{
        The convergence of the threshold expansion at $\Mpi=340$\;MeV, shown for various components of $\mKdf^\NLO$ (LO is omitted) in the symmetric basis using the kinematic configuration described in \cref{app:families}.
        Lines represent the full $\mKdf^\NLO$, obtained numerically.
        Note that the ``BH'' contribution is $\mKdf^\BH=-\mDBH$.
        Colored bands show the difference between these and the corresponding results obtained from the threshold expansion, up to the order calculated in this work.
        Thus, a narrow band indicates good convergence of the threshold expansion, whereas a wide one shows how much and in which direction it differs from the full $\mKdf^\NLO$ values.
        The dashed vertical line indicates the $5\pi$ threshold.}
    \label{fig:convergence}
\end{figure}

\Cref{fig:convergence} (cf.\ fig.~6 in \rcite{Baeza:2023ljl}) shows the convergence for six of the 15 nonzero components of $\mKdf$.%
\footnote{
    Namely, one component for $I=3$, four for $I=2$, nine for $I=1$ (one for SS, four for DD, and two each for SD and DS), and one for $I=0$.}
The remaining components (i.e., $\mKdfI{1,\DS}$, the second component of $\mKdfI{1,\SD}$, etc.) are related to these via permutations of the initial and final states and are qualitatively similar.
The plots show very good convergence across all components for $\Delta\lesssim1$, and, in many cases, the convergence of the total $K$-matrix is significantly better than that of the individual parts ($\nonOPE$, $\BH$, etc.).
In particular, the formal failure of the convergence of the $\sOPE$ contribution beyond $\Delta=8/9$ (recall \cref{sec:LO-sOPE}) does not seem to notably affect the overall convergence:
The $\sOPE$ contribution is relatively small and converges poorly rather than diverging.

\newcommand{\plotIthreeOPEPWmin}{-1}
\newcommand{\plotIthreeOPEPWmax}{+21}

\newcommand{\plotItwooneoneOPEPWmin}{-4}
\newcommand{\plotItwooneoneOPEPWmax}{+2}
\newcommand{\plotItwoonetwoOPEPWmin}{-1}
\newcommand{\plotItwoonetwoOPEPWmax}{+11}
\newcommand{\plotItwotwooneOPEPWmin}{-8}
\newcommand{\plotItwotwooneOPEPWmax}{+4}
\newcommand{\plotItwotwotwoOPEPWmin}{-7}
\newcommand{\plotItwotwotwoOPEPWmax}{+4}

\newcommand{\plotIzeroOPEPWmin}{-0.5}
\newcommand{\plotIzeroOPEPWmax}{+3}

\newcommand{\plotIoneDDoneoneOPEPWmin}{-4}
\newcommand{\plotIoneDDoneoneOPEPWmax}{+3}
\newcommand{\plotIoneDDonetwoOPEPWmin}{-1}
\newcommand{\plotIoneDDonetwoOPEPWmax}{+9}
\newcommand{\plotIoneDDtwooneOPEPWmin}{-7}
\newcommand{\plotIoneDDtwooneOPEPWmax}{+4}
\newcommand{\plotIoneDDtwotwoOPEPWmin}{-6}
\newcommand{\plotIoneDDtwotwoOPEPWmax}{+4}

\newcommand{\plotIoneSDoneOPEPWmin}{-0.2}
\newcommand{\plotIoneSDoneOPEPWmax}{+0.5}
\newcommand{\plotIoneSDtwoOPEPWmin}{-1.5}
\newcommand{\plotIoneSDtwoOPEPWmax}{+1.5}
\newcommand{\plotIoneDSoneOPEPWmin}{-0.5}
\newcommand{\plotIoneDSoneOPEPWmax}{+1.5}
\newcommand{\plotIoneDStwoOPEPWmin}{-1.5}
\newcommand{\plotIoneDStwoOPEPWmax}{+1.5}
\newcommand{\plotIoneSSOPEPWmin}{-14}
\newcommand{\plotIoneSSOPEPWmax}{+1}

\begin{figure}
    % \begin{comment} % Uncomment for fast version with pre-rendered plots
    \leftfudge
    \phantom{\footnotesize$-$}%
    \includegraphics[scale=\figscale]{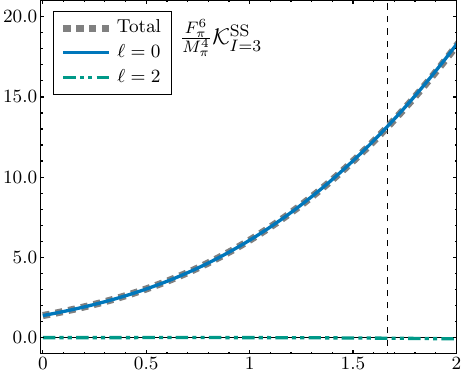}
    \includegraphics[scale=\figscale]{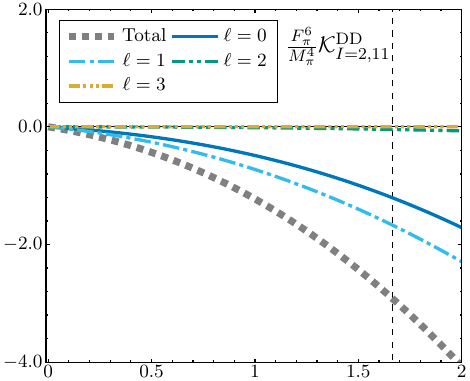}

    \leftfudge
    \includegraphics[scale=\figscale]{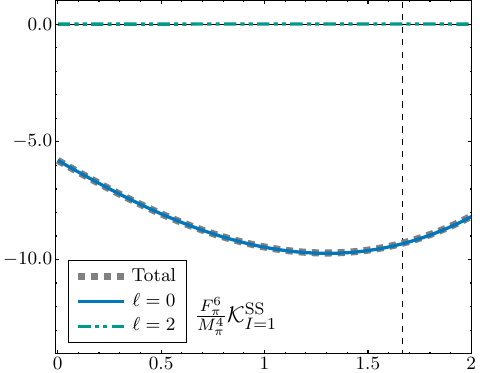}
    \includegraphics[scale=\figscale]{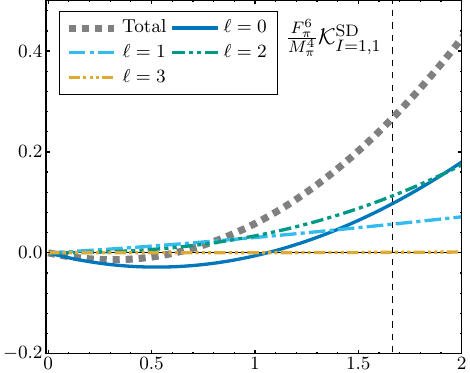}

    \leftfudge
    \phantom{\footnotesize0}%
    \includegraphics[scale=\figscale]{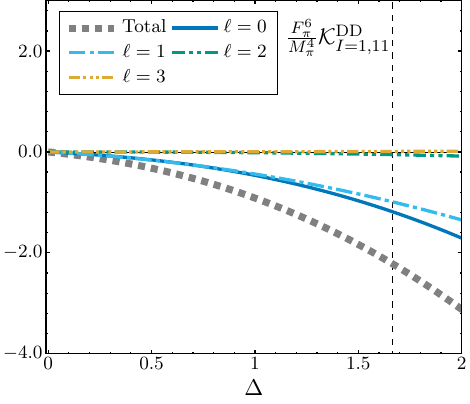}
    \phantom{\footnotesize$-$}%
    \includegraphics[scale=\figscale]{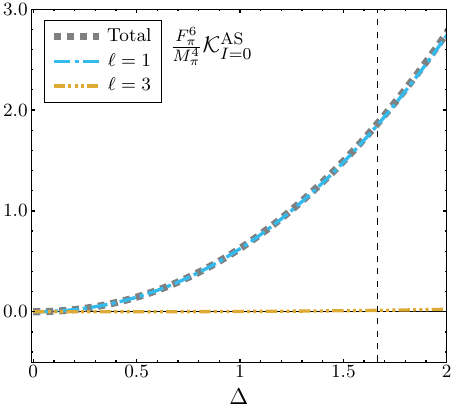}
    \caption{
        Comparison of contributions to $\mKdf^{\NLO,\OPE}$ from different interacting-pair partial waves in $\m M_2^\NLO$,
        numerically evaluated at \mbox{$\Mpi=340$\;MeV} using the same kinematic configuration as in \cref{fig:convergence}.
        Lines are drawn so that the number of dots equals $\ell$.
        Partial waves that are identically zero, as well as the negligibly small $\ell>3$ contributions, have been omitted.}
    \label{fig:partial-waves}
\end{figure}

For the NLO OPE contribution, we also study the relative contribution of higher partial waves of the interacting pair, as shown in \cref{fig:partial-waves} (cf.\ fig.~7 in \rcite{Baeza:2023ljl}).
We see that the result is dominated by the lowest partial waves, with negligible contributions from those above $\ell=3$, the highest that appears in our threshold expansion.
This partly motivates the good convergence seen in \cref{fig:convergence}.

\section{Conclusions and outlook}\label{sec:conclusions}

We have computed the NLO ChPT result for the three-particle $K$-matrix, $\mKdf$, in all three-pion isospin channels, thereby extending the maximum-isospin results of \rcite{Baeza:2023ljl}.
The LO result is given in \cref{tab:LO-results}, while the NLO expressions are provided in \cref{tab:NLO-results}.
Numerical values of the LO and NLO three-pion $K$-matrix at the physical point are given in \cref{tab:numeric-results}.

In order to gain insight in the behavior of the $K$-matrix, we have performed numerical investigations of its convergence.
In \cref{fig:results-zoomed,fig:results}, we observe that NLO contributions are not necessarily small compared to LO contributions, especially not at larger-than-physical pion masses.
Thus, the poor convergence of the chiral expansion for the three-particle $K$-matrix, previously observed at maximal isospin~\cite{Baeza:2023ljl}, appears to be generic for all three-pion systems.

On a more positive note, we find that the threshold expansion, truncated at quadratic order (cubic order for $I=0$), provides a good description of the full NLO result up to the five-pion inelastic threshold.
This is illustrated in \cref{fig:convergence}.
We also find that the OPE contribution to $\mKdf$ is largely dominated by $s$-, $p$- and $d$-waves, as shown in \cref{fig:partial-waves}.

With the list of $K$-matrix components now complete for three-pion systems up to the given orders, it will be interesting to compare future lattice results to these predictions;
currently, no results exist beyond those already discussed in \rcite{Baeza:2023ljl}.
We also note that, although the technical complexity of the calculation has increased compared to that at maximum isospin, there have been no new hurdles that could not be dealt with effectively
using the techniques developed in \rcite{Baeza:2023ljl}.

This work provides an encouraging outlook towards further developments within the program of multi-hadron dynamics at the interface of lattice QCD and effective theories.
Next steps involve the inclusion of particles other than pions.
In particular, consideration of systems with kaons and pions would allow comparison with available lattice results~\cite{Draper:2023boj}.
Another generalization would be to use ChPT to determine the form of the intermediate short-distance $K\to3\pi$ decay quantity, $A_{K3\pi}^{\rm PV}$, introduced in \rcite{Hansen:2021ofl}.
This quantity plays an analogous role in three-pion decays to that of $\mKdf$ in three-to-three scattering.
This connection to ChPT has already been exploited at LO in another version of the three-body formalism~\cite{Pang:2023jri}.

\subsection*{Acknowledgments}

The work of JBB was supported by the Spanish MU grant FPU19/04326. Additionally, JBB received support from the European project H2020-MSCA-ITN-2019//860881-HIDDeN and the staff exchange grant 101086085-ASYMMETRY, and from the Spanish Ministerio de Ciencia e Innovaci\'on project PID2020-113644GB-I00.
The work of JB and MS was supported by the Swedish Research Council grants contract numbers 2016-05996 and 2019-03779;
MS also received support from the French government under the France 2030 investment plan, as part of the ``Initiative d'Excellence d'Aix-Marseille Université -- A\!*MIDEX'' under grant AMX-22-RE-AB-052.
The work of FRL was supported in part by the U.S.\ Department of Energy (USDOE), Office of Science, Office of Nuclear Physics, under grant Contract Numbers DE-SC0011090 and DE-SC0021006. FRL also acknowledges financial support by the Mauricio and Carlota Botton Fellowship.
TH was supported by the Czech Science Foundation grant 23-06770S, by the Charles University grant PRIMUS 23/SCI/025, and by the MSCA Fellowships CZ -- UK2 project No.~CZ.02.01.01/00/22\_010/0008115 financed by Programme Johannes Amos Comenius (OP JAK).
The work of SRS was supported in part by the USDOE grant No.~DE-SC0011637.

JBB and FRL would like to thank the Physics Department at the University of Washington for its hospitality during a visit in which this work was initiated.

\appendix

\section{Details on the bull's head subtraction}\label{app:bullhead}
% Wherein we rehash appendix A from the previous paper, which could be interesting
% since many of the new coefficients seem to have much worse analytic approximations
% and therefore possibly stronger cutoff dependence

In \cref{sec:NLO-BH}, we presented the results of the bull's head calculation using the analytic approximation, resulting in the compact but highly scheme-dependent values of \cref{tab:BH,tab:NLO-results}.
Without using the approximation, the exact results can be stated in terms of the following $14$ cutoff-dependent integrals, defined in \cref{eq:Hmnp}:
\begin{equation}\label{eq:Hmnp-list}
    \begin{gathered}
        H_{0,0,0}\,,\quad
        H_{0,1,0}\,,\quad
        H_{0,2,0}\,,\quad
        H_{2,0,0}\,,\quad
        H_{4,0,0}\,,\quad
        H_{6,0,0}\,,\\
        H_{-4,0,0}\,,\quad
        H_{-2,0,0}\,,\quad
        H_{0,0,1}\,,\quad
        H_{0,0,2}\,,\quad
        H_{0,0,3}\,,\quad
        H_{0,1,1}\,,\quad
        H_{0,1,2}\,,\quad
        H_{0,2,1}\,;\quad
    \end{gathered}
\end{equation}
only those in the first line are present at maximum isospin.
The result is presented in terms of the $H_{m,n,p}$ (and $H_{m,n}\equiv H_{m,n,0}$, corresponding to the notation of \rcite{Baeza:2023ljl}) in \cref{tab:BH-exact}.
Note that this form is reached by applying \cref{eq:alg-rel} as well as the integration-by-parts relation derived in \rcite{Baeza:2023ljl}, namely,
\begin{equation}
    H_{m,n+1} + H_{m-2,n+1} = \tfrac{1}{6}\big[(2-m)H_{m,n,0} - (m+1)H_{m+2,n}\big]\,,
\end{equation}
which [unlike \cref{eq:alg-rel}] is made invalid by setting $H(x)=1$.
Therefore, \cref{tab:BH} is \emph{not} obtained directly from \cref{tab:BH-exact}, but from longer intermediate expressions (see \rcite{Baeza:2023ljl} for details).

\newcommand{\breaktbl}[1]{\\[-1mm]\multicolumn{3}{c|}{}&\qquad#1\;}
\begin{table}[tp]
    \centering
    % Please ignore the error message here; Overleaf doesn't understand the array package
    \begin{tabular}{cR@{$\tFM6$}L|>{$}p{0.8\textwidth}<{$}}
        \toprule
        \multirow{5}{*}{\rotatebox{90}{$I=3$}}
        &&\Kiso            &   - \frac{27}{2} H_{0,0} + \frac{9}{4} H_{2,0}        \\
        &&\Kisoone         &   - \frac{117}{4} H_{0,0} - \frac{189}{4} H_{0,1} + \frac{21}{8} H_{2,0} - \frac{3}{4} H_{4,0}         \\
        &&\Kisotwo         &   - \frac{243}{160} H_{0,0} - \frac{5751}{64} H_{0,1} - \frac{567}{8} H_{0,2} - \frac{2241}{320} H_{2,0} + \frac{423}{160} H_{4,0} + \frac{369}{1280} H_{6,0}         \\
        &&\KA              &   \frac{891}{64} H_{0,0} - \frac{1161}{128} H_{2,0} + \frac{45}{64} H_{4,0} + \frac{9}{128} H_{6,0}    \\
        &&\KB              &   - \frac{81}{320} H_{0,0} - \frac{297}{640} H_{2,0} - \frac{27}{160} H_{4,0} + \frac{27}{640} H_{6,0}    \\
        \midrule
        \multirow{5}{*}{\rotatebox{90}{$I=2$}}
        &&\KT              &   \frac{13}{2} H_{-4,0} + \frac{85}{24} H_{-2,0} - \frac{857}{288} H_{0,0} + \frac{4}{9} H_{0,0,1} + \frac{83}{128} H_{2,0} - \frac{1}{128} H_{4,0}     \\
        &&\KT[1]           &   - \frac{62}{5} H_{-4,0} + \frac{19}{24} H_{-2,0} - \frac{3407}{2880} H_{0,0} + \frac{7}{9} H_{0,0,1} - \frac{4}{9} H_{0,0,2} - \frac{259}{144} H_{0,1} + \frac{4}{9} H_{0,1,1} \breaktbl{+} \frac{2783}{2304} H_{2,0} - \frac{623}{2560} H_{4,0} + \frac{9}{2560} H_{6,0}      \\
        &&\KT[2]           &   - \frac{1341}{40} H_{-4,0} - \frac{5289}{160} H_{-2,0} - \frac{1193}{640} H_{0,0} - 4 H_{0,0,1} - \frac{8469}{2560} H_{2,0} - \frac{423}{2560} H_{4,0} \breaktbl{+} \frac{27}{1280} H_{6,0}         \\
        &&\KT[3]           &   \frac{9}{10} H_{-4,0} - \frac{1}{8} H_{-2,0} - \frac{539}{640} H_{0,0} + \frac{473}{1536} H_{2,0} + \frac{39}{1280} H_{4,0} - \frac{7}{7680} H_{6,0}     \\
        &&\KT[4]           &   - \frac{3}{40} H_{-4,0} - \frac{47}{160} H_{-2,0} - \frac{109}{640} H_{0,0} + \frac{349}{7680} H_{2,0} - \frac{11}{3840} H_{4,0} + \frac{1}{7680} H_{6,0}     \\
        \midrule
        \multirow{15}{*}{\rotatebox{90}{$I=1$\hspace{2cm}}}
        &&\KSS             &   - 36 H_{0,0} - \frac{45}{2} H_{0,0,1} + \frac{207}{32} H_{2,0}          \\
        &&\KSS[1]          &   - \frac{249}{8} H_{0,0} - \frac{225}{4} H_{0,0,1} + \frac{45}{2} H_{0,0,2} - \frac{4437}{32} H_{0,1} - \frac{45}{2} H_{0,1,1} + \frac{3}{64} H_{2,0} - \frac{69}{32} H_{4,0}          \\
        &&\KSS[2]          &   \frac{243}{40} H_{0,0} - \frac{1215}{32} H_{0,0,1} + \frac{1665}{32} H_{0,0,2} - \frac{45}{2} H_{0,0,3} - \frac{50463}{512} H_{0,1} - \frac{1665}{32} H_{0,1,1} \breaktbl{+} \frac{45}{2} H_{0,1,2} - \frac{13671}{64} H_{0,2} - \frac{45}{4} H_{0,2,1} - \frac{41733}{2560} H_{2,0} + \frac{44001}{5120} H_{4,0} + \frac{8487}{10240} H_{6,0}          \\
        &&\KSSA            &   \frac{135}{16} H_{0,0} - \frac{23895}{1024} H_{2,0} - \frac{171}{64} H_{4,0} + \frac{207}{1024} H_{6,0}          \\
        &&\KSSB            &   \frac{27}{160} H_{0,0} - \frac{13851}{5120} H_{2,0} - \frac{7047}{2560} H_{4,0} + \frac{621}{5120} H_{6,0}          \\
        \cmidrule{3-4}
        &&\KSD             &   \frac{45}{8} H_{-2,0} - 3 H_{0,0} - 3 H_{0,0,1} - \frac{57}{32} H_{2,0} + \frac{9}{64} H_{4,0}      \\
        &&\KSD[1]          &   - \frac{54}{5} H_{-2,0} - \frac{489}{640} H_{0,0} - \frac{69}{16} H_{0,0,1} + 3 H_{0,0,2} - \frac{1383}{64} H_{0,1} - 3 H_{0,1,1} - \frac{20193}{2560} H_{2,0} \breaktbl{+} \frac{1665}{1024} H_{4,0} - \frac{81}{1024} H_{6,0}        \\
        &&\KSD[2]          &   \frac{459}{80} H_{-2,0} + \frac{189}{80} H_{0,0} + \frac{189}{5120} H_{2,0} - \frac{729}{1280} H_{4,0} + \frac{27}{1024} H_{6,0}         \\
        &&\KSD[3]          &   - \frac{243}{320} H_{-2,0} - \frac{351}{160} H_{0,0} - \frac{2673}{1024} H_{2,0} + \frac{297}{2560} H_{4,0} + \frac{27}{1024} H_{6,0}         \\
        &&\KSD[4]          &   \frac{81}{320} H_{-2,0} - \frac{513}{320} H_{0,0} - \frac{2241}{640} H_{2,0} - \frac{27}{80} H_{4,0} + \frac{27}{640} H_{6,0}         \\
        \cmidrule{3-4}
        &&\KDD             &   - \frac{5}{2} H_{-4,0} + \frac{139}{24} H_{-2,0} + \frac{1825}{288} H_{0,0} + \frac{4}{9} H_{0,0,1} - \frac{153}{128} H_{2,0} + \frac{11}{128} H_{4,0}     \\
        &&\KDD[1]          &   \frac{47}{10} H_{-4,0} - \frac{859}{120} H_{-2,0} + \frac{12343}{2880} H_{0,0} + \frac{7}{9} H_{0,0,1} - \frac{4}{9} H_{0,0,2} - \frac{259}{144} H_{0,1} + \frac{4}{9} H_{0,1,1} \breaktbl{-}\frac{25541}{11520} H_{2,0} + \frac{813}{2560} H_{4,0} - \frac{99}{2560} H_{6,0}         \\
        &&\KDD[2]          &   \frac{441}{40} H_{-4,0} + \frac{11721}{160} H_{-2,0} + \frac{17329}{640} H_{0,0} -13 H_{0,0,1} - \frac{24849}{2560} H_{2,0} - \frac{11907}{2560} H_{4,0} \breaktbl{-} \frac{297}{1280} H_{6,0}          \\
        &&\KDD[3]          &   - \frac{9}{20} H_{-4,0} + \frac{33}{40} H_{-2,0} + \frac{1913}{1920} H_{0,0} - \frac{1471}{7680} H_{2,0} + \frac{13}{3840} H_{4,0} + \frac{77}{7680} H_{6,0}      \\
        &&\KDD[4]          &   \frac{3}{40} H_{-4,0} + \frac{23}{160} H_{-2,0} + \frac{191}{1920} H_{0,0} - \frac{31}{7680} H_{2,0} - \frac{139}{3840} H_{4,0} - \frac{11}{7680} H_{6,0}      \\
        \midrule
        \multirow{4}{*}{\rotatebox{90}{$I=0$}}
        &&\KAS             &   - \frac{27}{2} H_{-2,0} - \frac{27}{2} H_{0,0}         \\
        &&\KAS[1]          &   \frac{3321}{80} H_{-2,0} + \frac{4563}{320} H_{0,0} - \frac{621}{320} H_{2,0}         \\
        &&\KAS[3]          &   - \frac{729}{560} H_{-4,0} + \frac{26\,811}{1120} H_{-2,0} + \frac{87\,399}{4480} H_{0,0} + \frac{8343}{4480} H_{2,0}         \\
        &&\KAS[4]          &   \frac{2187}{2240} H_{-4,0} - \frac{243}{2240} H_{-2,0} - \frac{7047}{2240} H_{0,0} - \frac{4617}{2240} H_{2,0}         \\
        \bottomrule
    \end{tabular}
    \caption{
        Exact expressions for the bull's head contribution $\mKdf^\BH=-\mDBH$; cf.\ \cref{tab:BH}, and also eq.~(4.31) in \rcite{Baeza:2023ljl}.
        The $H_{m,n,p}$ are defined in \cref{eq:Hmnp}, and $H_{m,n}\equiv H_{m,n,0}$.}
    \label{tab:BH-exact}
\end{table}

\subsection{Cutoff dependence}\label{app:cutoff}

\begin{figure}[tp]
    % \begin{comment} % Uncomment for fast version with pre-rendered plots
    \leftfudge
    \includegraphics[scale=\figscale]{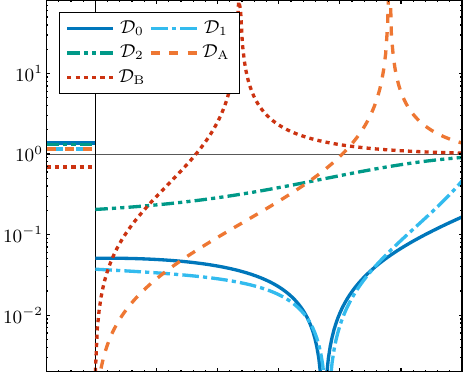}
    \includegraphics[scale=\figscale]{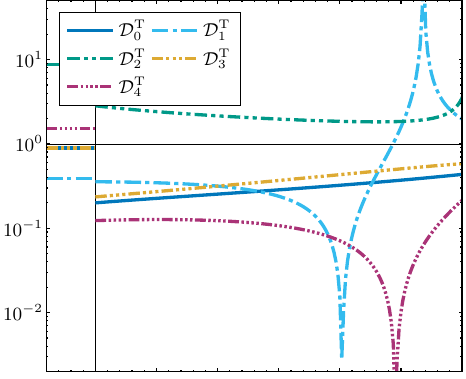}

    %\phantom{\footnotesize$^-$}%
    \leftfudge
    \includegraphics[scale=\figscale]{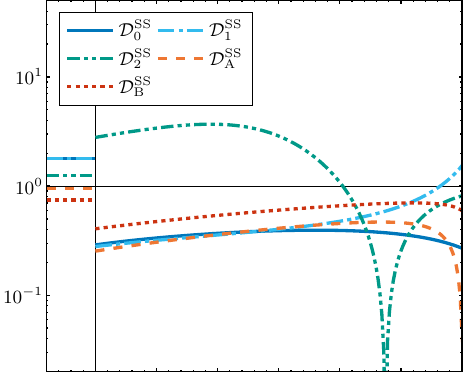}
    \includegraphics[scale=\figscale]{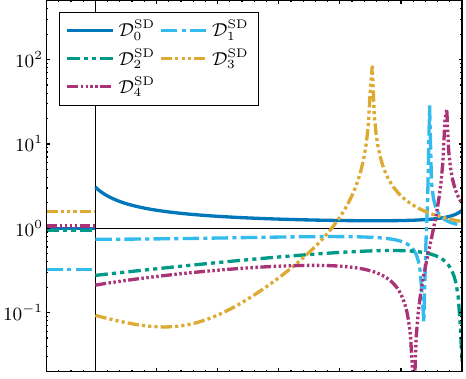}

    \leftfudge
    \includegraphics[scale=\figscale]{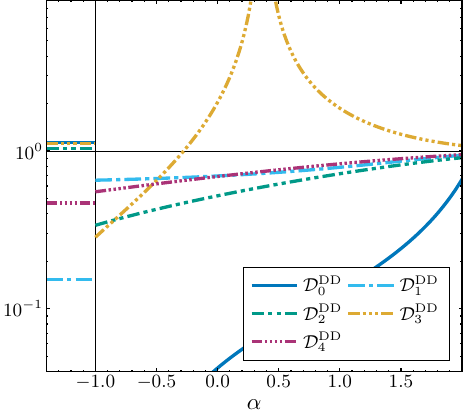}
    \midfudge
    \includegraphics[scale=\figscale]{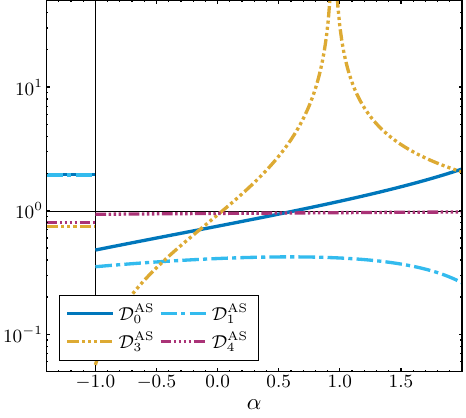}
    \caption{
        Illustration of the size of the numerical bull's head remainders $\cD_X$.
        Each line shows $|\cD_X/\cK^\BH_X|$, where $\cK_X^\BH$ is the complete bull's head contribution to $\cK_X$, plotted logarithmically as a function of the parameter $\alpha$ defined in \cref{eq:alpha}.
        The standard cutoff, \cref{eq:H-standard}, is recovered at $\alpha=-1$; the result using \cref{eq:H-symmetric} is shown to the left of that.
        The coefficients are grouped and displayed similarly to \cref{fig:results}.
        Horizontal lines are drawn at the ratio $1$, roughly indicating the border between `small remainders' and `large remainders'.
        Positive spikes indicate $\cK^\BH_X=0$, and negative spikes indicate $\cD_X=0$.}
    \label{fig:cutoff}
\end{figure}

The cutoff function $H(x)$ is arbitrary as long as it smoothly interpolates between \mbox{$H(x\leq0)=1$} and \mbox{$H(x\geq 1)=0$}, with the standard choice for $\Kdf$ being
\begin{equation}
    H(x) = \exp\Bigl[-\tfrac1x\, \exp\bigl(-\tfrac{1}{1-x}\bigr)\Bigr]\,,   \qquad 0 < x < 1\,.
    \label{eq:H-standard}
\end{equation}
A generalization corresponds to the replacement \cite{Briceno:2017tce}
\begin{equation}
    x \to 1+\frac4{3-\alpha}(x-1)\,,\qquad
    -1\leq \alpha < 3\,,
    \label{eq:alpha}
\end{equation}
with $\alpha=-1$ recovering~\cref{eq:H-standard}.
Another choice is the symmetric function
introduced in \rcite{Baeza:2023ljl}, 
\begin{equation}
    H(x) = \Big[1 + \exp\bigl(\tfrac{1}{x} - \tfrac{1}{1-x}\bigr)\Big]^{-1}\,,\qquad 0 < x < 1\,.
    \label{eq:H-symmetric}
\end{equation}
The numerical bull's head remainders $\cD_X$ defined in \cref{eq:DX} are the only cutoff-dependent terms in the threshold expansion of $\mKdf$.
\Cref{fig:cutoff} shows their dependence on the
choice of cutoff; the upper left panel displays similar information to fig.~11 in \rcite{Baeza:2023ljl}.

\newcommand{\addcutoffplot}[2]{%
    % \addplot[D#2line, mark=none, wide legend,%
    %         solid, line width=2mm, opacity=.1, %
    %         unbounded coords=jump, y filter/.expression={y > 0 ? nan : y}]%
    %     table[x=alpha, y expr={abs(\thisrow{K#1#2})}, col sep=tab]%
    %     {PlotData/\cutoff_remainder_rel.dat};
    \addplot[D#2line, mark=none, wide legend]%
        table[x=alpha, y expr={abs(\thisrow{K#1#2})}, col sep=tab]%
        {PlotData/\cutoff_remainder_rel.dat};%
    }

\section{Group-theoretical enumeration of operators}\label{app:group}
% Wherein Steve's operator-counting results find a home

In this appendix, we describe how group-theoretical considerations can be used to determine the number of kinematic operators at each order in the threshold expansion.

It follows from \cref{eq:threxp-def,eq:threxp-rel} that operators in this expansion can be written as products of the quantities $t_{ij}= (p_i - k_j)^2$, where we recall that $\{p_i\}$ and $\{k_j\}$ are, respectively, the final and initial momenta.
We are interested here in operators that are linear, quadratic and cubic in the $t_{ij}$.
Such products can be decomposed into irreps of the group $S_3'\times S_3$, where $S_3'$ and $S_3$ act, respectively, on the outgoing and incoming particle momenta.
As explained in \rcite{\isospin}, and recalled in the main text, operators of a given isospin lie in (in general a sum of) particular irreps of $S_3'\times S_3$.
By counting the number of different irreps that appear in products of the $t_{ij}$, we can determine the number of independent operators for each isospin at each order in the threshold expansion.
This is a more systematic approach than an explicit enumeration, and, indeed, has led to the discovery of additional operators, as noted in the main text.

In fact, there is an additional symmetry that must be considered, namely the PT symmetry that interchanges initial and final momenta (and which holds exactly in QCD).
Thus the operators must be decomposed into irreps of the group $G \equiv (S'_3\times S_3) \rtimes Z_2$, which involves a semidirect product.
To see this, we consider the defining representation, which acts on the vectors $\{k_1, k_2, k_3,p_1,p_2,p_3\}$.
The matrices forming the individual subgroups are, in block form,
\begin{equation}
S_3 \to \begin{pmatrix} S_3 & 0 \\ 0 & 1 \end{pmatrix},\quad
S'_3 \to \begin{pmatrix} 1 & 0 \\ 0 & S_3 \end{pmatrix},\quad
Z_2 \to  \begin{pmatrix} 0 & 1 \\ 1 & 0 \end{pmatrix}.
\end{equation}
Thus, we have
\begin{equation}
 \begin{pmatrix} 0 & 1 \\ 1 & 0 \end{pmatrix} \begin{pmatrix} S_3 & 0 \\ 0 & 1 \end{pmatrix}
  \begin{pmatrix} 0 & 1 \\ 1 & 0 \end{pmatrix} =
  \begin{pmatrix} 1 & 0 \\ 0 & S_3 \end{pmatrix},
\end{equation}
showing that the $Z_2$ acts nontrivially.
Our tasks are thus to determine the character table of $G$, and then to decompose operators of a given order in the $t_{ij}$ into irreps using the standard character decomposition.

We first recall some results for the permutation group.
The character table of $S_3$ is given in \cref{tab:S3}.
\begin{table}[htp]
\centering
\begin{tabular}{C||C|C|C}
\toprule
\text{Class} & 1 & (12) & (231) \\
\midrule
\text{Dim} & 1 & 3 & 2 \\
\midrule
1 & 1 & 1 & 1 \\
-1 & 1 & -1 & 1 \\
D & 2 & 0 & -1 \\
\bottomrule
\end{tabular}
\caption{
    Character table of $S_3$.}
\label{tab:S3}
\end{table}
Here, we label the irreps $1$, $-1$ (the sign or alternating irrep), and D, the standard or doublet irrep.
The character table of $S'_3\times S_3$ is then given in the standard way for tensor products, leading to 9 classes and 9 irreps.
Classes are given simply by combining classes for the individual $S_3$s, e.g., $\{(12)',(231)\}$, while irreps are products of the individual irreps, e.g., $(-1)'\otimes D$.
Characters or product irreps are simply the products of the characters of the individual irreps.

The inclusion of $Z_2$, which interchanges $S'_3$ and $S_3$, leads to some of the conjugacy classes of $S'_3\times S_3$ being combined, and introduces additional classes.
The combined classes are
\begin{align}
(12)_S &= \{ (12)',\;1\}+ \{1',\;(12)\}\,,
\\
(231)_S &= \{(231)',\;1\} + \{1',\;(231)\}\,,
\\
(12)(231)_S &= \{(12)',\;(231)\} + \{(231)',\;(12)\}\,,
\end{align}
which reduces the 9 classes of $S'_3\times S_3$ down to 6.
There are three additional classes, which involve the $Z_2$ element in combination with other transformations.
In the defining irrep, these are represented by elements of the form
\begin{align}
z_1 &= \left\{\begin{pmatrix} 0 & S_3 \\ S_3^{-1} & 0 \end{pmatrix} \right\},
\\
z_2 &= \left\{\begin{pmatrix} 0 & \text{even} \\ \text{odd} & 0 \end{pmatrix},
 \begin{pmatrix} 0 & \text{odd} \\ \text{even} & 0 \end{pmatrix} \right\},
\\
z_3 &= \left\{\begin{pmatrix} 0 & \text{even}_1 \\ \text{even}_2^{-1} &0\end{pmatrix},
 \begin{pmatrix} 0 & \text{odd}_1 \\ \text{odd}_2^{-1} & 0 \end{pmatrix} \right\}.
\end{align}
Here, $S_3$ means any element of the group, ``even" and ``odd" refer to an arbitrary even and odd element.
In particular, in the class $z_3$, ``even$_1$” and ``even$_2$” are arbitrary but different even elements, and similarly for ``odd${_1}$'' and ``odd${_2}$.''
These three classes have $6$, $18$, and $12$ elements, respectively.

\begin{table}[htp]
    \centering
\begin{tabular}{C||C|C|C|C|C|C|C|C|C}
\toprule
\text{Class} & 1',1 & (12)_S & (231)_S & (12)',(12)  
& (12)(231)_S & (231)',(231) & z_1 & z_2 & z_3 \\
\midrule
\text{Dim} & 1 &6 & 4 & 9 & 12 & 4 & 6 & 18 & 12\\
\midrule
1^+_+ & 1 & 1 & 1 & 1 & 1 & 1 & 1 & 1 & 1 \\
1^-_+ & 1 &1 & 1 & 1 &1 & 1 & -1 & -1 & -1 \\
1_-^+ & 1 & -1 & 1 & 1 & -1 & 1 & 1 & -1 & 1\\
1_-^- & 1 & -1 & 1 & 1 & -1 & 1 & -1 & 1 & -1\\
2^+_+ & 2 & 0 & 2  & -2 & 0 & 2 & 0 & 0 & 0\\
\SD^+_+ & 4 & 2 & 1 & 0 & -1 & -2 & 0 & 0 & 0\\
\SD^+_- & 4 & -2 & 1 & 0 & 1 & -2 & 0 & 0 & 0\\
\DD^+_+ & 4 & 0 & -2& 0 & 0 & 1 & 2 & 0 & -1\\
\DD^-_+ & 4 & 0 & -2 & 0 & 0 & 1 & -2 & 0 & 1 \\
\bottomrule
\end{tabular}
\caption{
    Character table of $(S'_3\times S_3)\rtimes Z_2$.}
\label{tab:charfull}
\end{table}

The character table of $G$ is given in \cref{tab:charfull}.
The notation for irreps is as follows: 
SD is a combination of a singlet from $S_3'$ and doublet from $S_3$, together with the PT-conjugate;
DD is the combination of two doublets;
the superscript $\pm$ indicates the sign obtained under the action of $Z_2$;
and the subscript $\pm$ indicates the sign obtained if the combined parity of the $S_3'\times S_3$ permutation is odd.
The notation SD and DD mirrors that used in the main text, while the singlet (here called ``$1$”) is denoted as SS in the main text.

The mapping from isospin irreps to those of $G$ has been explained in \rcite{\isospin}, and is recalled in \cref{sec:threxp}.
Operators with $I=3$ lie in the singlet irrep, $1^+_+$, those with $I=2$ lie in the $\DD^+_+$ irrep, those with $I=1$ lie in the singlet, $\SD^+_+$ and $\DD^+_+$ irreps, and those with $I=0$ lie in the $1^+_-$ irrep.

We now decompose operators composed of the $t_{ij}$.
At linear order, there are 9 such operators, and the character vector is $\{9,3,0,1,0,0,3,1,0\}$, which decomposes as
\begin{equation}
(1^+_+) + (\SD^+_+) + (\DD^+_+)\,.
\end{equation}
It follows that, at this order, there is a single contribution to $I=3$ [that given by the $\Kisoone$ term in \cref{eq:threxp-I3}], a single contribution to $I=2$ [that given by the $\KT[0]$ term in \cref{eq:threxp-I2}], and three contributions to $I=1$ [given by the $\KSS[1]$ term in \cref{eq:threxp-I1SS}, the $\KDD[0]$ term in \cref{eq:threxp-I1DD}, and the $\KSD[0]$ term in \cref{eq:threxp-I1SD}].
There are no contributions to $I=0$ at this order.

Moving now to quadratic order, there are $9\times 10/2=45$ distinct terms of the form $t_{ij} t_{k\ell}$.
We find the character vector to be $\{45,9,0,5,0,0,9,1,0\}$\,, which decomposes as
\begin{equation}
3(1^+_+) + (1^-_+) + (1^+_-) + 4(\SD^+_+) + 4(\DD^+_+)  + (\SD^+_-) + (\DD^-_+)\,.
\end{equation}
Thus there are three singlets, leading to the the $\Kisotwo$, $\KA$, and $\KB$ ($I=3$) terms in \cref{eq:threxp-I3}, and the corresponding three $I=1$ SS terms in \cref{eq:threxp-I1SS}.
Similarly, there are four DD terms in $I=2$ and $I=1$, given by the $\KT[1,2,3,4]$ terms in \cref{eq:threxp-I2}, and the corresponding terms in \cref{eq:threxp-I1DD}.
There are four SD terms in $I=1$, given by the $\KSD[1,2,3,4]$ terms in \cref{eq:threxp-I1SD}.
Terms of this order were not considered in \rcite{\isospin}.
Finally, there is a single $I=0$ contribution, with coefficient $\KAS$ in \cref{eq:threxp-I0}.

Moving lastly to cubic order, there are $9\times 10 \times 11/6=165$ distinct terms that are cubic in the $t_{ij}$.
With some effort, one finds that the character vector is $\{165,19,3,5,1,3,19,1,1\}$.
For example, in the class $(12)(231)_S$, picking the element $(12)'(231)$, only the term $t_{31} t_{32}  t_{33}$ is invariant.
The most tricky case is the class $z_3$.
Picking the element where ${\rm even}=1$ and ${\rm even}'=(231)$, the single invariant term is $ t_{12} t_{31} t_{23}$.

The decomposition of this character vector is
\begin{equation}
7(1^+_+) + 3(1^-_+) + 3 (1^+_-) + 4 (2^+_+) +
12(\SD^+_+) + 12(\DD^+_+)  + 6(\SD^+_-) + 6(\DD^-_+)\,.
\end{equation}
Given the large numbers of irreps, we focus on the $I=0$ case, for which we learn that there are three independent $1^+_-$ irreps, and thus three coefficients at this order.
These correspond to the coefficients $\KAS[1]$, $\KAS[3]$, and $\KAS[4]$, in \cref{eq:threxp-I0}.
The final coefficient was missed in the enumeration of \rcite{\isospin}.
As a side note, we observe that there are seven independent $1^+_+$ irreps, one less than the eight explicit cubic forms given in \rcite{\dwave}.
We have confirmed that there is one linear relation between these eight forms.

\section{Families of single-parameter kinematic configurations}\label{app:families}

In \rcite{Baeza:2023ljl}, we performed numerical checks using several families of single-parameter kinematic configurations described in appendix~D of that paper.
These families have symmetries that lead to vanishing results when combined with the momentum-exchange antisymmetry present in some non-maximal isospin channels.
Therefore, different families are needed for the analysis carried out here in \cref{sec:validity} and numerical cross-checks of the non-OPE and bull's head subtraction results.

A simple way to choose three incoming momenta satisfying $\bm k_1+\bm k_2+\bm k_3=\bm 0$ with all $|\bm k_i|$ different is, for $n>2$,
\begin{equation}
    \bm k_1=p\,(1,0,0)\,,\qquad
    \bm k_2=p\,\bigl(-1+\tfrac 1n,1,0\bigr)\,,\qquad
    \bm k_3=p\,\bigl(-\tfrac 1n,-1,0\bigr)\,,
\end{equation}
where $p$ is the single continuous parameter that governs the kinematics.
The outgoing momenta $\bm p_i$ can be generated similarly and then rotated using some orthogonal matrix $U$.
This matrix can also include reflections, e.g., swapping the $y$ and $z$ components of the momenta.
Various choices of $n$ and $U$ result in a range of families that are sufficiently distinct and non-symmetric to study all components of $\mKdf$ as functions of $p$.

To obtain \cref{fig:convergence,fig:partial-waves}, we used the kinematic configuration obtained with $n=4$ and
\begin{equation}
    U = 
    \renewcommand*{\arraystretch}{1.5}
    \begin{pmatrix}
     \frac{1}{2} & -\frac{\sqrt{3}}{2} & 0 \\
     \frac{3}{4} & \frac{\sqrt{3}}{4} & -\frac{1}{2} \\
     \frac{\sqrt{3}}{4} & \frac{1}{4} & \frac{\sqrt{3}}{2} \\
    \end{pmatrix};
\end{equation}
this reads
\begin{alignat}{3}
\bm k_1&=p\,(1,0,0)\,,\qquad&\bm p_1&=p\,\bigl(\tfrac{1}{2},\tfrac{3}{4},\tfrac{\sqrt{3}}{4}\bigr)\,,\notag\\
\bm k_2&=p\,\bigl(-\tfrac 34,1,0\bigr)\,,\qquad&\bm p_2&=p\,\bigl(-\tfrac{\sqrt{3}}{2}-\tfrac{3}{8},\tfrac{\sqrt{3}}{4}-\tfrac{9}{16},-\tfrac{3\sqrt{3}}{16}+\tfrac{1}{4}\bigr)\,,\label{eq:P1}\\
\bm k_3&=p\,\bigl(-\tfrac 14,-1,0\bigr)\,,\qquad&\bm p_3&=p\,\bigl(\tfrac{\sqrt{3}}{2}-\tfrac{1}{8},-\tfrac{\sqrt{3}}{4}-\tfrac{3}{16},-\tfrac{\sqrt{3}}{16}-\tfrac{1}{4}\bigr)\,.\notag
\end{alignat}
This configuration is rather general since the outgoing momenta do not lie in the same plane as the incoming ones.

On the other hand, to numerically calculate the threshold expansion at NLO as a cross-check, three or four independent configurations need to be used, depending on the isospin channel.
It is rather straightforward to obtain two additional independent sets of momenta by changing a sign in the $y$ component or swapping the $y$ and $z$ components of the incoming momenta,
\begin{equation}
\begin{alignedat}{3}
\bm k_2'&=p\,\bigl(-\tfrac 34,-1,0\bigr)\,,\qquad&
\bm k_2''&=p\,\bigl(-\tfrac 34,0,1\bigr)\,,\\
\bm k_3'&=p\,\bigl(-\tfrac 14,1,0\bigr)\,,&
\bm k_3''&=p\,\bigl(-\tfrac 14,0,-1\bigr)\,,
\end{alignedat}
\end{equation}
while keeping $\bm p_1$, $\bm p_2$, $\bm p_3$, and $\bm k_1$ as in \cref{eq:P1}.
Naturally, the resulting three configurations will have many kinematic variables in common.
To extract all the threshold parameters, these need to be complemented with another configuration that generates a set of invariants that does not overlap with the rest to such an extent.
We thus used a variant of the $n=3$ case with $U$ being simply the rotation by $30^\circ$ around the $z$-axis, which, for completeness, reads
\begin{alignat}{3}
\bm k_1&=p\,(1,0,0)\,,\qquad&\bm p_1&=p\,\bigl(\tfrac{\sqrt{3}}{2},\tfrac{1}{2},0\bigr)\,,\notag\\
\bm k_2&=p\,\bigl(-\tfrac 13,1,0\bigr)\,,\qquad&\bm p_2&=p\,\bigl(-\tfrac{1}{2}-\tfrac{1}{2 \sqrt{3}},-\tfrac{1}{6}+\tfrac{\sqrt{3}}{2},0\bigr)\,,\\
\bm k_3&=p\,\bigl(-\tfrac 23,-1,0\bigr)\,,\qquad&\bm p_3&=p\,\bigl(\tfrac{1}{2}-\tfrac{1}{\sqrt{3}},-\tfrac{1}{3}-\tfrac{\sqrt{3}}{2},0\bigr)\,.\notag
\end{alignat}

\addcontentsline{toc}{section}{References}
\renewcommand\raggedright{}
%\bibliography{references}

\providecommand{\href}[2]{#2}\begingroup\raggedright\endgroup

\end{document}